\documentclass[12pt]{book}
\usepackage[a4paper, top=1in, bottom=1in, left=1in, right=1in, includeheadfoot]{geometry}
\usepackage[english]{babel}
\usepackage{fancyhdr}
\usepackage{url}
\usepackage{graphicx}
\usepackage{amssymb,amsmath}
\usepackage{parskip}
\usepackage{titlepic}

\begin{document}

\author{Luigi Foschini}

\title{{\bf Yet Another Introduction to}\\
{\bf Relativistic Astrophysics}}

\titlepic{\includegraphics[width=\textwidth]{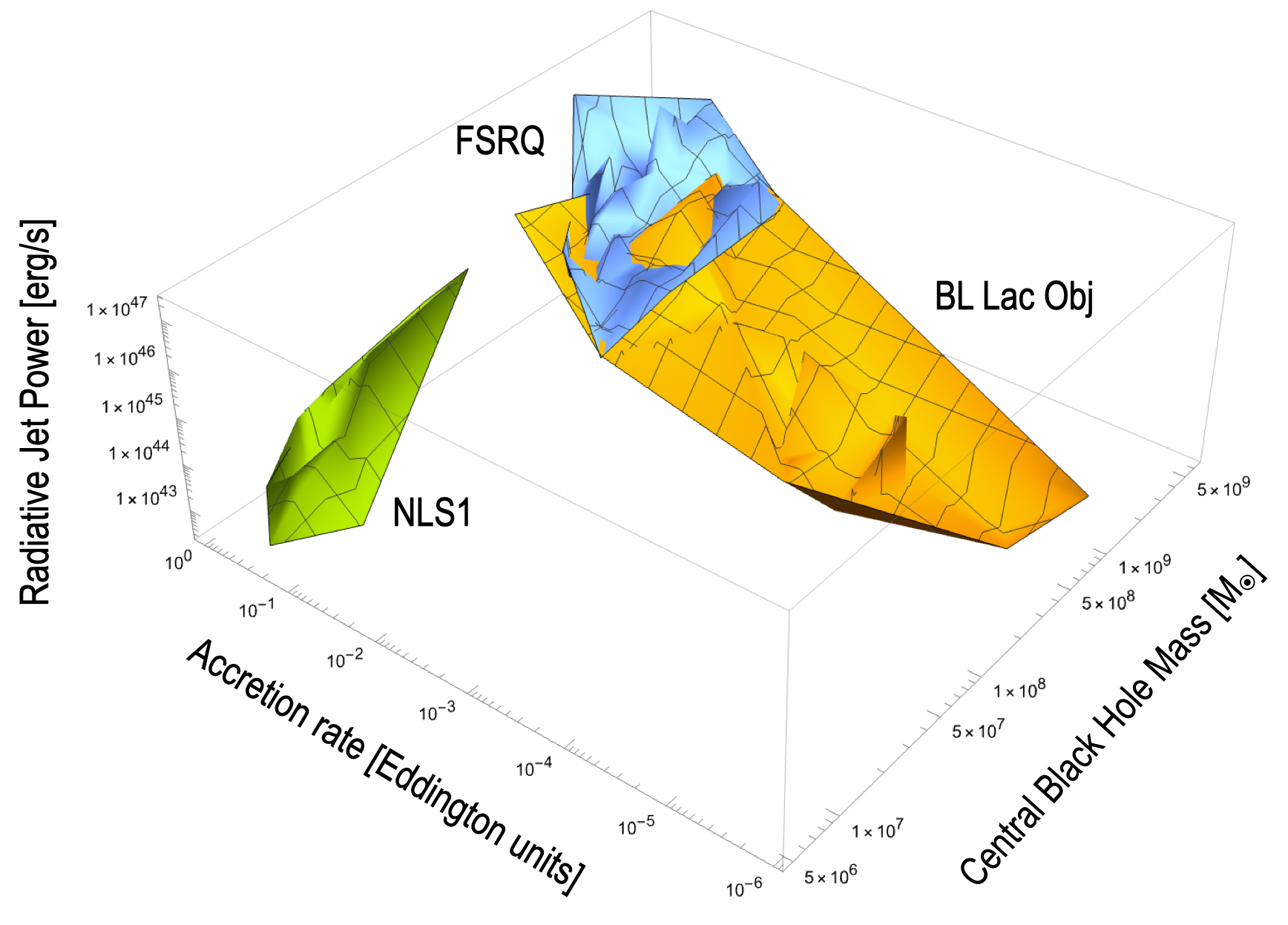}}

\date{}

\maketitle

\chapter*{Foreword\markboth{Foreword}{}}

\emph{Relativistic Astrophysics deals with what we can say about the Universe by using the language of the theory of relativity. This peculiar type of language is strongly required when some conditions are verified, such as velocities close to that of light (for example, relativistic jets), and extreme energies or gravitational fields (for example, spacetime around black holes). This research field has wide overlaps with other fields, such as high-energy astrophysics and cosmology. Therefore, it is necessary to make a selection of topics. These notes have been prepared for a series of short lessons for PhD students held at the Department of Physics and Astronomy of the University of Padova in March 2017. Ten hours of lessons can give only a taste of what is available. Given my researches, I selected mainly topics about black holes and jets, particularly about supermassive objects in the centre of galaxies. Anyway, the large bibliography should give enough information to widen the necessarily limited horizon of these notes. In addition, I have not imported many figures from other works, but I cited them, so that I hope the reader could be further motivated to read more works.}

\emph{A large part of these notes deals with well-grounded knowledge. For relativity theory, I made use of many well-known textbooks, particularly Rindler \cite{RINDLER}, Cheng \cite{CHENG}, Good \cite{GOOD}, and d'Inverno \cite{INVERNO}. I refer to them for more details. I felt the need to write these notes to emphasize the relationship between science and language, which is often hidden or neglected in standard textbooks. I think that this could be of great help in understanding some physical concepts, particularly in fields outside our everyday life, such as relativity and quantum mechanics. I have written a book on these topics, ``Scienza e linguaggio'' \cite{FOSCHINI2}, where is possible to find more details. For those who do not know Italian language, I refer to the three short essays in English \cite{FOSCHINI3,FOSCHINI4,FOSCHINI5}. Please note that part of these notes is a renewed elaboration of previous notes: chapters 1-4 have been adapted from my lecture notes of the Summer School on ``General Relativity and Relativistic Astrophysics'' held at the Observatory of the Valle d'Aosta Region (Lignan, Aosta, Italy) on July 18-22, 2016; chapter 8 derived from the notes of my lessons at the Summer School on ``Gamma-ray Astrophysics and Multifrequency: Data Analysis and Astroparticle Problems'', held at the Department of Physics of the University of Perugia (Italy) on July 3-7, 2006.} 

\emph{The present version (third) has been updated according to the most recent research about jetted AGN. As known, the error is intrinsic with doing and these notes are certainly not an exception. Therefore, I invite the reader, who will find errors or typos, to write me.}

\emph{I would like to thank Stefano Ciroi for having invited me to hold the course in the beautiful University of Padova, permeated by history (it was established in 1222). Many thanks to Valentina Braito and Paola Grandi for useful hints, and to Marco Berton for a critical reading of the draft. I would also like to thank Alessandra Zorzi for her invaluable help with the administrative side. Thanks for comments also to Enrico Congiu, Igor Kulikov, and Giovanni La Mura.}

\emph{Merate, \today}
\vskip 12pt

\noindent
dr. Luigi Foschini\\
Brera Astronomical Observatory\\
National Institute of Astrophysics (INAF)\\
Via E. Bianchi, 46\\
23807 Merate (LC), Italy\\
email: \texttt{luigi.foschini@inaf.it}\\
home page: \url{http://www.brera.inaf.it/utenti/foschini/}\\


\chapter{Classical Relativity (Galilean)}

Classical mechanics was born with Galileo and Newton. It marks several novelties with respect to the science before Renaissance, particularly the insertion of quantity and measurement. The laws of mechanics were defined within reference frames, in order to allow the definition of physical quantities such as velocity, acceleration, impulse, ... and, in turn, their measurability. These frames were represented by a set of three Cartesian orthogonal axes $(x,y,z)$. The position of one point within one of these frames is indicated by the distances from the three axes. For example, $(10\,\mathrm{cm},15\,\mathrm{cm},40\,\mathrm{cm})$. It is also necessary to add a fourth number, the time of measurement. 

Frames could move one with respect to the other either with constant and uniform velocity (\emph{inertial frames}) or with acceleration (\emph{non-inertial frames}). Newton made also the hypothesis of an absolute reference frame, located in the centre of mass of the Solar System. Also time was absolute, and did not depend on any frame.

The existence of different reference frames made it necessary to understand if and how physical laws change with respect to frames. In case of changing laws, that is were physics depending on space and time, then there would be no science. Therefore, it is necessary that equations of physics are \emph{invariant} with respect to reference frames.

\begin{figure}[t]
\begin{center}
\includegraphics[scale=0.35]{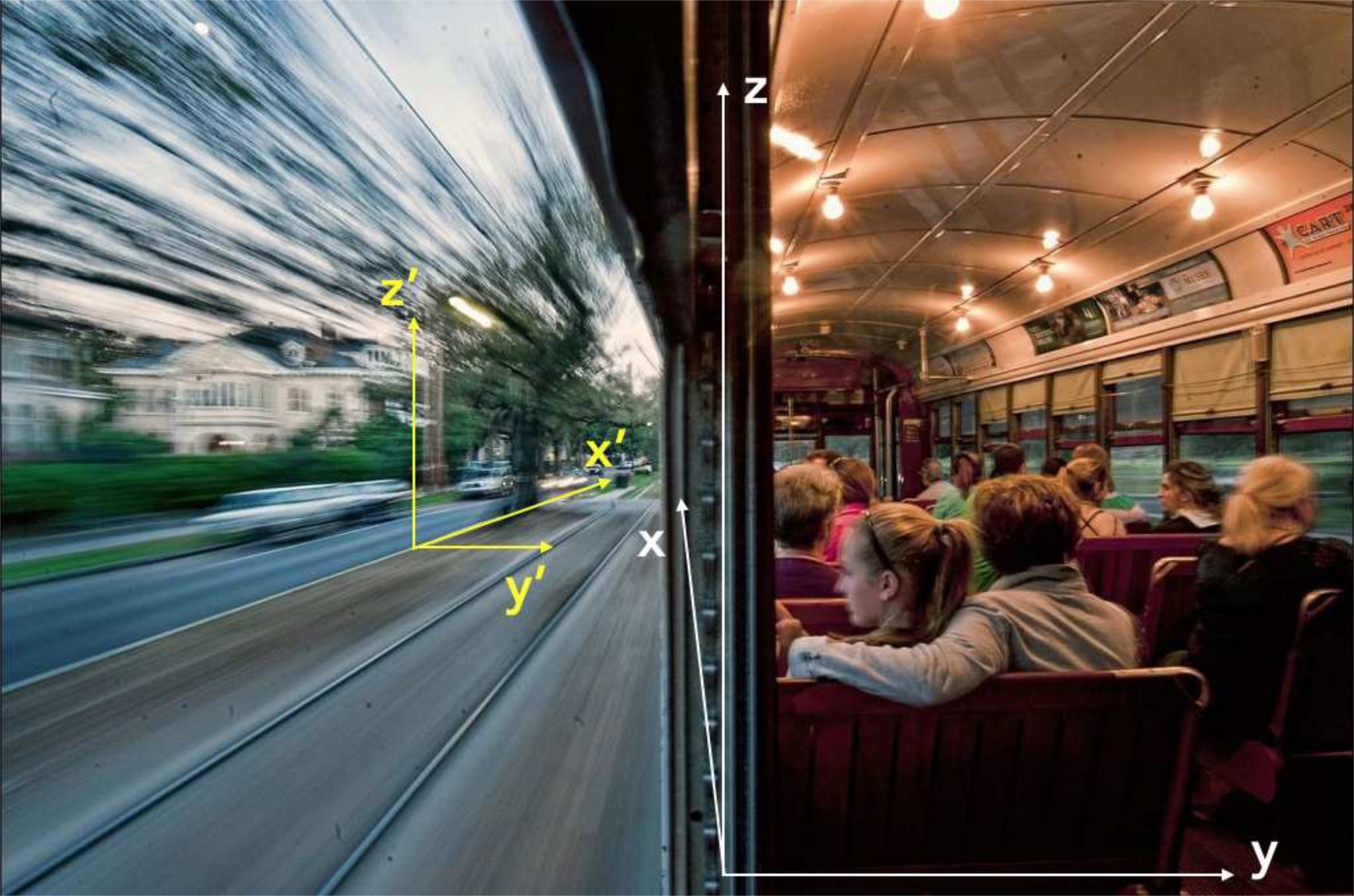}
\caption{Classical relativity. Two reference frames moving one with respect to the other with constant and uniform velocity are equivalent. (Adapted from an image found on \emph{Reddit}, user \texttt{kevlarr}, 2015)}
\label{fig:relagal}
\end{center}
\end{figure}

Let us consider two reference frames, one in motion with respect to the other with constant and uniform velocity, that is two inertial frames. We are free to set our main reference frame, because the transformation from one frame to the other is symmetrical: one frame is equal to the other, no preference is required on physical basis, just our easiness to calculate transformation under certain conditions. One example is shown in Fig.~\ref{fig:relagal}: one tram is moving with constant and uniform velocity. The road is often adopted as main reference frame and the tram is the moving frame. However, I have selected this image because it shows the opposite case, that is the passengers on the tram are at rest while the road is moving. There is nothing strange from the physical point of view, because -- as stated above -- the transformations between two inertial reference frames are symmetrical. Were the windows of the tram perfectly obscured, were the vehicle moving slow without accelerations, without sudden movements of any kind, then the passengers would not have any feeling of motion.

Let us consider two inertial frames $S=(x,y,z,t)$ and $S'=(x',y',z',t')$. The latter is moving with respect to the former with velocity $v$ costant, uniform, and directed along the $x$ axis (Fig.~\ref{fig:relagal}). The coordinates of one frame can be transformed into the other by taking into account the space covered during a certain time with the velocity $v$:

\begin{equation}
\begin{aligned}
x' & = x - vt\\
y' & = y\\
z' & = z\\
t' & = t\\
\end{aligned}
\label{eq:galileo}
\end{equation}

This is the simplest case, with the velocity exactly directed along one of the orthogonal axes of the reference frame. In the case of an arbitrary direction, it is necessary to decompose the velocity in three components $(v_x,v_y,v_z)$, each one directed along the three axes. Therefore, Eqs.~(\ref{eq:galileo}) become:

\begin{equation}
\begin{aligned}
x' & = x - v_xt\\
y' & = y - v_yt\\
z' & = z - v_zt\\
t' & = t\\
\end{aligned}
\label{eq:galileo2}
\end{equation}

It is worth noting that the time never changes, because it is the only absolute reference in classical mechanics. One could also think that the two frames are not parallel each other, so that the transformations become even more difficult, by adding also different spatial components in the change of coordinates. As usual, complex details are left to the students as ``easy'' exercise...

Accelerated frames are different, and an asymmetry is present in changing from one frame to the other. According to classical mechanics, the effect of a force applied to a mass is to accelerate it. Written in mathematical language:

\begin{equation}
\mathbf{F} = m\mathbf{a}
\label{eq:newton}
\end{equation}

The bold-face symbols indicate a \emph{vector}\footnote{Another common notation for vectors is to use an arrow above the letter, that is $\vec{F}=m\vec{a}$.}, a mathematical entity characterised by magnitude, direction, and sign\footnote{It is worth noting that in English, the word direction is also associated with a sign of motion. Therefore, on English textbooks is common to find that a vector has a magnitude and a direction only. In Italian, the word ``direzione'' means only the position of a line in a space. A particle moving along a ``direzione'' could go in one or in the opposite way. To define which way, there is need of a third information, which is the ``verso'' that I translated as sign to keep the whole information.}, to be compared with a \emph{scalar}, which has only magnitude (example, temperature). Back to the velocity definition written above, instead of writing about a velocity directed along the $x$ axis, I could have written a vector $\mathbf{v}$ with one individual component $v_x$ along the $x$ axis, that is $\mathbf{v}=(v_x,0,0)$. In the case of velocity with arbitrary direction, it is possible to write $\mathbf{v}=(v_x,v_y,v_z)$. Also the position of a body can be indicated by a position vector, which -- for example -- indicate the distance from the origin of the three orthogonal axes of the frame: $\mathbf{r}=(x,y,z)$. Thus, the vector notation allows to write in a compact language the Eqs.~(\ref{eq:galileo}) and (\ref{eq:galileo2}):

\begin{equation}
\mathbf{r'} = \mathbf{r} + \mathbf{v}t
\label{eq:galileo3}
\end{equation}

Time remains still outside, because it is a scalar quantity and does not depend on any reference frame. I stress that the vector is a mathematical entity in a three dimensional Euclidean space and does not take into account the time. Vectors can change with time, but the latter is not included into the definition of the mathematical entity. This is an important point, which will return again later when dealing with special and general relativity.

If an object is moving with respect to one frame with velocity (for example, one passenger in the tram is walking on the aisle), then the transformation of the velocity from one frame to the other is obtained by differentiating with respect to time the Eq.~(\ref{eq:galileo3}). By writing $\mathbf{u} = d\mathbf{r}/dt$ the velocity of the passenger on the tram ($S$), then the velocity measured on the road ($S'$) is:

\begin{equation}
\mathbf{u'} = \mathbf{u} + \mathbf{v}
\label{eq:galileo4}
\end{equation}

By differentiating again with respect to time, one obtains the transformation of accelerations, which is exactly null, because all the velocities are constant and uniform.

Let us back to accelerated reference frames and to the Eq.~(\ref{eq:newton}). As written above, the acceleration inserts an asymmetry, because it is the effect of a force on a mass. Let us consider the tram accelerating: passengers -- that in the inertial case did not feel the motion -- now are compressed against their own seats. They understand that they -- and not the road -- are moving: asymmetry! This time, the change from one frame to the other is based on the acceleration:

\begin{equation}
\frac{d\mathbf{u'}}{dt} = \frac{d\mathbf{u}}{dt} + \mathbf{a}
\label{eq:newton2}
\end{equation}

where $\mathbf{a}$ is the acceleration of the tram frame. By multiplying for the mass $m$ of the physical object (for example, the passenger), one obtains:

\begin{equation}
m\frac{d\mathbf{u'}}{dt} = m\frac{d\mathbf{u}}{dt} + m\mathbf{a}
\label{eq:newton3}
\end{equation}

which, in turn, is a sum of forces according to the Eq.~(\ref{eq:newton}). Therefore, the passenger experiences an additional force equal to $m\mathbf{a}$, named \emph{inertial force}, which feels like a compression against the seat. An observer on the road sees the tram accelerating, but does not feel any force.


\chapter{Special Relativity}

\section{FitzGerald-Lorentz transformations}
Classical relativity underwent a crisis with the development of the James Clerk Maxwell's theory of electromagnetism during the XIX century. As Galilean transformations were applied to the electromagnetism, some problems popped up. G. F. FitzGerald (1889) and H. A. Lorentz (1892), independently, found a new set of transformations conserving the invariance of Maxwell's equations, but the price was the loss of insight. Transformations of Eq.~(\ref{eq:galileo}) became as follows:

\begin{equation}
\begin{aligned}
x' & = \frac{x - vt}{\sqrt{1-(\frac{v}{c})^2}}\\
y' & = y\\
z' & = z\\
t' & = \frac{t - \frac{v}{c^2}x}{\sqrt{1-(\frac{v}{c})^2}}\\
\end{aligned}
\label{eq:lorentz}
\end{equation}

The counterintuitive effect were that, as the velocity was comparable with that of the light in the empty space, the spatial and temporal lengths were not conserved. Einstein was the first to understand that there was nothing wrong. He developed the special theory of relativity on the basis of two postulates:

\begin{enumerate}
\item \emph{Postulate of relativity:} inertial reference frames are equivalent each other;
\item \emph{Postulate of the constant light velocity:} the light has a velocity in the empty space of $c\sim 3\times 10^5$~km/s, which is constant in any direction and in any inertial reference frame\footnote{Today, the speed of light in vacuum is a fundamental constant equal to $299792458$ m/s. There is no error, because it is no more measured. The metre is now a derived quantity, defined as the path of light in vacuum during one second. }.
\end{enumerate}

If one observer (reference frame) $S$ sees one particle in inertial motion (that is with constant and uniform velocity), then another inertial observer $S'$ sees the particle in constant and uniform motion (Postulate 1). One can transform the particle position in $S$ into the position in $S'$, and vice versa. For the sake of simplicity, let us suppose that the relative velocity of the two frames was directed only along the $x$ axis. Now, let us use the Postulate 2. At a certain time $t=t'=0$, the two frames coincide. The particle emits an electromagnetic wave, which in turn moves with velocity $c$ independently on the reference frame. Therefore, its wave front is a sphere with equation:

\begin{equation}
x^2 + y^2 + z^2 - c^2t^2 = constant
\label{eq:sfera1}
\end{equation}

The same holds for $S'$:

\begin{equation}
x'^2 + y'^2 + z'^2 - c^2t'^2 = constant
\label{eq:sfera2}
\end{equation}

When the two frames coincide, $v=0$, and the two constants are the same, which implies:

\begin{equation}
x^2 + y^2 + z^2 - c^2t^2 = x'^2 + y'^2 + z'^2 - c^2t'^2
\label{eq:sfera3}
\end{equation}

According to the hypothesis of a velocity directed only along the $x$ axis, $v=v(x,t)$ (cf Eq. \ref{eq:lorentz}), the previous equation becomes:

\begin{equation}
x^2 - c^2t^2 = x'^2 - c^2t'^2
\label{eq:sfera4}
\end{equation}

By changing variables as follows:

\begin{equation}
\begin{aligned}
T & = ict\\
T' & = ict'\\
\end{aligned}
\label{eq:cambiovar}
\end{equation}

and by substituting Eqs~(\ref{eq:cambiovar}) into Eq.~(\ref{eq:sfera4}), one obtains:

\begin{equation}
x^2 + T^2 = x'^2 + T'^2
\label{eq:sfera5}
\end{equation}

which represents the distance from the origin of one point in a two-dimensional space $(x,T)$. It is invariant with respect to a rotation. Let us write $\theta$ the counterclockwise rotation angle of $(x',T')$ with respect to $(x,T)$. Then, the coordinate transformation after a rotation is:

\begin{equation}
\begin{aligned}
x' & = x\cos\theta + T\sin\theta\\
T' & = -x\sin\theta + T\cos\theta\\
\end{aligned}
\label{eq:rotazione}
\end{equation}

If $x'=0$, then, from Eqs. (\ref{eq:lorentz}) and (\ref{eq:cambiovar}), one obtains $x=vt=vT/ic$. The first equation of the system ~(\ref{eq:rotazione}) becomes:

\begin{equation}
0 = \frac{vT}{ic}\cos\theta + T\sin\theta
\label{eq:rotazione1}
\end{equation}

By dividing for $T\cos\theta$, one has:

\begin{equation}
0 = \frac{v}{ic} + \tan\theta \, \rightarrow \, \tan\theta = i\frac{v}{c}
\label{eq:rotazione2}
\end{equation}

By using trigonometry, one can write $\cos\theta$ as a function of $\tan\theta$:

\begin{equation}
\cos\theta = \frac{1}{\sec\theta} = \frac{1}{\sqrt{1+\tan^2\theta}} = \frac{1}{\sqrt{1-\frac{v^2}{c^2}}}
\label{eq:trigo}
\end{equation}

The factor $\cos\theta$ is commonly indicated by using the Greek letter $\Gamma$ or $\gamma$, and it is called \emph{Lorentz factor}. Moreover, as the quantity $v/c$ is often used, it is usual to adopt the Greek letter $\beta$. Therefore, the Eq.~(\ref{eq:trigo}) can now be rewritten as:

\begin{equation}
\gamma = \frac{1}{\sqrt{1-\beta^2}}
\label{eq:florentz}
\end{equation}

Therefore, it is possible to rewrite the system~(\ref{eq:rotazione}) as follows:

\begin{equation}
\begin{split}
x' & = x\cos\theta + T\sin\theta =\\
   & = \cos\theta (x+T\tan\theta) =\\
   & = \gamma(x+T\frac{iv}{c}) =\\
   & = \gamma(x+ict\frac{iv}{c}) = \\
   & = \gamma(x-vt)
\end{split}
\label{eq:trasfo1}
\end{equation}

and:

\begin{equation}
\begin{split}
T' & = -x\sin\theta + T\cos\theta =\\
   & = \cos\theta (-x\tan\theta + T) \\
ict' & = \gamma(-x\frac{iv}{c} + ict) = \\
  t' & = \gamma(t-\frac{xv}{c^2})
\end{split}
\label{eq:trasfo2}
\end{equation}

Equations~(\ref{eq:trasfo1}) and (\ref{eq:trasfo2}) are just two of the four equations of FitzGerald-Lorentz (Eq.~\ref{eq:lorentz}), where I have emphasised the two new parameters: $\gamma$ (or $\Gamma$) and $\beta$.

\section{Properties of Lorentz transformations}
\begin{figure}[t]
\begin{center}
\includegraphics[angle=270,scale=0.5]{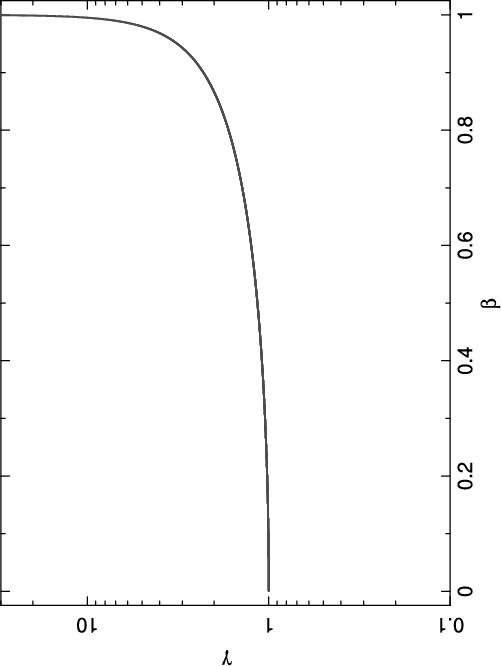}
\caption{Lorentz factor $\gamma$ as a function of the normalised velocity $\beta$.}
\label{fig:lorentzgrap}
\end{center}
\end{figure}

The Lorentz factor $\gamma$ is of paramount importance, because it shows how much the special relativity is different from the classical one. Fig.~\ref{fig:lorentzgrap} displays the value of $\gamma$ as a function of $\beta$, which in turn is included between $0$ (null velocity) and $1$ (velocity equal to that of light in the empty space).

In the case of velocity much smaller than that of light ($\beta\rightarrow0$), which is a typical velocity on human scale\footnote{It is worth noting that $\beta\rightarrow 0$ can be obtained by thinking either $v\rightarrow 0$ or $c\rightarrow \infty$, as some scientists living at the Newton's epoch thought.}, the Lorentz transformations (\ref{eq:trasfo1}) and (\ref{eq:trasfo2}) become the equations of classical relativity (\ref{eq:galileo}). Therefore, it is apparent that classical mechanics is an approximation of special relativity, when velocity is much smaller than that of light. The new theory does not cancel the old one, but it enlarges the domain of validity.

As $\beta$ and $\gamma$ increase, the time is no more the same when changing reference frame, as it is clear looking at Eq.~(\ref{eq:trasfo2}). Simultaneous events in one reference frame, do not have the same time coordinate in another frame (\emph{relativity of simultaneity}). Space and time coordinates are now interwoven into a four-dimensional spacetime, named after Hermann Minkowski. A change of position along the $x$ axis with velocity $v\rightarrow c$ means a rotation within the spacetime $(x,T)$ with an angle $\theta$ defined as $\tan\theta = iv/c$. I would like to stress that the Minkowski spacetime is an Euclidean (flat) four-dimensional space, unlike in the case of general relativity. Time is taken into account as a fourth dimension by using the light speed in empty space as a conversion factor ($ct$). This implies a symmetry: it is possible to change $x$ with $ct$ in the Lorentz transformations, having care of substituting $v$ with $-v$. 

If $\beta \rightarrow 1$, then $\gamma \rightarrow \infty$, which has significant implications to be explored in the next sections. If $\beta > 1$ ($v>c$), then $\gamma$ has no more real solutions. It has complex solutions as the result of the square root of a negative number\footnote{Some people speculated about the existence of particles moving faster than light called tachyons (e.g. \cite{RECAMI}).}.

\section{Side effects of Lorentz transformations}
Counterintuitive effects were the main obstacle hampering the understanding of Lorentz transformations. One was the \emph{length contraction}. Let us consider an inertial reference frame $S'$ moving with velocity $v$ with respect to another inertial frame $S$. A pole in $S'$ has a length $l_0$ as measured between the two ends of coordinates $x_{\rm A}'$ and $x_{\rm B}'$ (as usual, for the sake of simplicity, I will consider one coordinate only). The Lorentz transformations between the two frames $S$ and $S'$ are:

\begin{equation}
\begin{aligned}
x_{\rm A}' & = \gamma(x_{\rm A} - vt_{\rm A})\\
x_{\rm B}' & = \gamma(x_{\rm B} - vt_{\rm B})
\end{aligned}
\label{eq:contraz1}
\end{equation}

The rest frame length, as measured in $S'$, is:

\begin{equation}
l_0 = x_{\rm A}' - x_{\rm B}'
\label{eq:contraz2}
\end{equation}

while the length of the pole as measured in $S$ at a time $t=t_{\rm A}=t_{\rm B}$ is:

\begin{equation}
l = x_{\rm A} - x_{\rm B}
\label{eq:contraz3}
\end{equation}

By using the Eqs.~(\ref{eq:contraz1}):

\begin{equation}
\begin{split}
l_0 & = x_{\rm A}' - x_{\rm B}' =\\
    & = \gamma(x_{\rm A} - vt_{\rm A}) -  \gamma(x_{\rm B} - vt_{\rm B}) =\\
    & = \gamma(x_{\rm A} - vt_{\rm A} -  x_{\rm B} + vt_{\rm B}) = \\
    & = \gamma(x_{\rm A} - x_{\rm B}) = \gamma l
\end{split}
\label{eq:contraz4}
\end{equation}

because $vt_{\rm A} = vt_{\rm B}$. Since $\gamma > 1$, it results that $l<l_0$, which means that the pole length is reduced by a factor $\gamma$ in the direction of the motion. In all the other directions different from the velocity direction, there is no change.

\emph{Time dilation} is another counterintuitive effect. It is a misleading term, because it implies that time is just another spatial coordinate. Time cannot be dilated or contracted, because it is not a spatial dimension. Time is different from space, even in relativity, as proved by Kurt G\"odel \cite{GODEL1,GODEL2}\footnote{He found a solution of the Einstein's equations of the gravitational field (see Chapter 4) in the case of a universe static, but rotating. One consequence was to find closed timelike curves (see Sect. 2.5), which means that it would be possible to travel back in time, just as it is possible to move in space. Indeed, closed timelike curve is the scientific jargon for a time machine. However, some astrophysicists told G\"odel that the universe is expanding. Thus, in his second paper \cite{GODEL2}, the Austrian mathematician added the cosmological expansion and found that it removed the closed timelike curves. Therefore, the time cannot be regarded as simply the fourth dimension. It is something different from spatial coordinates. Read more about the physics of time in \cite{FOSCHINI10,FOSCHINI11}.} 
Therefore, it is better to write about a change in the clock rhythm or pace.

Let us consider the two inertial reference frames $S$ and $S'$, moving one with respect to the other with constant and uniform velocity $v$. A clock in the position $x_{\rm A}'$ reveals two consecutive events separated by a time interval $T_0$, that is $(x_{\rm A}',t_{A}')$ and $(x_{\rm A}',t_{A}'+T_0)$. In the system $S$, one has:

\begin{equation}
\begin{aligned}
t_{\rm A} & = \gamma(t_{\rm A}' + \frac{vx_{\rm A}'}{c^2})\\
t_{\rm B} & = \gamma(t_{\rm A}' + \frac{vx_{\rm A}'}{c^2} + T_0)
\end{aligned}
\label{eq:dilat1}
\end{equation}

and, therefore, the time interval measured in $S$ is then:

\begin{equation}
\begin{split}
T & = t_{\rm B} - t_{\rm A} =\\
  & = \gamma(t_{\rm A}' + \frac{vx_{\rm A}'}{c^2} + T_0) - \gamma(t_{\rm A}' + \frac{vx_{\rm A}'}{c^2}) =\\
  & = \gamma(t_{\rm A}' + \frac{vx_{\rm A}'}{c^2} + T_0 - t_{\rm A}' - \frac{vx_{\rm A}'}{c^2}) = \\
  & = \gamma T_0
\end{split}
\label{eq:dilat2}
\end{equation}

The rhythm of the moving clock is increased by a factor $\gamma$. 
\begin{figure}[ht]
\begin{center}
\includegraphics[scale=0.4]{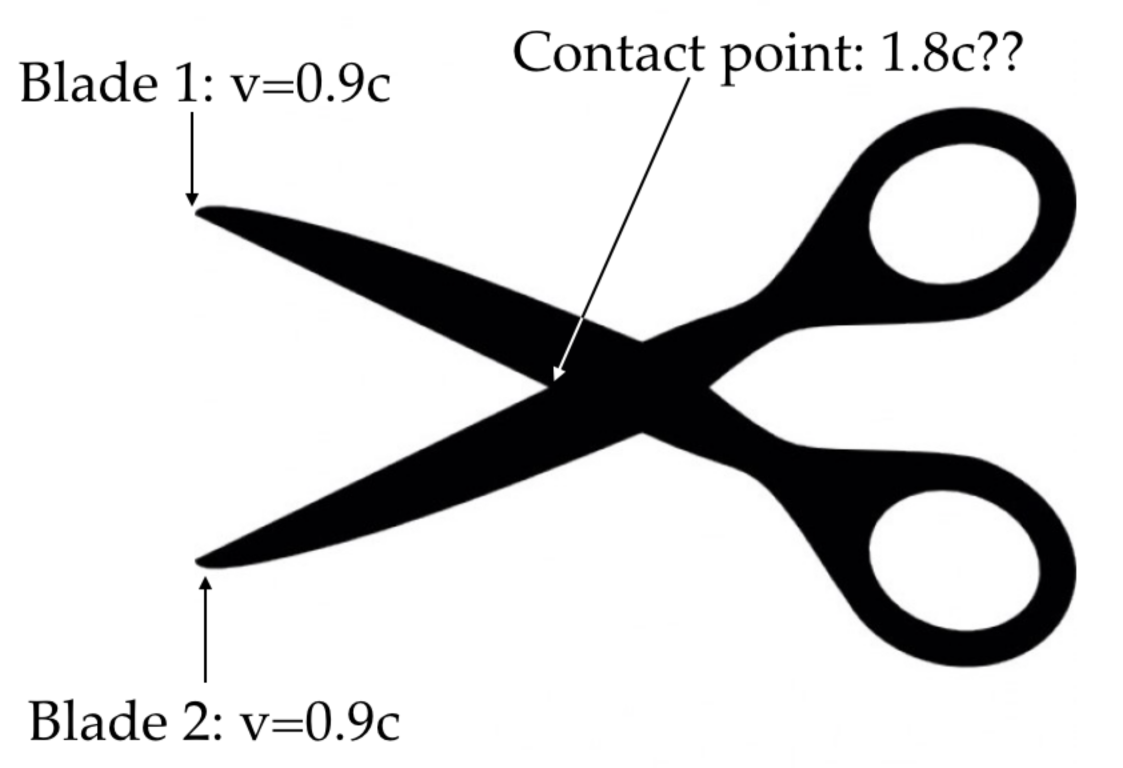}
\caption{Scissors paradox. From \cite{GOOD}.}
\label{fig:paradox2}
\end{center}
\end{figure}

It is also possible to write the transformations for velocity (see the next section) and acceleration (see the reference books). However, now I would like to show an important outcome, paradoxical at first glance. Let us consider the blades of the scissors in Fig.~\ref{fig:paradox2} closing at relativistic speed (for example, $0.9c$; example taken from \cite{GOOD}). Then, according to classical relativity, blades in proximity of the fulcrum should move at superluminal speed one with respect to the other ($0.9c+0.9c=1.8c$), which in turn implies $\beta>1$ and $\gamma$ a complex number. This does not happen in special relativity, where the addition of velocities results always in $\beta<1$, as shown in the next section.

This example shows the impossibility to have solid bodies when $v \rightarrow c$. It looks paradoxical, but only because we think to apply our knowledge on human level to domains outside the human experience. The fastest spacecraft is the Juno probe, launched by NASA in 2011: on July 4th, 2016, it reached the speed of $\sim 74$~km/s, which is just $\sim 2.4\times 10^{-4}c$. No human being ever lived at relativistic speeds, except in sci-fi movies... Therefore, there is no direct human experience we can apply. Paradoxes occur when we want to force analogies based on everyday life into the relativity domain. 

\section{Velocity transformation -- Beaming}
\begin{figure}[t]
\begin{center}
\includegraphics[scale=0.4]{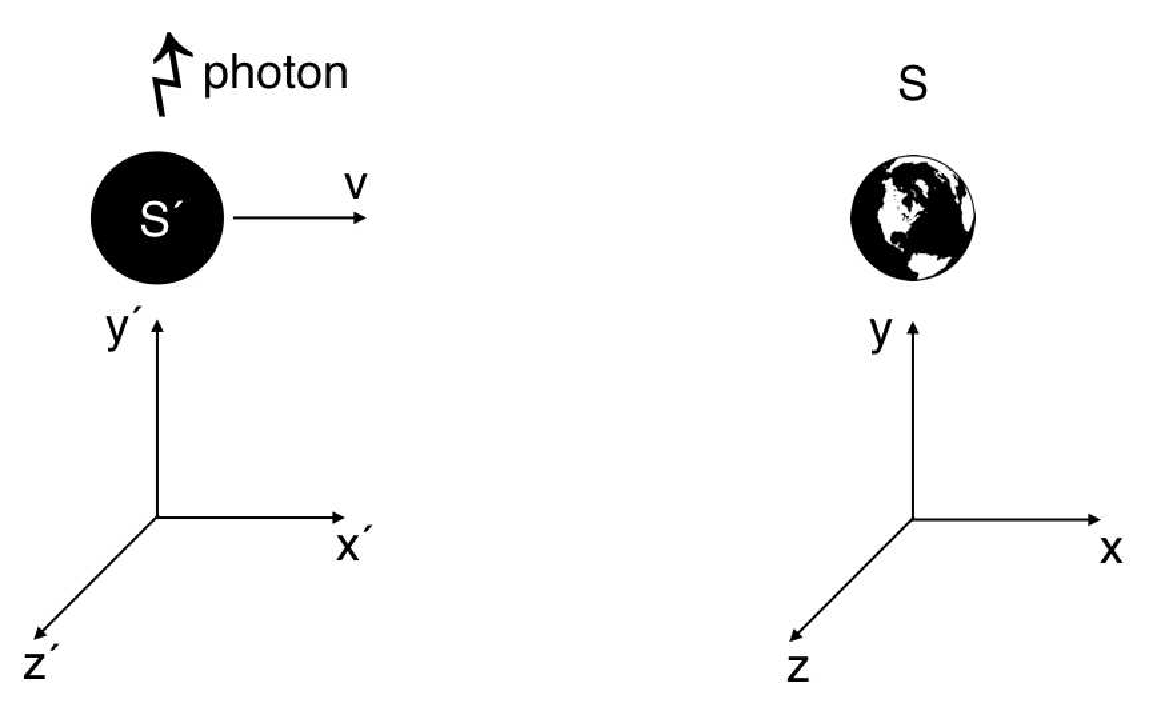}
\caption{Reference frames for velocity transformation.}
\label{fig:velo7}
\end{center}
\end{figure}

The transformation of velocity has interesting applications in astrophysics. In a relativistic jet, a blob of plasma, identified with the reference frame $S'$, is moving with bulk velocity $v$ with respect to the observer frame $S$ (Fig.~\ref{fig:velo7}). Let us to consider a generic particle with velocity $u'$ with respect to $S'$ and we want to calculate the velocity $u$ as measured in $S$. Let us define:

\begin{equation}
\begin{aligned}
(u_1,u_2,u_3) & = (\frac{dx}{dt},\frac{dy}{dt},\frac{dz}{dt}) \quad {\rm in\, S}\\
(u'_1,u'_2,u'_3) & = (\frac{dx'}{dt'},\frac{dy'}{dt'},\frac{dz'}{dt'}) \quad {\rm in\, S'}
\end{aligned}
\label{eq:velo1}
\end{equation}

Again, for the sake of simplicity, let us consider the bulk velocity $v$ directed along the $x$ axis only. The differential Lorentz transformations are:

\begin{equation}
\begin{aligned}
dx & = \gamma(dx' + vdt')\\
dy & = dy'\\
dz & = dz'\\
dt & = \gamma(dt' + \frac{vdx'}{c^2})
\end{aligned}
\label{eq:velo2}
\end{equation}

and we derive the velocity components:

\begin{equation}
\begin{aligned}
u_x & = \frac{dx}{dt} = \frac{u'_x + v}{1+\frac{u'_x v}{c^2}} \\
u_y & = \frac{dy}{dt} = \frac{u'_y}{\gamma(1+\frac{u'_x v}{c^2})}\\
u_z & = \frac{dz}{dt} = \frac{u'_z}{\gamma(1+\frac{u'_x v}{c^2})}
\end{aligned}
\label{eq:velo3}
\end{equation}

It is interesting to understand what happens at a photon emitted at the right angle respect to the direction of motion, which is the $y'$ axis in the present configuration (Fig.~\ref{fig:velo7}). Its velocity vector in $S'$ has components $(u'_{x},u'_{y},u'_{z})=(0,c,0)$. By inserting these values into Eqs.~(\ref{eq:velo3}), we obtain that we measure on the Earth $S$:

\begin{equation}
\begin{aligned}
u_x & = \frac{0 + v}{1+\frac{0 v}{c^2}} = v\\
u_y & = \frac{c}{\gamma(1+\frac{0 v}{c^2})} = \frac{c}{\gamma}\\
u_z & = \frac{0}{\gamma(1+\frac{0 v}{c^2})} = 0
\end{aligned}
\label{eq:velo4}
\end{equation}

It is worth noting the direction from which we see the photon arriving:

\begin{equation}
\tan \theta = \frac{u_y}{u_x} = \frac{1}{\gamma \beta}
\label{eq:velo5}
\end{equation}

\begin{figure}[t]
\begin{center}
\includegraphics[scale=0.4]{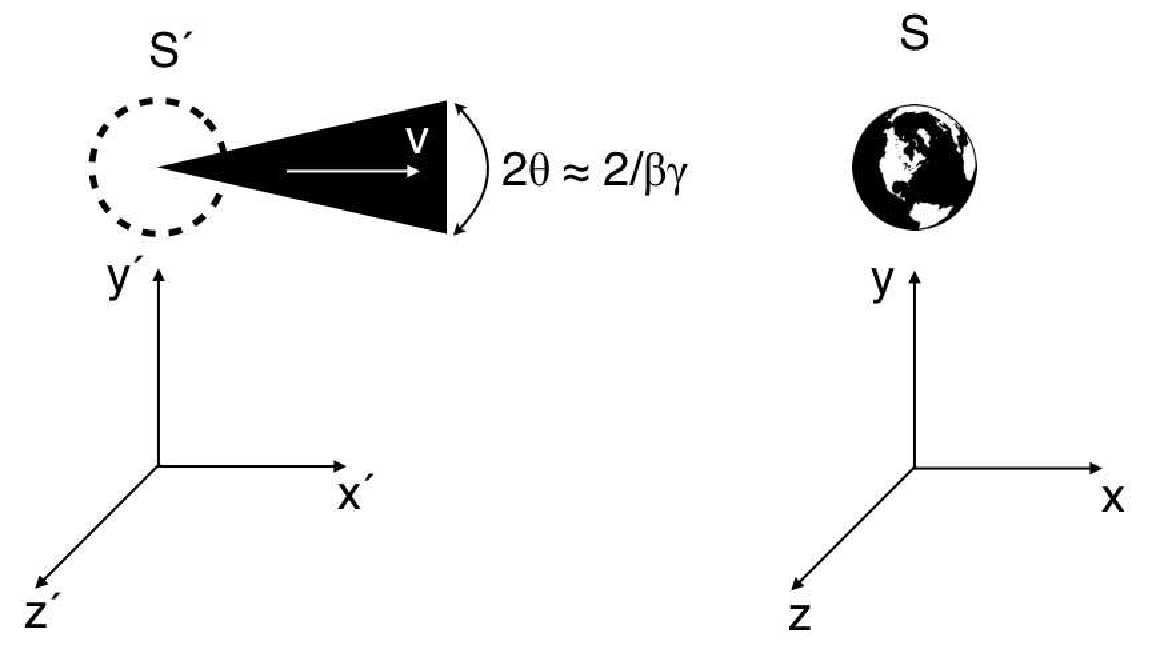}
\caption{Relativistic aberration.}
\label{fig:velo6}
\end{center}
\end{figure}

The photons emitted by a moving source, in its hemisphere directed toward the observer, seem to come from a cone with half opening angle equal to $\theta \sim (\gamma\beta)^{-1}$. This is known as \emph{relativistic aberration} \cite{REES}. In the case of relativistic speed, that is when $\beta \sim 1$ and $\gamma >> 1$, the angle is very small, literally a beam of light (hence the phenomenon is called \emph{beaming}, Fig.~\ref{fig:velo6}). The source seems to be much brighter. Particularly, the observed luminosity $L_{\rm obs}$ in the direction of the beam is $\delta^4$ times the intrinsic one, $L$, where:

\begin{equation}
\delta = \frac{1}{\gamma \sqrt{1 - \beta\cos\theta}}
\label{eq:doppler}
\end{equation}

is called \emph{Doppler factor}, just like in the case of the sound waves studied by Christian Doppler. In the latter case, the frequency of approaching sound waves is increasing, while the opposite happens in the case of the sound source moving away (classical examples are moving sirens of ambulance, police car, firefighter truck, train, ...). Also in the case of light, the frequency $\nu$ of an electromagnetic wave approaching the observer shifts to higher values by a factor $\delta$, while the opposite happens when moving away. It is possible to summarise, in the following table, the effects on the electromagnetic waves emitted by a cosmic source moving toward the observer:

\begin{equation}
\begin{aligned}
\nu_{\rm obs} & = \delta \nu \\
t_{\rm obs}  & = \delta^{-1} t\\
E_{\rm obs} & = \delta E\\
L_{\rm obs} & = \delta^4 L\\
\end{aligned}
\label{eq:doppler1}
\end{equation}

Common astrophysical values for $\delta$ are $\sim 10$ (relativistic jets from active galactic nuclei), which implies an observed luminosity $\sim 10^4$ times greater than the intrinsic one. 

\subsection{Superluminal motion}
\begin{figure}[t]
\begin{center}
\includegraphics[scale=0.4]{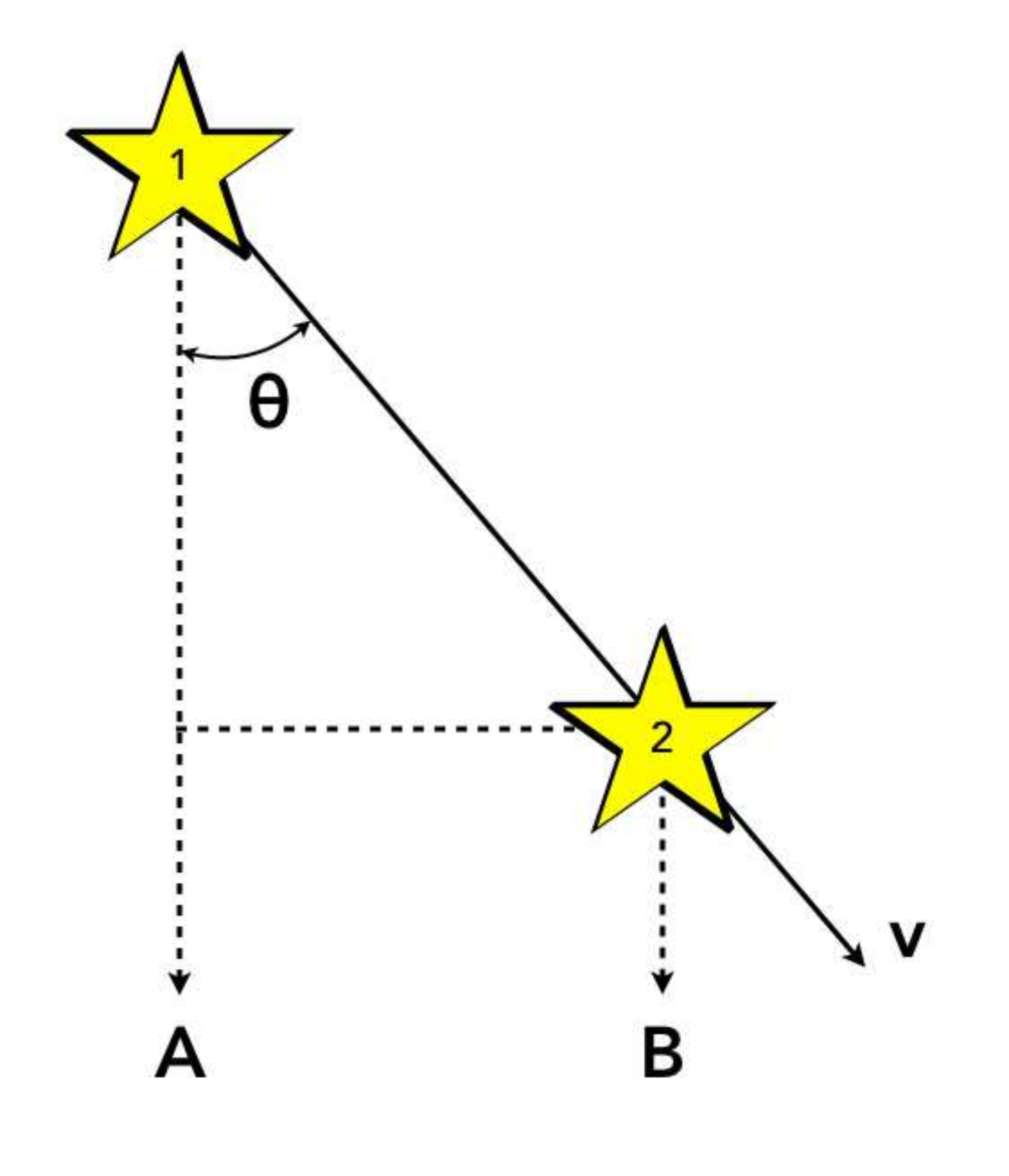}
\caption{Superluminal motion.}
\label{fig:superluminale}
\end{center}
\end{figure}

Another observed effect of special relativity is the so-called \emph{superluminal motion}, that is some cosmic sources seem to move with velocity greater than $c$. Obviously, it is just an apparent speed and the proper calculations show that the intrinsic speed is always smaller than $c$. Let us consider the case depicted in Fig.~\ref{fig:superluminale}. 

A moving source (for example, the blob of matter of a relativistic jet) is traveling with speed $v=\beta c$ and a direction making an angle $\theta$ with respect to the observer, placed in $AB$. When the blob is in the position $1$, it emits a photon, which covers a distance $d$ to arrive at the observer in $A$. Then, the blob moves ahead and reaches the position $2$, where it emits a second photon, which arrives at the obsersver in $B$. By taking into account trigonometry, the distance covered by the second photon is equal to $d-v\Delta t \cos\theta$. The time difference as measured by the observer is:

\begin{equation}
\begin{aligned}
\Delta t_{\rm obs} & = (t_{2} + \frac{d - v\Delta t \cos\theta}{c}) - (t_{1} + \frac{d}{c}) =\\
                   & = t_2 - t_1 + \frac{d}{c} - \frac{v\Delta t \cos\theta}{c} - \frac{d}{c} = \\
                   & = \Delta t - \frac{v\Delta t \cos\theta}{c} = \\
                   & = \Delta t (1 - \beta \cos\theta)
\end{aligned}
\label{eq:superlum1}
\end{equation}

that is the observed time difference is smaller than the intrinsic one by a factor $(1 - \beta \cos\theta)$, which is always smaller than $1$. The space covered by the cosmic source as measured by the observer is a projection on the sky and is equal to $v\Delta t \sin \theta$. Therefore, the observed velocity is:

\begin{equation}
v_{\rm obs} = \frac{v\Delta t \sin \theta}{\Delta t_{\rm obs}} = \frac{v \sin\theta}{1 - \beta \cos\theta}
\label{eq:superlum2}
\end{equation}

having taken into account the Eq.~(\ref{eq:superlum1}). When $\theta$ is small and $v\rightarrow c$, it results $v_{\rm obs}>c$.
 
\section{Four-dimensional spacetime}
One fundamental difference between special and classical relativity is that, in the former, space and time are merged into a four-dimensional continuum, named Minkowski spacetime. The points $(x,y,z,t)$ are \emph{events}. There are no more physical objects independent on time, as it could be in the three-dimensional Euclidean space. The common analogy of the fabric with a physical object that warps it is misleading, because the fabric is something static, while the spacetime is dynamic. I stress that the points of the Minkowski spacetime are events, not geometrical points as in the Euclidean space. 

After Riemann, a space is defined by its \emph{metric}, which is the relationship between its points. Intuitively, it is possible to look at the metric as a calculator, which gives you the distance between two points after the input of the their coordinates. 

The metric of the three-dimensional Euclidean space is:

\begin{equation}
dr^2 = dx^2 + dy^2 + dz^2
\label{eq:minko1}
\end{equation}

while the Minkowski space has the following metric:

\begin{equation}
ds^2 = c^2dt^2 - dx^2 - dy^2 - dz^2
\label{eq:minko2}
\end{equation}

It is worth noting the convention on the signs of the metric (\ref{eq:minko2}): in the present case, I adopted the sequence $(+,-,-,-)$, but it is possible to find in some textbook also the case $(-,+,+,+)$. Both are correct, but obviously it is necessary that the mathematical inferences derived from one choice or the other must be consistent. 

\begin{figure}[t]
\begin{center}
\includegraphics[scale=0.4]{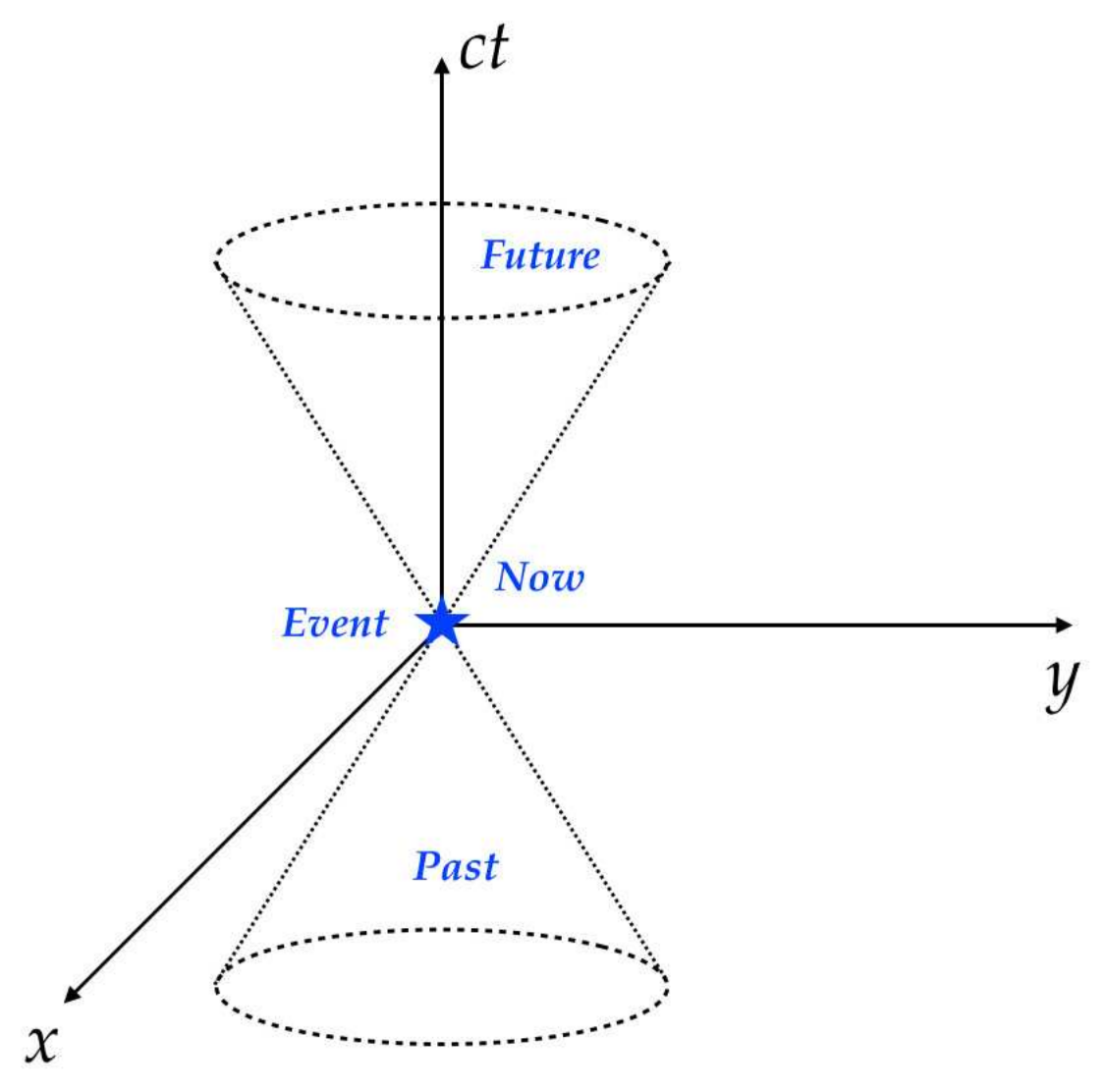}
\caption{Light cone in the Minkowski spacetime.}
\label{fig:minko1}
\end{center}
\end{figure}

There is one main difference between the two spaces characterised by the metrics (\ref{eq:minko1}) and (\ref{eq:minko2}). The Euclidean space is an absolute space, with properties (for example, lengths, angles,...) common to any observer. On the contrary, this is not the case of the Minkowski spacetime, where lengths and clock paces depend on the observer. Only the metric $ds^2$ is invariant under change of reference frame. Particularly, it is possible to define a fundamental structure constituted by the location of events that could send or receive light (thus information) to/from a certain event:

\begin{equation}
ds^2 = c^2dt^2 - dx^2 - dy^2 - dz^2 = 0
\label{eq:cono1}
\end{equation}

The Eq.~(\ref{eq:cono1}) represents a sphere in the Minkowski spacetime, although is best known as \emph{light cone}. Given the impossibility to represent a four-dimensional sphere on a sheet of paper, it is common use to remove one dimension (for example, $z$), so that the Eq.~(\ref{eq:cono1}) becomes: 

\begin{equation}
c^2dt^2 - dx^2 - dy^2 = 0
\end{equation}

which is just an infinite cone (Fig.~\ref{fig:minko1}). Events inside the cone are characterised by $ds^2>0$ and are named \emph{timelike}, while the events outside have $ds^2<0$ (\emph{spacelike}). Events on the cone surface have $ds^2=0$ (\emph{null} or \emph{lightlike}). Only photons can be on the cone surface, because they move at the light speed. Real events are inside the cone, moving from the past to the future while obeying to causality. The lines of real events are named \emph{worldlines}. Therefore, these cones are bundles of plausible worldlines, that is tied to causality. It is worth noting that special relativity does not have a preferential verse of time, as we noted earlier with the possibility to swap the Lorentz transformations. Spacelike events are not linked to the event at the origin of the cone, because any information should travel with $v>c$.

As shown earlier, vectors are mathematical entities in a three-dimensional Euclidean space. A vector can be defined by means of its components along the Cartesian axes of the reference frame. For example, the position vector $\mathbf{r} = (x,y,z)$, which means that the three components are the projections of the vector on the three orthogonal axes:

\begin{equation}
\mathbf{r} = x \mathbf{e_x} + y \mathbf{e_y} + z \mathbf{e_z}
\label{eq:vector1}
\end{equation}

where the three vectors $\mathbf{e_x}, \mathbf{e_y}, \mathbf{e_z}$ are unit vectors directed along the three Cartesian axes. It is necessary to define an important operation, the \emph{scalar product}\footnote{Also known as dot product, or inner product, or projection, just to understand its geometrical meaning.}, as:

\begin{equation}
\mathbf{A}\cdot\mathbf{B} = AB \cos \theta
\end{equation}

where $\mathbf{A},\mathbf{B}$ are two vectors and $\theta$ is the angle between their directions. It is immediate to understand that the scalar product of a vector by itself gives the square of its length. Moreover, in the case of unit vectors in an Euclidean (flat) space, $\mathbf{e}_i\cdot\mathbf{e}_j=1$ if $i=j$, otherwise $\mathbf{e}_i\cdot\mathbf{e}_j=0$. They form a \emph{basis vector}.  The Eq.~(\ref{eq:vector1}) in a Minkowski spacetime becomes:

\begin{equation}
\mathbf{r} = - x \mathbf{e_x} - y \mathbf{e_y} - z \mathbf{e_z} - ict \mathbf{e_t}
\label{eq:vector2}
\end{equation}

It is necessary to adopt a complex number for the time coordinate in order to have a minus sign when squared, because $i^2 = -1$. This is just a convention, but it is not much adopted today, because it can be misleading when one has to include quantum mechanics. Generally, one defines a four vector as:

\begin{equation}
\mathbf{r} = x_1 \mathbf{e_1} + x_2 \mathbf{e_1} + x_3 \mathbf{e_3} + x_4 \mathbf{e_4}
\label{eq:vector3}
\end{equation}

All the systems of acceptable coordinates must satisfy the Lorentz transformations, which implies that only inertial frames are valid. Therefore, it is necessary to explicitly define the square of any four vector as:

\begin{equation}
\mathbf{r^2} := x_4^2 - x_1^2 - x_2^2 - x_3^2 
\label{eq:vector4}
\end{equation}

where the symbol $:=$ means \emph{by definition} and taking into account that the square of a unit vector is equal to unity. Each four vector is a segment of a worldline in a Minkowski spacetime. I would like to underline that it is not fixed, as it happens in the three-dimensional Euclidean space, but it is a segment of spacetime! 

For the sake of simplicity, I will continue using the coordinates $(x,y,z,ct)$. Let us define the \emph{proper time}:

\begin{equation}
d\tau^2 := \frac{ds^2}{c^2} = dt^2 - \frac{dx^2+dy^2+dz^2}{c^2}
\label{eq:vector5}
\end{equation}

Then, it is possible to define the four-vector velocity as the 4-vector tangent to a worldline of a particle in the Minkowski spacetime:

\begin{equation}
\mathbf{U} = \frac{d\mathbf{r}}{d\tau} = (\frac{dx}{d\tau},\frac{dy}{d\tau},\frac{dz}{d\tau},\frac{dct}{d\tau})
\label{eq:vector6}
\end{equation}

By setting $u$ the particle velocity, and taking into account Eq.~(\ref{eq:vector5}), then:

\begin{equation}
\frac{d\tau^2}{dt^2} = 1 - \frac{u^2}{c^2}
\label{eq:vector7}
\end{equation}

which gives:

\begin{equation}
\frac{dt}{d\tau} = \gamma(u)
\label{eq:vector8}
\end{equation}

and it follows that:

\begin{equation}
\frac{dx}{d\tau} = \frac{dx}{dt}\frac{dt}{d\tau} = u_1 \gamma(u)
\label{eq:vector9a}
\end{equation}

With similar equations, it is possible to write the other components of the velocity. To summarise, the 4-vector $\mathbf{U}$ can be written as follows:

\begin{equation}
\mathbf{U} = \gamma(u)(u_1,u_2,u_3,c) = \gamma(u)(\mathbf{u},c)
\label{eq:vector9}
\end{equation}

where $\mathbf{u}$ is a 3-vector. With similar inferences, one could write also the 4-vector of the acceleration:

\begin{equation}
\mathbf{A} = \frac{d\mathbf{U}}{d\tau} = \frac{d^2\mathbf{r}}{d\tau^2}
\label{eq:vector10}
\end{equation}

By setting $\dot{\gamma} = d\gamma/dt$, one has:

\begin{equation}
\mathbf{A} = \gamma \frac{d\mathbf{U}}{dt} = \gamma \frac{d}{dt}(\gamma\mathbf{u},\gamma c) = \gamma(\dot{\gamma}\mathbf{u}+\gamma\mathbf{a},\dot{\gamma} c)
\label{eq:vector11}
\end{equation}

If $u=0$, that is the particle is at rest with respect to the reference frame, one has $\mathbf{A}=(\mathbf{a},0)$, because $\gamma(u)=1$ and $\dot{\gamma}=0$. This is the proper acceleration, which in turn implies that the 4-acceleration is equal to zero if and only if the proper acceleration is zero. On the contrary, the 4-velocity can never be completely null, because when $\mathbf{u}=0$, it remains always $\mathbf{U}=(\mathbf{0},c)$.

\section{Relativistic Mechanics}
In classical mechanics, a force $\mathbf{F}$ acting on an object with mass $m$, results in a change of velocity of the latter. Although the Newton's law is written as $\mathbf{F}=m\mathbf{a}$, what is changing is the impulse $\mathbf{p}=m\mathbf{v}$. Since the mass is constant, the derivative of the impulse results in the above formula:

\begin{equation}
\mathbf{F} = \frac{d\mathbf{p}}{dt} = \frac{d m\mathbf{v}}{dt} = m\frac{d\mathbf{v}}{dt} = m\mathbf{a}
\label{eq:newton1}
\end{equation}

In special relativity, there is the 4-impulse:

\begin{equation}
\mathbf{P} = m_0\mathbf{U}
\label{eq:mecca1}
\end{equation}

where $m_0$ is the Newtonian or proper or \emph{rest frame mass}. By taking into account Eq.~(\ref{eq:vector9}), the above equation is often rewritten as:

\begin{equation}
\begin{split}
\mathbf{P} & = m_0\mathbf{U} =\\
        & = m_0 \gamma(u)(\mathbf{u},c) =\\
        & = m (\mathbf{u},c)
\end{split}
\label{eq:mecca2}
\end{equation}

where $m=m_0\gamma(u)$ is the \emph{relativistic mass} and is changing with the velocity, while the Newtonian mass is inert. By writing the explicit definition of relativistic mass and expanding the Lorentz factor in series, one obtains:

\begin{equation}
\begin{split}
m & = m_0\gamma(u) =\\
  & = m_0 \frac{1}{\sqrt{1-\frac{u^2}{c^2}}} \sim\\
  & \sim m_0 (1 - \frac{1}{2}(-\frac{u^2}{c^2})) = \\
  & = m_0 + \frac{1}{c^2}\frac{1}{2}m_0u^2
\end{split}
\label{eq:mecca3}
\end{equation}

that is:

\begin{equation}
mc^2 = m_0c^2 + \frac{1}{2}m_0u^2
\label{eq:mecca4}
\end{equation}

The second term on the right-hand side is the kinetic energy $\mathcal{E}_{\rm k}$: 

\begin{equation}
\mathcal{E}_{\rm k} = \frac{1}{2}m_0u^2 = m_0c^2(\gamma -1)
\label{eq:mecca5}
\end{equation}

According to Einstein, the relativistic mass is a measure of the total energy, in agreement with the well-known equation $\mathcal{E}=mc^2$. Eq.~(\ref{eq:mecca4}) applied to photons, which have null rest frame mass ($m_0=0$), results in $\mathcal{E}=\mathcal{E}_{\rm k}$: all their energy is kinetic. It is often written that photons have no mass: now, it is evident that this is valid only in classical physics, while it is not correct in relativity, keeping in mind the different concept of mass and energy. 

This interpretation, however, generates many troubles\footnote{It would be better to definitely remove the concept of mass in relativistic and quantum physics, while it still makes sense in classical physics \cite{FOSCHINI21}.}. The most obvious refers to the deflection of light from a gravitational field. One could think in terms of masses: photons have no rest mass, but they have relativistic mass; therefore, they can be deflected by another massive body. However, the deflection is actually a gravitational redshift (e.g. \cite{CHENG}). By using the relativistic mass interpretation, one is mixing classical and relativistic concepts: newtonian mass and attractive gravitational force, with the relativistic mass of a photon, which is the result of $E=mc^2$, but it is not a newtonian mass. The correct treatment of the gravitational redshift requires a change of the rhythm of time, which depends on the place of the particle emitting light in the gravitational potential (see also Sect.~4.1). The velocity of a particle is the time rate of change of position, where time here is the proper time of Eqs.~(\ref{eq:vector5},\ref{eq:vector8}). Therefore, in the Eq.~(\ref{eq:mecca2}), instead of considering $\gamma(u)$ as operating on $m_0$ and transforming it into $m=m_0\gamma(u)$, one keeps it operating on the velocity vector, which in turn means that velocities are measured with the proper time. The relativistic impulse is still $\mathbf{P}=m_0\mathbf{U}=m_0[\gamma(u)(\mathbf{u},c)]$, like the classical equation, although with different meaning. When time is taken properly into account, there is no need to invoke a relativistic mass and the impulse definition of classical mechanics remains unchanged (see also, for example, \cite{LANCZOS,REFIORENTIN,VECCHIATO}). 

These results made it clear the need of a revision of some fundamental principles of classical physics: conservation of impulse and energy. In relativity, these two principles are merged. It is no more possible to speak about inert mass, but one has to adopt the term mass-energy, or, simply, energy. As stated by Einstein, the mass is the energy content of a body\footnote{See \cite{JAMMER} for an extended analysis of the concept of mass.}. This is consistent with what has been shown in Sect.~$3.2$ about solid objects moving at $v\sim c$. The definition of relativistic mass shows that as $v\rightarrow c$, then $m\rightarrow \infty$. Therefore, the only way to travel at the light speed is to have null rest mass, as in the case of photons.

\section{Toward general relativity...}
Special relativity deals with inertial reference frames. The following endeavour of Einstein was to deal with accelerated frames, which were divided into two types at his epoch, as indicated by the use of two different types of mass: gravitational and inertial, although both had the same numerical value. Baron von E\"otv\"os performed some experiments between 1889 and 1922, where he proved that the two types of mass were the same (\emph{equivalence principle}). Einstein understood the impact of these results when he realised that one person in free fall does not feel his own weight. This means that gravitation can be considered as any other accelerated system in a space without gravity. It is just a matter of reference frames. However, the accelerated frames added one more geometrical issue. On one side, as established by the mass-energy relationship, the light can be warped by a gravitational field. On the other side, a \emph{gedankenexperiment} suggested to Einstein that the Minkowski spacetime was no more the best language to use. The length contraction and clock pace change in special relativity can be understood by taking into account the relative motion, which is constant and uniform. However, if the frame is accelerated -- the relative motion is continuously changing -- then sizes and paces are continuously changing as the velocity changes. This implies that the coordinates of an accelerated reference frame have no more meaning. It is necessary to search for another language to speak about accelerated frames. This language is that of tensors.

\chapter{Tensors}

\section{Tensors in Euclidean space}
As in classical mechanics we adopted vectors in three-dimensional Euclidean space, in special relativity we used still vectors but in a four-dimensional Minkowski spacetime. In both cases, the structure of the space is fixed as flat. Already G.~B. Riemann noted that the distribution of matter and forces determined the geometry. G. Ricci Curbastro\footnote{Curiously, the Italian mathematician Gregorio Ricci Curbastro signed his well-known paper on tensors as Ricci only. Therefore, the tensor named after him was simply called Ricci tensor and not Ricci Curbastro tensor.} and T. Levi-Civita developed the new Riemann geometry by elaborating the tensor calculus. The power of tensors is that there is no need of coordinate system, because the tensor itself defines the space properties. 

There are also tensors in the three-dimensional Euclidean space. Therefore, I would like to introduce tensors by making some example of this type. The stress tensor is adopted in the theory of constructions, in order to speak about the behaviour of solid structures to different pressures. Let us consider an infinitesimal surface $dS$ and a unit vector $\mathbf{n}$ orthogonal to it. The sign of $\mathbf{n}$ is selected so that traction forces are positive and compression forces are negative. Then, the surface forces $d\mathbf{F}$ acting on the surface $dS$ are defined as:

\begin{equation}
d\mathbf{F} = \mathbf{f}dS
\end{equation}

where $\mathbf{f}$ are the forces per unit surface. It is possible to prove\footnote{It is one of the many theorems proved by A. Cauchy...} that there is one and only one $3\times 3$ symmetric tensor $T_{ij}$ such that:

\begin{equation}
d\mathbf{F} = T_{ij}\mathbf{n}dS
\end{equation}

$T_{ij}$ is called \emph{stress tensor} and it is represented by a $3\times 3$ matrix (hence the subscripts $i,j$), because it is an object in a three-dimensional Euclidean space:

\begin{equation}
T_{ij} = 
\begin{bmatrix}
\sigma_{11} & \tau_{12} & \tau_{13}\\
\tau_{21} & \sigma_{22} & \tau_{23}\\
\tau_{31} & \tau_{32} & \sigma_{33}\\
\end{bmatrix}
\label{eq:sforzo1}
\end{equation}

where I used different Greek letters for the elements of the diagonal, and outside it. The physical meaning of the stress tensor is that of a pressure. The elements along the diagonal, that is those for which the subscripts $i$ and $j$ are equal, indicate the normal stresses (pressures). The other elements of the matrix ($i\neq j$) indicate the shear stresses. Since this tensor is an entity in an Euclidean space, the time is external to the space and there is one stress tensor for a certain time. Therefore, changing time could require a change also in the stress tensor. 

\begin{figure}[t]
\begin{center}
\includegraphics[scale=0.4]{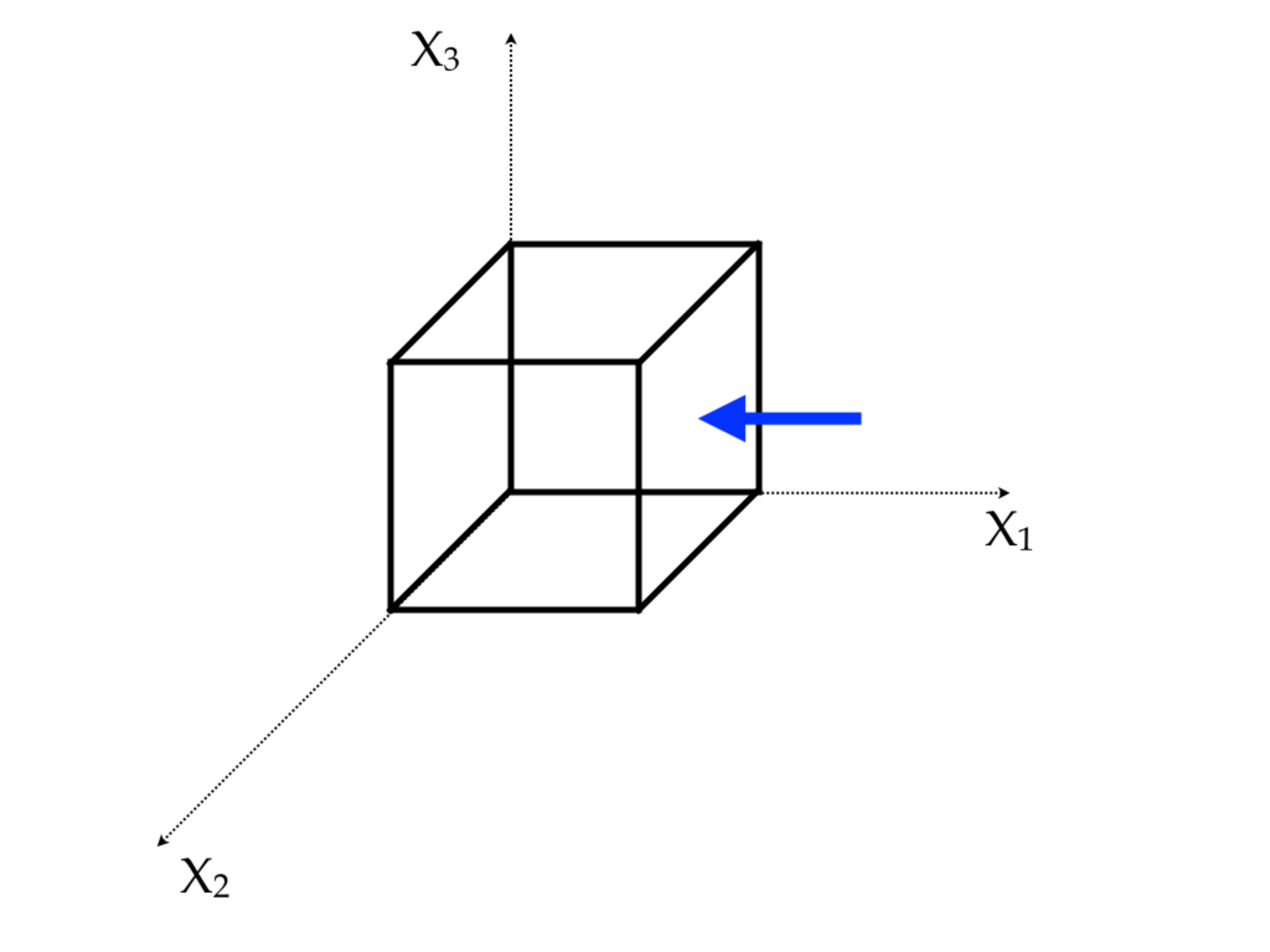}
\caption{Stress tensor and pressure.}
\label{fig:sforzo1}
\end{center}
\end{figure}

Fig.~\ref{fig:sforzo1} displays a solid cube in a reference frame $(X_1,X_2,X_3)$. A pressure directed along the $X_1$ axis is applied on one face of the cube. Just to set some number, let us say $10$~Pa. Therefore, the stress tensor of Eq.~(\ref{eq:sforzo1}) becomes:

\begin{equation}
\begin{bmatrix}
-10 & 0 & 0\\
0 & 0 & 0\\
0 & 0 & 0\\
\end{bmatrix}
\label{eq:sforzo2}
\end{equation}

\begin{figure}[t]
\begin{center}
\includegraphics[scale=0.4]{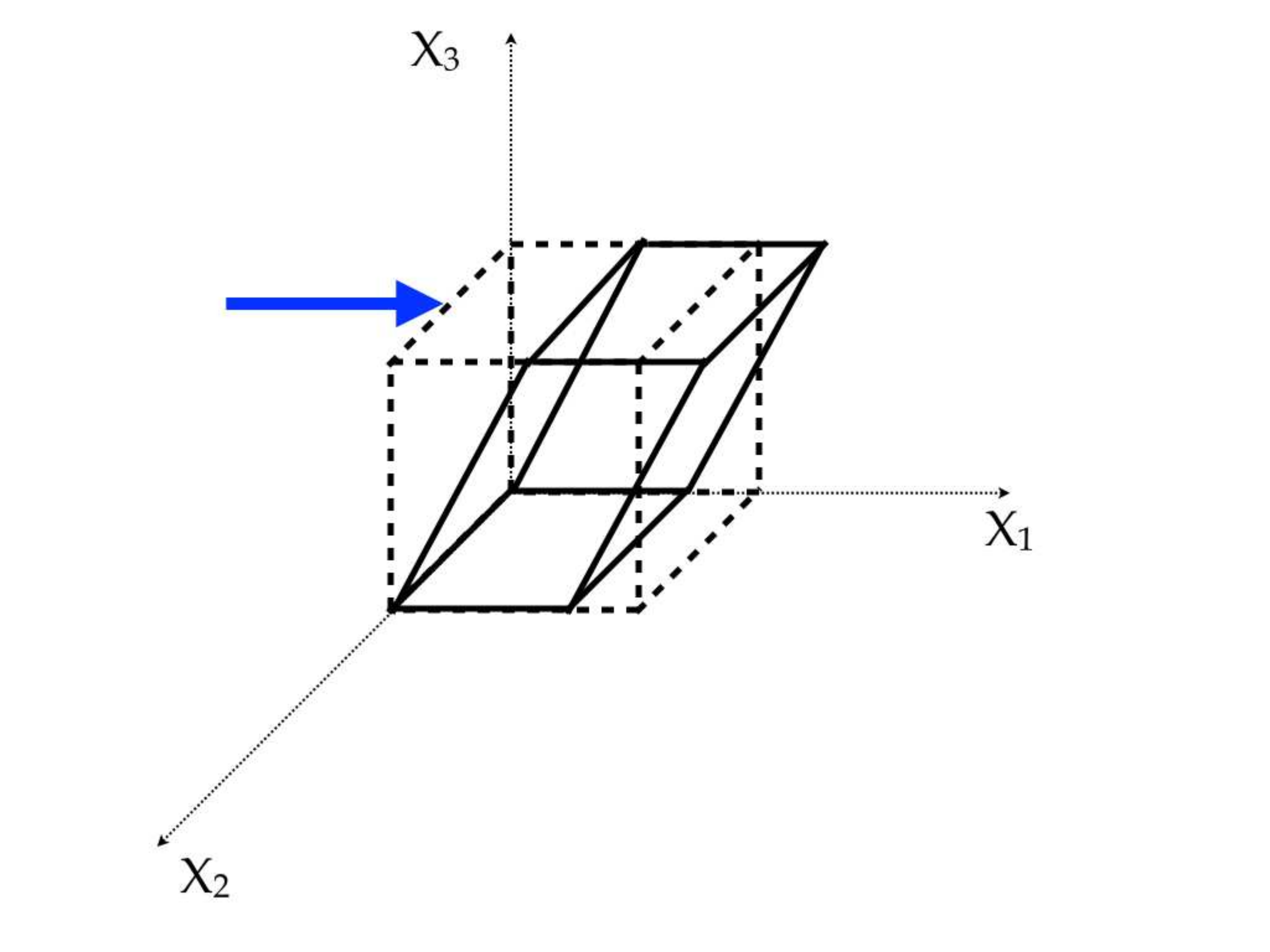}
\caption{Stress tensor and shear stress.}
\label{fig:sforzo2}
\end{center}
\end{figure}

The negative sign indicates that the pressure is directed along the $X_1$, but toward the opposite side (so, it is a compression). If now a pure shear stress, again with value $10$~Pa, is applied to one face of the cube, parallel to the plane $X_{1}X_{2}$, as shown in Fig.~\ref{fig:sforzo2}, the matrix~(\ref{eq:sforzo1}) becomes:

\begin{equation}
\begin{bmatrix}
0 & 5 & 0\\
5 & 0 & 0\\
0 & 0 & 0\\
\end{bmatrix}
\label{eq:sforzo3}
\end{equation}

By using another tensor, it is possible to express the small movements generated by the forces in a solid or a fluid. In this case, the matrix elements on the diagonal indicate the displacements in the direction of the three Cartesian axes, while the other elements are the tangential changes of position. 

The main difference between a tensor and a vector is about the change of directions. For example, let us consider one vector $\mathbf{v}$:

\begin{equation}
\mathbf{v}=
\begin{bmatrix}
1 \\
2 \\
3 \\
\end{bmatrix}
\label{eq:sforzo4}
\end{equation}

It is clear from Eq.~(\ref{eq:sforzo4}) that a vector could be represented also by using the tensor language and it is just a tensor of rank one. The numerical values are completely arbitrary, just to fill the slots. Let us think the vector $\mathbf{v}$ as the representation of the velocity of an object. If there is a change in magnitude, but not in direction and sign, it is sufficient to multiply the vector by a scalar $k$ equal to the ratio between the new and the old magnitude value. That is:

\begin{equation}
k\mathbf{v}=k
\begin{bmatrix}
1 \\
2 \\
3 \\
\end{bmatrix}
=
\begin{bmatrix}
1k \\
2k \\
3k \\
\end{bmatrix}
\label{eq:sforzo5}
\end{equation}

The vector components simply increased (or decreased) by a factor $k$. If $k$ is negative, then it is possible to change sign to the vector, while the direction remains constant. The only way to change a direction of a vector is to apply the vector product $\otimes$ defined as follows:

\begin{equation}
\mathbf{C}=\mathbf{A}\otimes\mathbf{B} = AB \sin \theta \mathbf{c}
\label{eq:sforzo6}
\end{equation}

$\mathbf{C}$ is the result of the vector product of $\mathbf{A}$ and $\mathbf{B}$. Its direction is perpendicular to the plane containing both $\mathbf{A}$ and $\mathbf{B}$ (direction represented by the unit vector $\mathbf{c}$). Its magnitude is equal to the product of the magnitudes of the two vectors multiplied by the sine of the angle $\theta$ between the directions of $\mathbf{A}$ and $\mathbf{B}$. Therefore, the vector product can change the direction, but only by $90^{\circ}$. The only way to change direction by different angles is to use tensors of rank greater or equal than two, because -- as shown -- the off-diagonal elements indicate directions different from the three Cartesian axes.

\section{Metric tensor}
It is possible to write again the line element of the Minkowski spacetime, given in the Eq.~(\ref{eq:minko2}), by using the language of tensors:

\begin{equation}
ds^2 = g_{ij}dx^{i}dx^{j}
\label{eq:tenso1}
\end{equation}

where $x^{i}=(t,x,y,z)=(x^0,x^1,x^2,x^3)$ and $g_{ij}$ is the \emph{metric tensor} (or, simply, \emph{metric}), which sets the properties of the space so to measure distances. The reason of subscripts and superscripts will be understood later. 
Reminding Eq.~(\ref{eq:vector3}), which indicates how to build a vector through a basis, it is possible to understand the metric as the dot product between two bases, that is:

\begin{equation}
g_{ij}=
\begin{bmatrix}
g_{00} & g_{01} & g_{02} & g_{03}\\
g_{10} & g_{11} & g_{12} & g_{13}\\
g_{20} & g_{21} & g_{22} & g_{23}\\
g_{30} & g_{31} & g_{32} & g_{33}\\
\end{bmatrix}
= \mathbf{e}_i\cdot\mathbf{e}_j=
\begin{bmatrix}
\mathbf{e}_{0}\cdot\mathbf{e}_{0} & \mathbf{e}_{0}\cdot\mathbf{e}_{1} & \mathbf{e}_{0}\cdot\mathbf{e}_{2} & \mathbf{e}_{0}\cdot\mathbf{e}_{3}\\
\mathbf{e}_{1}\cdot\mathbf{e}_{0} & \mathbf{e}_{1}\cdot\mathbf{e}_{1} & \mathbf{e}_{1}\cdot\mathbf{e}_{2} & \mathbf{e}_{1}\cdot\mathbf{e}_{3}\\
\mathbf{e}_{2}\cdot\mathbf{e}_{0} & \mathbf{e}_{2}\cdot\mathbf{e}_{1} & \mathbf{e}_{2}\cdot\mathbf{e}_{2} & \mathbf{e}_{2}\cdot\mathbf{e}_{3}\\
\mathbf{e}_{3}\cdot\mathbf{e}_{0} & \mathbf{e}_{3}\cdot\mathbf{e}_{1} & \mathbf{e}_{3}\cdot\mathbf{e}_{2} & \mathbf{e}_{3}\cdot\mathbf{e}_{3}\\
\end{bmatrix}
\label{eq:metrica1}
\end{equation}
 
When the spacetime is flat, as in the case of Minkowski, then $\mathbf{e}_i\cdot\mathbf{e}_j = 0$ when $i\neq j$, because the basis vectors are orthogonal each other. In addition, the scalar product between the same basis vectors has to fulfil the definition Eq.~(\ref{eq:vector4}). The resulting metric is:

\begin{equation}
g_{ij}=
\begin{bmatrix}
1 & 0 & 0 & 0\\
0 & -1 & 0 & 0\\
0 & 0 & -1 & 0\\
0 & 0 & 0 & -1\\
\end{bmatrix}
\label{eq:metrica2}
\end{equation}

which means that Eq.~(\ref{eq:tenso1}) results in the Eq.~(\ref{eq:minko2}). If we use a polar coordinates system $x^{i}=(t,r,\theta,\phi)$ defined as follows:

\begin{equation}
\begin{aligned}
x & = r\sin\theta\cos\phi\\
y & = r\sin\theta\sin\phi\\
z & = r\cos\theta\\
t & = t\\
\end{aligned}
\label{eq:tenso2}
\end{equation}

then the metric becomes:

\begin{equation}
g_{ij}=
\begin{bmatrix}
1 & 0 & 0 & 0\\
0 & -1 & 0 & 0\\
0 & 0 & -r^2 & 0\\
0 & 0 & 0 & -r^2\sin^2\theta\\
\end{bmatrix}
\label{eq:metrica2bis}
\end{equation}

and the line segment is:

\begin{equation}
ds^2 = dt^2 - dr^2 - r^2d\theta^2 - r^2\sin^2\theta d\phi^2
\label{eq:tenso3}
\end{equation}

The metric of Eq.~(\ref{eq:metrica1}) is a general expression independent on a specific space. It defines the space, through the scalar product of the basis vectors. If the space is not flat, then the basis vectors are not orthogonal each others and could even be of different length. This implies that the bases and the metric are different from their inverse quantities. Let us to write the inverse basis vectors with $\mathbf{e}^{j}$. Their relationship with the basis vector is:

\begin{equation}
\mathbf{e}_{i}\cdot\mathbf{e}^{j} = \delta_{i}^{j}
\end{equation}

As the metric is $g_{ij}=\mathbf{e}_{i}\cdot\mathbf{e}_{j}$, its inverse is $g^{ij}=\mathbf{e}^{i}\cdot\mathbf{e}^{j}$. Both are related by $g_{ik}g^{kj}=\delta_{i}^{j}$. As there are two basis vectors, it is possible to write any vector in term of both bases. 

\begin{equation} 
\begin{cases}
\mathrm{Contravariant} & V^{i}=\mathbf{V}\cdot \mathbf{e}^{i}\\
\mathrm{Covariant} & V_{i}=\mathbf{V}\cdot \mathbf{e}_{i}
\end{cases}
\label{eq:vectorexp}
\end{equation}

\begin{figure}[t]
\begin{center}
\includegraphics[scale=0.3]{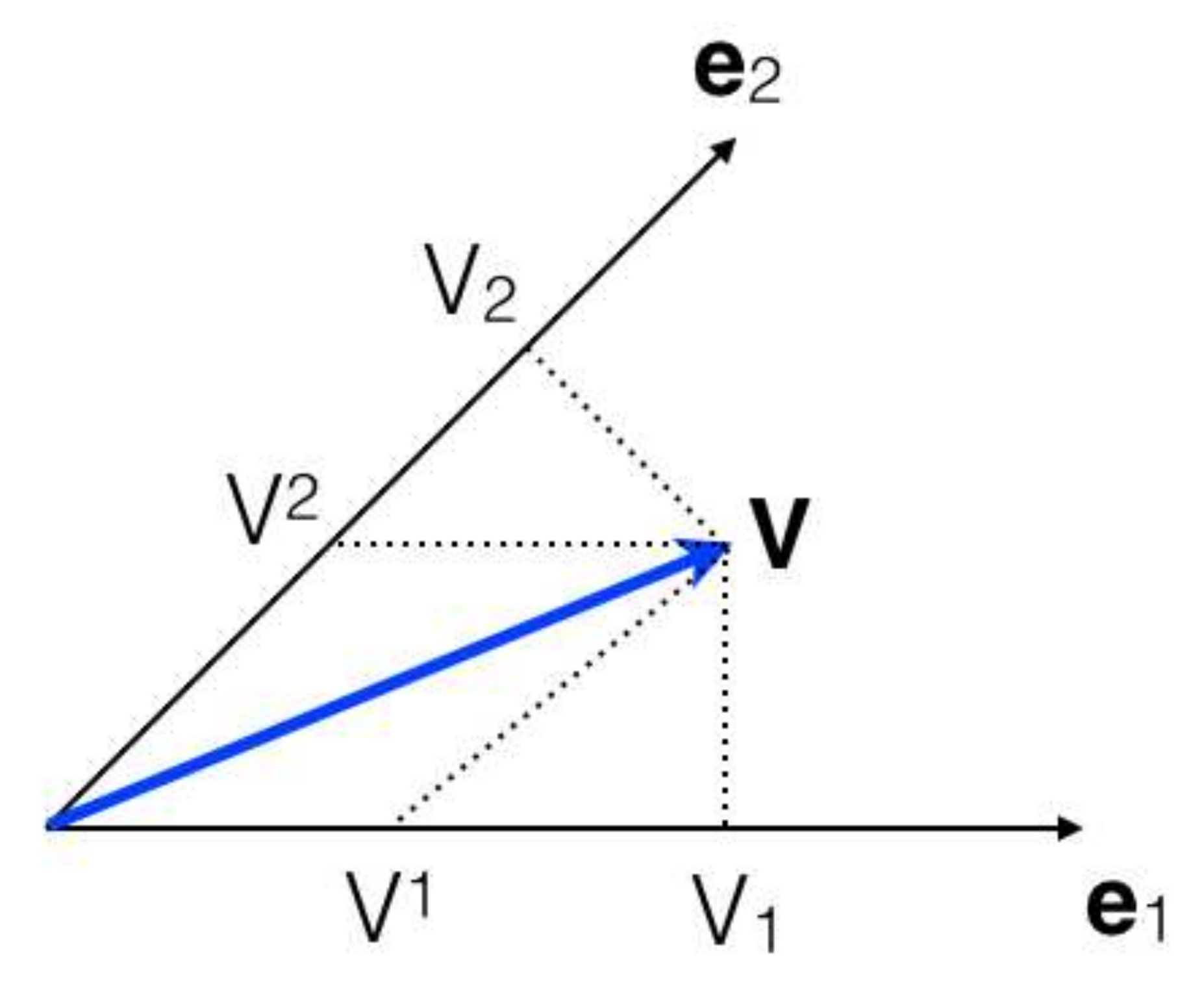}
\caption{Covariant and contravariant components of a vector. See the text for details.}
\label{fig:bases}
\end{center}
\end{figure}

The geometrical meaning of the two representations in Eq.~(\ref{eq:vectorexp}) can be understood by looking at Fig.~\ref{fig:bases}, where a two-dimensional space case is shown. The contravariant components $V^{i}$ are the parallel projections of the vector $\mathbf{V}$ onto the basis $(\mathbf{e}_1,\mathbf{e}_2)$, while the covariant components $V_{i}$ are the orthogonal projections on the same basis (or the parallel projections on the inverse basis). It is easy to understand that in a flat space, where the basis vectors are orthogonal each other, the contravariant and covariant components are the same. 

If one knows either the covariant and the contravariant vectors, then it is not necessary to know the metric, as:

\begin{equation}
\mathbf{x}\cdot\mathbf{y} = x^{i}\mathbf{e}_{i}\cdot y_{j}\mathbf{e}^{j} = x^{i}y_{j}(\mathbf{e}_{i}\cdot \mathbf{e}^{j}) = x^{i}y_{j}
\end{equation}

On the other side, if one knows only either one or the other type of vector, then it is necessary to use the metric:

\begin{equation}
\mathbf{x}\cdot\mathbf{y} = g_{ij}x^{i}y^{j} = g^{ij}x_{i}y_{j}
\end{equation}

It should be now evident the reasons of writing Eq.~(\ref{eq:tenso1}) in that specific form, or:

\begin{equation}
d\mathbf{x}\cdot d\mathbf{x}=ds^2 = g_{ij}dx^{i}dx^{j} = g^{ij}dx_{i}dx_{j}
\label{eq:tenso4}
\end{equation}

When the metric is constant and does not depends on the coordinates, this means that changes of coordinates are always of the same type. This is possible in a flat space, such as a three-dimensional Euclidean space. If the metric depends on the coordinates, such as in a curved space, then the transformation is no more linear. For example, in an accelerated frame, a change of coordinate is non-linear with time: 

\begin{equation}
x \rightarrow x' = x + vt + \frac{at^2}{2}
\end{equation}

\chapter{Equations of the gravitational field}
In the fall of 1915, Albert Einstein developed the equations of the gravitational field by using the tensor language. The field equations are:

\begin{equation}
R_{ij} - \frac{1}{2}g_{ij}R = k T_{ij}
\label{eq:campo1}
\end{equation}

where: 

\begin{equation}
k = \frac{8\pi G}{c^4} = 2.073\times 10^{-48} \, \mathrm{s}^2\mathrm{cm}^{-1}\mathrm{g}^{-1}
\label{eq:campo2}
\end{equation}

is the \emph{Einstein gravitational constant}. $G$ is the Newton universal gravitational constant, equal to $6.67428\times 10^{-8}\, \mathrm{cm}^3\mathrm{g}^{-1}\mathrm{s}^{-2}$. The term on the left-hand side of the Eq.~(\ref{eq:campo1}) is also named \emph{Einstein tensor}:

\begin{equation}
G_{ij} = R_{ij} - \frac{1}{2}g_{ij}R
\label{eq:campo3}
\end{equation}

As one can see, it is composed of the \emph{Ricci (Curbastro) tensor} $R_{ij}$, of the \emph{curvature scalar} $R$, and of the metric $g_{ij}$. The Ricci Curbastro tensor and the curvature scalar both derived from the \emph{Riemann tensor} by means of a proper index gymnastics (for details, look at \cite{INVERNO,RINDLER}). The curvature indicates the how much the space deviates from an Euclidean one. For example, in the latter, the sum of internal angles of a triangle is equal to $\pi$. In a curved space, such as the Earth's surface at large scale, the sum of internal angles is greater than $\pi$. 
The tensor $T_{ij}$ on the right-hand side of the Eq.~(\ref{eq:campo1}) is the \emph{energy-impulse} or \emph{energy-momentum} tensor. As John A. Wheeler once wrote, the field equations (\ref{eq:campo1}) can be understood as the mass-energy tells the spacetime how to warp (read equations from right to left), while the spacetime tells mass-energy how to move (read the equations from left to right). In the case of no sources of mass-energy, Eqs.~(\ref{eq:campo1}) describe a Minkowski spacetime. If we do not set this constraint, the most general case is:

\begin{equation}
R_{ij} - \frac{1}{2}g_{ij}R + \Lambda g_{ij} = k T_{ij}
\label{eq:campo4}
\end{equation}

where $\Lambda$ is the well-known \emph{cosmological constant}, which Einstein considered his biggest blunder, although the observations of the latest decades have shown to be required.

To grab the physical meaning of the energy-impulse tensor, it is necessary to have clear that it is a mathematical entity in a spacetime (Fig.~\ref{fig:tensoreenergia}) and the mass-energy equivalence. Part of the tensor, specifically the $3\times3$ section with indexes $1,2,3$, has a physical meaning similar to the three-dimensional case of the stress tensor we have studied as example. In addition, there is the element $T_{00}$, which is an energy density, while the elements $T_{0,j}\; (j=1,2,3)$ indicate an energy flux (constant space and changing time). Lastly, the elements $T_{i,0}\; (i=1,2,3)$ indicate impulse density (changing space and constant time). The easiest example is the tensor in the case of a perfect fluid in thermodynamical equilibrium with the nearby environment:

\begin{figure}[t]
\begin{center}
\includegraphics[scale=0.3]{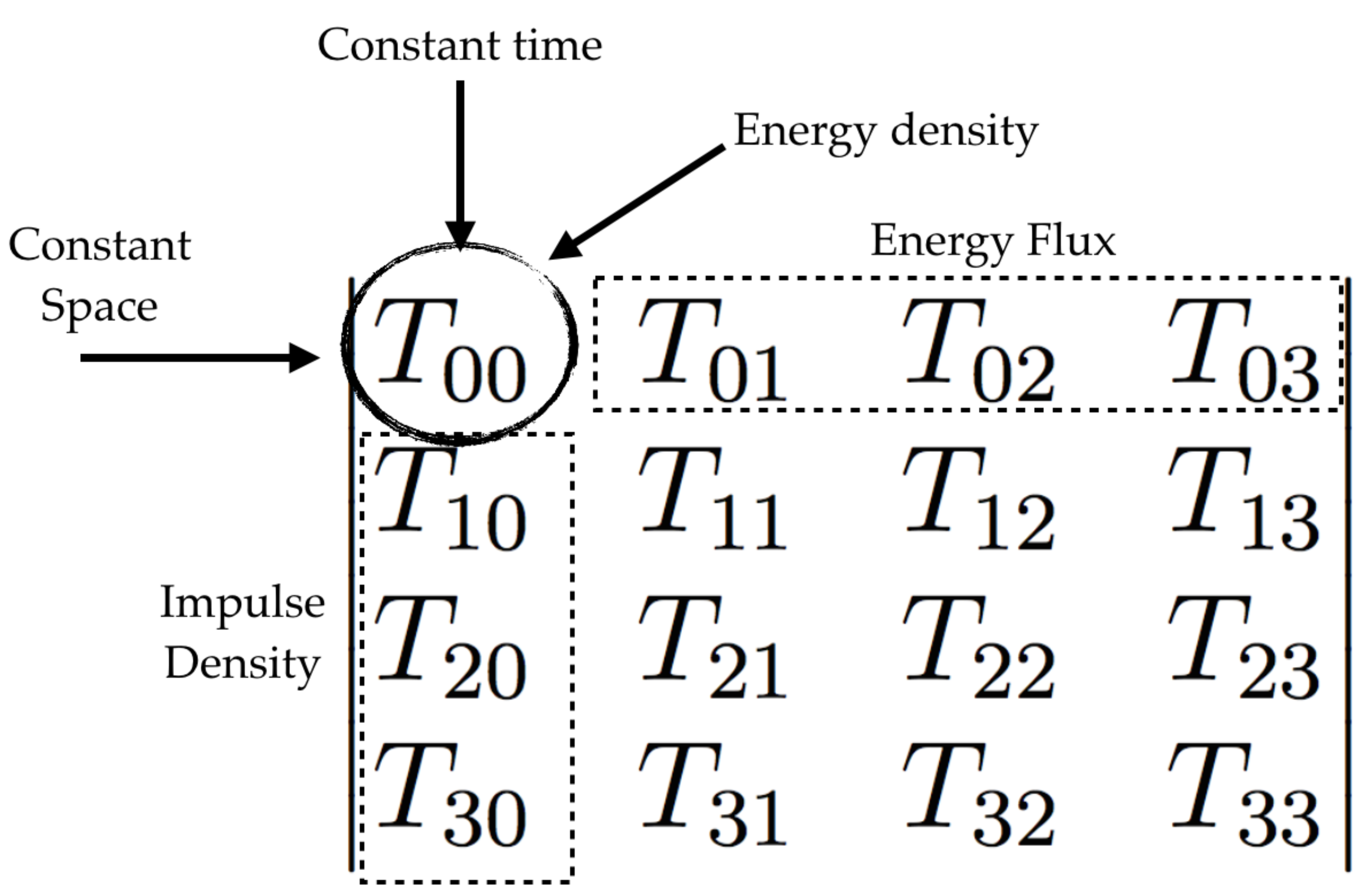}
\caption{Energy-impulse tensor.}
\label{fig:tensoreenergia}
\end{center}
\end{figure}

\begin{equation}
T_{ij}=
\begin{bmatrix}
\rho c^2 & 0 & 0 & 0\\
0 & p & 0 & 0\\
0 & 0 & p & 0\\
0 & 0 & 0 & p
\end{bmatrix}
\label{eq:fluido}
\end{equation}

That is the element $T_{00}=\rho c^2$ is the energy density in the fluid rest frame, and $\rho=nm$, with $n$ is the number density of particles with mass $m$. $T_{00}$ is simply the mass density in the non-relativistic case. All the elements $T_{0,j}=\rho c u^{j}$ (energy flux$/c$) and $T_{i,0}=\rho c u^{i}$ (impulse flux$\times c$) are null, because the fluid is in thermodynamical equilibrium, which means no energy or impulse exchange. There is only the isotropic pressure $T_{ij}=\rho u^{i}u^{j}=p$, if $i=j$ (no shear stress, $T_{ij}=0$, if $i\neq j$), which fills the elements on the diagonal of the remaining $3\times 3$ matrix ($i,j=1,2,3$).

A stream of relativistic particles moving along the $x$ axis in the positive direction, can be described by the following energy-impulse tensor:

\begin{equation}
T_{ij}=
\begin{bmatrix}
\rho c^2 & \rho c^2 & 0 & 0\\
\rho c^2 & \rho c^2 & 0 & 0\\
0 & 0 & 0 & 0\\
0 & 0 & 0 & 0
\end{bmatrix}
\label{eq:fluido2}
\end{equation}

If they move in the opposite direction:

\begin{equation}
T_{ij}=
\begin{bmatrix}
\rho c^2 & -\rho c^2 & 0 & 0\\
-\rho c^2 & \rho c^2 & 0 & 0\\
0 & 0 & 0 & 0\\
0 & 0 & 0 & 0
\end{bmatrix}
\label{eq:fluido3}
\end{equation}

In order to solve the field equations (\ref{eq:campo1}), one can try three different ways:

\begin{enumerate}
\item to calculate the metric, starting from the energy-impulse tensor;
\item to calculate the energy-impulse tensor, starting from the metric; 
\item to solve the equations, starting from all the available information, both from the metric and from the energy-impulse tensor.
\end{enumerate}

There are thousands of possible solutions. Last, but not least, these are non-linear equations: the energy-impulse generate the gravitational field, which in turn has energy, and can affect the energy-impulse tensor, which in turn affect the gravitational field, and so on. However, it is possible to make the equations simpler and calculate solutions with the help of external constraints, such as symmetry properties.

\section{Spacetime singularities (black holes)}
Einstein did not write about solutions of his equations. The first solution was proposed by Karl Schwarzchild just a few months later. It made use of the simplest symmetry: a static spherical mass $M$. Schwarzchild calculated the metric in polar coordinates ($t,r,\theta,\phi$), which is today named after him:

\begin{equation}
g_{\rm Schw}=
\begin{bmatrix}
-c^2(1-\frac{2r_{\rm g}}{r}) & 0 & 0 & 0\\
0 & (1-\frac{2r_{\rm g}}{r})^{-1} & 0 & 0\\
0 & 0 & r^2 & 0\\
0 & 0 & 0 & r^2\sin^2\theta
\end{bmatrix}
\label{eq:schw}
\end{equation}

where $r_{\rm g}=GM/c^2$ is the gravitational radius. In the static case, another important parameter is the Schwarzchild radius, which is just twice the gravitational radius ($r_{\rm S}=2r_{\rm g}$). Some points must be underlined:

\begin{enumerate}
\item if $r=r_{\rm S}=2r_{\rm g}$, then $g_{00}=0$ and $g_{11}\rightarrow \infty$: this is a singularity;
\item if $r=0$, then $g_{00}\rightarrow \infty$ and $g_{11}=g_{22}=g_{33}=0$: this is another singularity;
\item if $r\rightarrow \infty$, then $g_{00}=-c^2$ and $g_{11}=1$: the Schwarzchild metric is equal to that of Minkowski.
\end{enumerate}

The two singularities have different properties. The first one is a \emph{coordinate singularity}, which means that it depends on the coordinate choice. With a proper change of metric, it is possible to remove this singularity (for example, Kruskal metric). The second case is an \emph{essential singularity}, that is it cannot be eliminated. It remains whatever is the coordinate choice. The spacetime region inside $r=r_{\rm S}$ has another important feature, independent on the coordinates. According to the general relativity, the clock pace is slowed as the observer is closer and closer to a gravitational source. One photon traveling toward the above cited region, will experience a gravitational red shift equal to:

\begin{equation}
z = \frac{\lambda - \lambda_0}{\lambda} = \frac{1}{\sqrt{1-\frac{r_{\rm S}}{r}}}-1
\label{eq:schw2}
\end{equation}

If $r=r_{\rm S}$, then the denominator is null, and $z\rightarrow \infty$. This means that no radiation generated inside the surface $r=r_{\rm S}$ can propagate outside. Therefore, $r_{\rm S}$ is named \emph{event horizon}.

Several years later, in 1963, Roy Kerr published a new solution of the Einstein's equations. The mass still has a spherical shape, but it is rotating. The new metric is:

\begin{equation}
g_{\rm Kerr}=
\begin{bmatrix}
-c^2(1-\frac{2r_{\rm g}r}{\rho^2}) & 0 & 0 & -\frac{\omega \Sigma^2 \sin^2\theta}{\rho^2}\\
0 & \frac{\rho^2}{\Delta} & 0 & 0\\
0 & 0 & \rho^2 & 0\\
-\frac{\omega \Sigma^2 \sin^2\theta}{\rho^2} & 0 & 0 & \frac{\Sigma^2}{\rho^2}\sin^2\theta
\end{bmatrix}
\label{eq:kerr}
\end{equation}

In this case, by starting from the classical polar coordinates $t,r,\theta,\phi$, we adopted a new set of coordinates named after Boyer-Lindquist. The radial distance is now:

\begin{equation}
\rho = \sqrt{r^2 + a^2 r^2_{\rm g}\cos^2\theta}
\label{eq:bl1}
\end{equation}

which is modified by the presence of the angular momentum $J$ of the mass $M$, by means of the adimensional parameter $a$ (spin):

\begin{equation}
a = \frac{J}{GM^2/c}
\label{eq:bl2}
\end{equation}

which is the ratio between the angular momentum and its maximum value ($-1 < a < 1$). The quantity $\Sigma$ is linked to the equatorial area:

\begin{equation}
\Sigma = \sqrt{(r^2 + a^2r^2_{\rm g})^2 - a^2r_{\rm g}^2\Delta \sin^2\theta}
\label{eq:bl3}
\end{equation}

where:

\begin{equation}
\Delta = r^2 - 2r_{\rm g}r + r^2_{\rm g}a^2
\label{eq:bl4}
\end{equation}

In the end:

\begin{equation}
\omega = \frac{2r_{\rm g}^2rc}{\Sigma^2}a
\label{eq:bl5}
\end{equation}

indicates the angular velocity of the spacetime itself at different distances from the mass $M$. Let us comment on these results. If $a=0$ (non rotating mass), then $g_{\rm Kerr}=g_{\rm Schw}$. In the previous case, the singularity at the even horizon occurred when $g_{00}=0$ and $g_{11}\rightarrow \infty$. Now, there are two different surfaces. The first one, corresponding to $g_{11}\rightarrow \infty$, occurs if $\Delta=0$:

\begin{equation}
r_{\rm H} = r_{\rm g}(1+\sqrt{1-a^2})
\label{eq:kerr2}
\end{equation}

The second one, corresponding to $g_{00}=0$, is:

\begin{equation}
r_{\rm E} = r_{\rm g}(1+\sqrt{1-a^2\cos^2\theta})
\label{eq:kerr3}
\end{equation}

The two surfaces are tangent each other ($r_{\rm H}=r_{\rm E}$) only at the poles, when either $\theta = 0$ or $\theta = \pi$. Otherwise, $r_{\rm E}>r_{\rm H}$, and, particularly, $r_{\rm E}=2r_{\rm g}$ at the equator anyway. The region limited by the constraints $r_{\rm H}<r<r_{\rm E}$ is called \emph{ergosphere}. It has the peculiar property that \emph{anything} -- even the spacetime -- rotates in the same direction of the central mass, because of the elements of $g_{\rm Kerr}$ outside the main diagonal, $g_{04}$ and $g_{40}$. This frame dragging quickly drops as $r^{-3}$ just outside the ergosphere. It is still possible to escape from the gravitational pull when inside this region, although it requires extreme velocities. In the case of a maximally rotating mass, the escape velocity is $\sim 0.91c$ \cite{MEIER}.

Another rather curious point is the central singularity, when $\rho=0$. From the Eq.~(\ref{eq:bl6}):

\begin{equation}
\rho^2 = r^2 + a^2 r^2_{\rm g}\cos^2\theta = 0
\label{eq:bl6}
\end{equation}

which represents the equation of a ring with radius $a r_{\rm g}$. This means that the essential singularity of a Kerr metric is a ring, while is a point in the case of Schwarszchild. The radius of the ring depends on the rotation of the central mass.

Other solutions with the addition of electric charge have been proposed by Reissner and Nordstr\o m for the Schwarzschild metric and by Kerr and Newman for the rotating singularities (see \cite{INVERNO,MTW} for details). However, such cases are thought to be unlikely: any accumulation of charge over the horizon would give rise to electrostatic forces so strong to quickly attract opposite charges for neutralisation. 

\section{Orbits around spacetime singularities}
Orbits around a compact object, particularly around a spacetime singularity where the general relativity effects could play a significant role, are many more than what is dealt with Newtonian dynamics. A detailed analysis is beyond the aim of the present work and the interested reader could find extended studies in \cite{CHANDRA,MEIER,MTW}. Here, I would like to underline some peculiar differences of general relativity analysis. In classical mechanics (e.g. \cite{GOLDSTEIN}), the radial velocity of a particle in a central force field is given from the conservation of energy:

\begin{equation}
\dot{r}^2 = 2\mathcal{E} - V_{\rm CL}(r)
\label{eq:clpot}
\end{equation}

where the dot above the letter indicates the time derivative, $\mathcal{E}$ is the specific energy of the system and $V_{\rm CL}(r)$ is the potential, depending on the distance from the central mass. The potential can be written as:

\begin{equation}
V_{\rm CL}(r) = -\frac{1}{r^2}\left(l_{\rm z}^2 - \frac{2GM}{c^2}r\right) = -\frac{l_{\rm z}^2}{r^2} + \frac{2r_{\rm g}}{r}
\label{eq:clpot1}
\end{equation}

where $l_{\rm z}$ is the specific angular momentum along the $z$ axis (poles). By substituting Eq.~(\ref{eq:clpot1}) in Eq.~(\ref{eq:clpot}), it is possible to calculate the total energy of the system:

\begin{equation}
\mathcal{E} = \frac{\dot{r}^2}{2} + \frac{l_{\rm z}^2}{2r^2} - \frac{r_{\rm g}}{r}
\label{eq:clpot2}
\end{equation}

In the case of general relativity, and, specifically, by using the Schwarzchild metric of Eq.~(\ref{eq:schw}), the total specific\footnote{This time, specific means per unit rest mass, because in relativity energy and mass are related.} energy is \cite{RINDLER,SCHUTZ,VIETRI}:

\begin{equation}
\mathcal{E}^2 = \dot{r}^2 + \left(q + \frac{l_{\rm z}^2}{r^2}\right)\left(1-\frac{2r_{\rm g}}{r}\right)
\label{eq:grpot}
\end{equation}

where $q$ is either $1$ or $0$, in the case of particles or photons, respectively, and the dot above the letters indicates the derivation with respect to a parameter (usually the proper time) used to define the orbit. The Eq.~(\ref{eq:grpot}) is then split into two equations:

\begin{equation}
\begin{split}
\mathcal{E}_{\rm ph}^2 & = \dot{r}^2 + \frac{l_{\rm z}^2}{r^2}\left(1-\frac{2r_{\rm g}}{r}\right)\\
\mathcal{E}_{\rm p}^2 & = \dot{r}^2 + \left(1 + \frac{l_{\rm z}^2}{r^2}\right)\left(1-\frac{2r_{\rm g}}{r}\right)
\end{split}
\label{eq:grpot2}
\end{equation}

where is evident the analogous of the gravitational potential. Stable orbits correspond to minima of the potential function. In the Newtonian case the Eq.~(\ref{eq:clpot1}) has a minimum as:

\begin{equation}
r = l_{\rm z}^2/r_{\rm g}
\label{eq:clpot3}
\end{equation}

which means that if the particle has angular momentum, then there could be a stable orbit. In the case of general relativity, the minima for the particle potential are:

\begin{equation}
r = \frac{l_{\rm z}^2\pm \sqrt{l_{\rm z}^4-12l_{\rm z}^2r_{\rm g}^2}}{2r_{\rm g}}
\label{eq:grpot3}
\end{equation}

When the term under square root is negative -- i.e. if $l_{\rm z}<2r_{\rm g}\sqrt{3}$ -- then, there is no physical solution. Stable orbits can be present if $l_{\rm z} > 2r_{\rm g}\sqrt{3}$. In the case $l_{\rm z} = 2r_{\rm g}\sqrt{3}$, there is \emph{innermost stable orbit} corresponding to a circular orbit with radius: 

\begin{equation}
r_{\rm iso}=6r_{\rm g}= 3r_{\rm S}
\label{eq:grpot4}
\end{equation}

The energy of the orbit (for a particle) is given by inserting Eq.~(\ref{eq:grpot4}) in Eq.~(\ref{eq:grpot2}). Taking into account that the energy at infinity is $1$, the value at the innermost stable orbit is $\sqrt{8/9}$. As it will be shown in Chapter~6, the difference of energy is dissipated through radiative processes, which in turn implies an efficiency for Schwarzschild singularities:

\begin{equation}
\eta = 1 - \sqrt{\frac{8}{9}} \sim 0.06
\label{eq:grpot5}
\end{equation}

The case of a rotating Kerr singularity is much more complex and I refer to \cite{BARDEEN,CHANDRA,MEIER} for a complete mathematical derivation. I just report the radius of the innermost stable orbit \cite{MEIER}:

\begin{equation}
r_{\rm iso} = r_{\rm g}\left(d \mp \sqrt{\frac{7a^2-d^3+9d^2-3d(6-a^2)}{d-3}}\right)
\label{eq:grpot6}
\end{equation}

where $(-)$ refer to the prograde motion and $(+)$ to the retrograde one. The quantity $d$ is:

\begin{equation}
d = 3 + \sqrt{3+a^2+\sqrt[3]{1-a^2}\left((3-a)\sqrt[3]{1+a}+(3+a)\sqrt[3]{1-a} \right)}
\label{eq:grpot7}
\end{equation}

and the radius of the marginally bound orbit (not necessarily stable) \cite{MEIER}: 

\begin{equation}
r_{\rm mbo} = r_{\rm g}\left(1+\sqrt{1\mp a}\right)^2
\label{eq:grpot6a}
\end{equation}

In the case $a=0$, Eq.~(\ref{eq:grpot6}) reduces to the Schwarzschild case of Eq.~(\ref{eq:grpot4}). In the case of a maximally rotating singularity ($a=1$), $r_{\rm iso}=r_{\rm g}$ for a direct motion and $r_{\rm iso}=9r_{\rm g}$ for retrograde motion. In the former case, orbit corotating with the singularity, the efficiency is:

\begin{equation}
\eta = 1 - \frac{1}{\sqrt{3}} \sim 0.42
\label{eq:grpot8}
\end{equation}

It is worth mentioning that the condition $a=1$ is never satisfied, because the radiation lost by the particle and absorbed by the singularity results in a counteracting torque limiting the maximum spin value to $a\sim 0.998$ \cite{THORNE1}. This, in turn, limits the efficiency to $\eta \sim 0.30$.

\section{Gravitational waves}
As written above, Einstein proposed the equations of the gravitational field without any solution, which were instead offered later by Schwarzschild and Kerr. However, later in 1916, Einstein studied the solution in the case of weak field, that is he solved the Eqs.~(\ref{eq:campo1}) by making the hypothesis that the metric $g_{ij}$ was not too much different from that of Minkowski. Indicating the latter as $\eta_{ij}$, the weak field hypothesis can be written:

\begin{equation}
g_{ij} = \eta_{ij} + h_{ij}
\label{eq:gw1}
\end{equation}

where $h_{ij}$ is a metric as small as one likes. It is worth reminding that the gravitational field equations are non linear, because of the equivalence between mass and energy. The present technique is making a \emph{linearization}: it studies what happens when one uses as solution the Minkowski metric plus a small perturbation, by neglecting the higher order effects, as it could be the feedback of the gravitational energy on the elements of the energy-impulse tensor. Again, by operating on indexes (see for details \cite{RINDLER,SCHUTZ,WEBER}), it is possible to derive the following equation:

\begin{equation}
\square (h^{ij} - \frac{1}{2}\eta^{ij}h) = -\frac{16\pi G}{c^4}T^{ij}
\label{eq:gw2}
\end{equation}

where:

\begin{equation}
\square = \frac{\partial^2}{\partial t^2} - \frac{\partial^2}{\partial x^2} - \frac{\partial^2}{\partial y^2} - \frac{\partial^2}{\partial z^2} = \frac{\partial^2}{\partial t^2} - \nabla^2
\label{eq:gw3}
\end{equation}

This is called four-dimensional Laplace operator\footnote{The three-dimensional Laplace operator is the part of the Eq.~(\ref{eq:gw3}) indicated by the symbol $\nabla^2$.} or, simply, d'Alembert operator. It is also known as \emph{wave operator}, because it is used to speak about either electromagnetic or sound waves. While in the latter cases, there are changing electric and magnetic fields or a fluid displacement, in the former case, in the general relativity what is changing is the spacetime itself, because it is a perturbation of the metric. 

Gravitational waves were observed first by studying the orbital decay of a binary pulsar discovered in 1974 by Russell Hulse and Joseph Taylor\footnote{They obtained the Nobel prize for physics in 1993 for this work.}. The cumulative shift of the periastron time over about three decades of observations, clearly indicated that the two pulsars are gradually approaching toward a merger. The decay time is in agreement with what expected from the general relativity, implying the loss of energy through emission of gravitational waves (see \cite{WEISBERG}). Recently, there was also a direct observation by the \emph{Laser Interferometer Gravitational-Wave Observatory} (LIGO, \cite{LIGO}).

It is worth noting that sometimes this type of spacetime perturbation is referred as \emph{gravity wave} instead of \emph{gravitational wave}. This is quite a dumb mistake! The term gravity wave refers to waves generated at the contact surfaces between two fluids: for example, the waves of the sea (air-water interface).

\chapter{Radiative processes}

\section{Particle Acceleration}
The main source of radiation in relativistic astrophysics is composed of accelerated particles. Their distribution follows a power-law model:

\begin{equation}
dN \propto \mathcal{E}^{-s}d\mathcal{E}
\label{eq:powerlaw}
\end{equation}

where $N$ is the number of particles, $\mathcal{E}$ is the kinetic energy, and $s$ is the energy spectral index. This is a \emph{non-thermal} distribution, to be compared with the thermal particles (i.e. only random motion), which follows the Maxwell-Boltzmann distribution. When $s=2$, the particle number is dominated by low-energy particles, although the total energy is mostly due to the less high-energy particles. Some additional radiative loss or escape of particles from the acceleration site, will result in an increase of $s$ (steep spectrum), which in turn means that both the particle number and total energy are due to low-energy particles. Obviously, the opposite will happen in the case of a harder spectrum ($s<2$) \cite{VIETRI}. 

The problem of particle acceleration is divided into two issues: the acceleration and the injection \cite{VIETRI,MELIA}. There are many theories for both issues and are based on the simple physical principles of shocks (stochastic acceleration), and large scale electric and magnetic fields (direct acceleration). Stochastic acceleration is described by the Fermi mechanisms of the first and second order, although Enrico Fermi elaborated only the second order theory \cite{FERMI1,FERMI2}. Both theories refers to the interaction of a particle with a moving fluid (shock with velocity $v$), resulting in an average change of energy of the order of $v^2/c^2$ or $v/c$ for the second or first order acceleration, respectively \cite{VIETRI,MELIA}. Stochastic acceleration is thought to occur in supernovae. 

Large scale electric and magnetic fields could be generated in different ways, either via magnetic reconnection or via the motion of a magnetic field. We will return on these topics later, when dealing with winds and jets.

\section{Radiation from an accelerated charge}
As a charged particle is accelerated by an electric or a magnetic field, it emits electromagnetic radiation. It is worth noting that the word ``to accelerate'' means to increase the velocity, while in physics it means to change the velocity, either to increase or to decrease. It is also worth reminding that a velocity is referred as a vector, with a magnitude, a direction and a sign. Therefore, by changing velocity it means that there is a change either in magnitude, or direction, or sign. For example, the Lorentz force is the analogous of the Newton's law for the electric charge $q$ and is:

\begin{equation}
\mathbf{F} = \frac{d\mathbf{p}}{dt} = q(\mathbf{E} + \frac{\mathbf{v}}{c}\times\mathbf{B})
\label{eq:lorentzforce}
\end{equation}

where $\mathbf{E}$ is the electric field, $\mathbf{B}$ is the magnetic induction field, $\mathbf{v}$ is the particle velocity, and $\mathbf{p}$ its impulse. Eq.~\ref{eq:lorentzforce} shows that the effect of an electric field is to change the velocity full vector (magnitude, direction, sign), while the magnetic field can only change its direction (the magnetic field makes no work). Anyway, a change of the vector direction still means a change of velocity and, hence, an acceleration. 

Let us consider a distribution of non-relativistic accelerating electric charges. I just would like to recall some very basic concepts of classical electrodynamics and I refer to \cite{JACKSON,RYBICKI} for a detailed explanation. This is equivalent to a time-dependent electric current $\mathbf{J}(\mathbf{x},t)$, which in turn can be studied by using the Fourier series as $\mathbf{J}(\mathbf{x},t)=\mathbf{J}(\mathbf{x})e^{-i\omega t}$, where $\omega$ is the angular frequency of the charge oscillation. The solution for the (retarded) vector potential is (primed quantities refer to the source frame):

\begin{equation}
\mathbf{A}(\mathbf{x},t) = \frac{1}{c}\int d^3 x' \int dt' \frac{\mathbf{J}(\mathbf{x}',t')}{|\mathbf{x}-\mathbf{x}'|}\delta(t'-t+\frac{|\mathbf{x}-\mathbf{x}'|}{c})
\label{eq:potrit1}
\end{equation}

which is simplified by assuming a sinusoidal current density, as stated above:

\begin{equation}
\mathbf{A}(\mathbf{x}) = \frac{1}{c} \int \mathbf{J}(\mathbf{x}')\frac{e^{ik|\mathbf{x}-\mathbf{x}'|}}{|\mathbf{x}-\mathbf{x}'|}d^3x'
\label{eq:potrit2}
\end{equation}

where $k=\omega/c$ is the wave number. From Eq.~(\ref{eq:potrit2}), it is then possible to calculate the magnetic induction as $\mathbf{B}= \mathbf{\nabla}\times\mathbf{A}$ and the electric field as $\mathbf{E}=i\mathbf{\nabla}\times\mathbf{B}/k$ \cite{JACKSON}. If the emitting charges have size $d$, $\lambda=2\pi c/\omega$ is the wavelength, and $d\ll \lambda$ then the electromagnetic field could behave differently depending on the distance $r=|\mathbf{x}-\mathbf{x}'|$ from the source:

\begin{enumerate}
\item $d\ll r\ll \lambda$, near zone (static field);
\item $d\ll r\sim \lambda$, intermediate zone (induction field);
\item $d\ll \lambda \ll r$, far zone (radiation field).
\end{enumerate}

It is rather obvious that the point $3$ is of interest in astrophysical cases. In the far field, it is possible to adopt the electric dipole approximation, by taking into account that Eq.~(\ref{eq:potrit2}) is dominated by the exponential term. It is possible to conclude the emitted power (change of energy with time) as:

\begin{equation}
P = \frac{d\mathcal{E}}{dt} = \frac{2\mathbf{\ddot{d}}}{3c^3}
\label{eq:dipolepower}
\end{equation}

where $\mathbf{d}=\sum_i q_i \mathbf{r}_i$ is the electric dipole moment. The relativistic version of Eq.~(\ref{eq:dipolepower}) is:

\begin{equation}
P = \frac{2q^2\mathbf{a}'\cdot\mathbf{a}'}{3c^3} = \frac{2q^2}{3c^3}({a'}_{\perp}^2 + {a'}_{\parallel}^2) = \frac{2q^2}{3c^3}\gamma^4(a_{\perp}^2+\gamma^2 a_{\parallel}^2)
\label{eq:reldipolepower}
\end{equation}

where $\mathbf{a}'$ is the three-vector of the acceleration in the rest frame, while $a_{\perp}$ and $a_{\parallel}$ are the components perpendicular and parallel to the velocity, respectively.

The radiation from an accelerated charge is named after the German word \emph{bremsstrahlung}, which means braking radiation. As from Eq.~(\ref{eq:lorentzforce}), there are two types of bremsstrahlung: electric and magnetic. The former is just called bremsstrahlung or thermal bremsstrahlung, while the latter is known as cyclotron or synchrotron radiation, depending on the particle speed (non relativistic, relativistic, respectively).

Thermal bremsstrahlung occurs when there is collision between particles with different masses. If the particles are all equal, there is no bremsstrahlung, because there is no relative acceleration. In the case of electrons scattered by ions, the total emitted power per unit volume and unit frequency is \cite{VIETRI}:

\begin{equation}
P_{\rm br}(\nu) = 6.8\times 10^{-38} Z_{\rm i}^2 n_{\rm e}n_{\rm i} \sqrt{T_{\rm e}} e^{-\frac{h\nu}{kT_{\rm e}}} f_{\rm G} \, [{\rm erg\ s^{-1}cm^{-3}Hz^{-1}}] 
\label{eq:brem1}
\end{equation}

where $Z_{\rm i}$ is the electric charge of the ions, $n_{\rm e}$ and $n_{\rm i}$ are the number density of electrons and ions, respectively, $T_{\rm e}$ is the temperature of electrons, $\nu$ is the frequency, $f_{\rm G}$ is the Gaunt factor, to take into account quantum effects. Among the different factors in the Eq.~(\ref{eq:brem1}), it is worth noting: 

\begin{enumerate}
\item the term $\exp{(h\nu/kT_{\rm e})}$, which is representative of the thermal distribution (Maxwell-Boltzmann);
\item the term $\sqrt{T_{\rm e}}$, indicating that slowest interactions (i.e. with the largest deflection) have greatest weight. 
\end{enumerate}

In the case of relativistic electrons, Eq.~(\ref{eq:brem1}) has to be corrected as \cite{VIETRI}:

\begin{equation}
P_{\rm br,rel}(\nu) =  P_{\rm br}(\nu) \left( 1+\frac{T_{\rm e}}{2.3\times 10^9 \,\rm{K}} \right)
\label{eq:brem2}
\end{equation}

As shown by the constant at denominator of the temperature, the relativistic corrections are significant only at very high energies ($T_{\rm e}\sim 2.3\times 10^9 \,\rm{K} \sim 200 \,\rm{keV}$), but at these energies other processes (Compton scattering) dominates the collision between particles.

\section{Synchrotron}
Eq.~(\ref{eq:lorentzforce}) shows that the magnetic field change only the direction of the particle. This is an acceleration, but it results in a circular motion (in absence of particle drift) with frequency:

\begin{equation}
\omega_{\rm L} = \frac{qB}{\gamma mc} \sim 587 \frac{B}{\gamma} \,[\rm{rad/s}] \sim 1.8\times 10^{7}\frac{B}{\gamma} \,[\rm{rad/s}]
\label{eq:larmor}
\end{equation}

where the numerical approximations are valid for the electron. The first one is in the units of the International System ($B$ is in Tesla), and the second one is in the Gaussian system ($B$ in Gauss, $1\rm{G}=10^{-4}\rm{T}$). This frequency is called after \emph{Larmor} or \emph{gyrofrequency} or \emph{cyclotron frequency}. The emitted spectrum in the case of a non-relativistic particle is dominated by the emission lines at the \emph{gyrofrequency} and its harmonics, and it is called \emph{cyclotron}. As the particle velocity increases to relativistic values, the individual emission lines are so close that cannot be resolved and the spectrum is a quasi continuum. The process is now called \emph{synchrotron}.

The emitted power can be calculated by making use of Eqs.~(\ref{eq:lorentzforce}) and (\ref{eq:reldipolepower}). In the case of null electric field, Eq.~(\ref{eq:lorentzforce}) becomes:

\begin{equation}
\frac{d\mathbf{p}}{dt} = \gamma m\frac{d\mathbf{\mathbf{v_{\perp}}}}{dt} = q\frac{\mathbf{v_{\perp}}}{c}\times\mathbf{B}
\label{eq:lorentzforce2}
\end{equation}

There is no acceleration component parallel to the particle velocity, because of the vector product in the left-hand side. By substituting in Eq.~(\ref{eq:reldipolepower}), and taking into account that $a_{\perp}=\omega_{\rm L} v_{\perp}$:

\begin{equation}
P = \frac{2q^2}{3c^3}\gamma^4 a_{\perp}^2 = \frac{2q^2}{3c^3}\gamma^4 (\frac{qB}{\gamma mc}v_{\perp})^2 = \frac{2}{3}r_{0}^{2} c B^{2}\gamma^2 \beta_{\perp}^2
\label{eq:reldipolepower2}
\end{equation}

where $r_0 = q^2/mc^2$ is the classical electron radius. In the case of an isotropic distribution of velocities, the value of $\beta_{\perp}$ has to be averaged over all the angles between the velocity and the field, thus resulting in an additional factor $2/3$. Moreover, by taking into account the Thompson cross section: 

\begin{equation}
\sigma_{\rm T}=\frac{8}{3}\pi r_{0}^{2} \sim 6.65\times 10^{-25}~{\rm cm}^{2}
\label{eq:thompson}
\end{equation}

and the magnetic energy density $U_{\rm B}=B^2/8\pi$, the Eq.~(\ref{eq:reldipolepower2}) becomes:

\begin{equation}
P = \frac{4}{3}\sigma_{\rm T}c \beta^2 \gamma^2 U_{\rm B}
\label{eq:reldipolepower3}
\end{equation}

The emitted power is proportional to the magnetic energy density and to the Thompson cross section: this implies that it is significant for leptons, while negligible for hadrons. The cooling time for the single electron is defined as:

\begin{equation}
t_{\rm c,syn} = \frac{\mathcal{E}}{P} = \frac{\gamma mc^2}{P} \propto \frac{1}{\gamma B^2}
\label{eq:synchrocooling}
\end{equation}

that is, electrons with greatest energies in strong magnetic fields lose energy faster. 

\begin{figure}[ht]
\centering
\includegraphics[angle=270,scale=0.5]{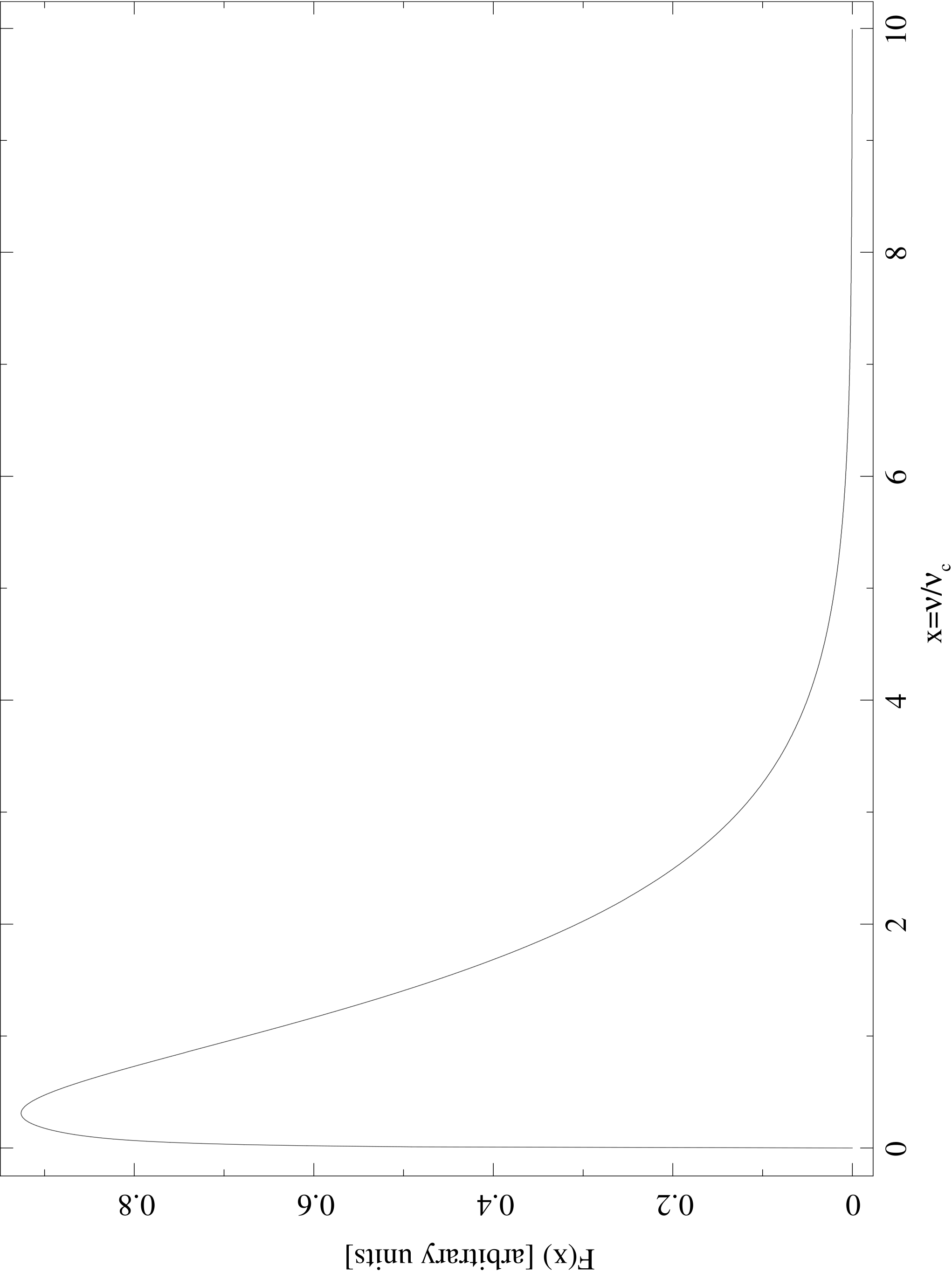}
\caption{Function $F(x)$ of Eq.~(\ref{eq:bessel1}).}
\label{fig:fsynchro}
\end{figure}

The spectrum of the individual particle at a certain time is given by \cite{JACKSON,VIETRI}:

\begin{equation}
dP = \frac{\sqrt{3}q^3B \sin \vartheta}{mc^2}F\left(\frac{\nu}{\nu_{\rm c}}\right)d\nu
\label{eq:spettroist}
\end{equation}

where $\vartheta$ is the angle between the velocity and the magnetic field $B$, and $\nu_{\rm c}$ is a critical frequency defined later. By indicating $x=\nu/\nu_{\rm c}$, $F(x)$ is a function defined as:

\begin{equation}
F(x) = x \int_{x}^{\infty} K_{5/3}(y)dy
\label{eq:bessel1}
\end{equation}
 
where $K_{5/3}$ is the modified Bessel function\footnote{See \cite{MF} for details about this kind of functions.} of order $5/3$. $F(x)$ can be approximated as:

\begin{equation}
F(x) \sim 
\begin{cases}
\frac{4\pi}{\sqrt{3}\Gamma(1/3)}\sqrt[3]{\frac{x}{2}} & x \ll 1\\
\sqrt{\frac{\pi x}{2}} \exp(-x) & x \gg 1
\end{cases}
\label{eq:bessel}
\end{equation}

where $\Gamma(1/3)$ is the Euler function\footnote{See \cite{MF} for details about this kind of functions.} $\Gamma$ calculated for $1/3$. The behaviour of $F(x)$ is displayed in Fig.~\ref{fig:fsynchro}. The peak $F(x)\sim 0.92$ is at $x=\nu/\nu_{\rm c}\sim 0.29$.

The emitted spectrum of Eq.~(\ref{eq:spettroist}), as written above, is broad band with a peak at the critical frequency:

\begin{equation}
\nu_{\rm c} = \frac{3\gamma^2 qB \sin \vartheta}{4\pi mc}
\label{eq:fcrit}
\end{equation}

which is different from the gyrofrequency of Eq.~(\ref{eq:larmor}) by a factor $\gamma^3$. The difference is due to the relativistic beaming, which confines the emission within a cone of angle $\sim \gamma^{-1}$ and compresses the time of arrival of photons by a factor $\sim \gamma^{-2}$ (detailed explanations can be found on \cite{COURVOISIER,RYBICKI}). 

Electrons in cosmic sources generally have power-law distributions of Eq.~(\ref{eq:powerlaw}). Therefore, the observed spectrum is the convolution of Eq.~(\ref{eq:spettroist}) with the electron distribution, where particles with higher energy contribute to the highest frequencies. Since the relativistic energy is $\mathcal{E}=\gamma mc^2$, the number particle distribution can be generally rewritten as:

\begin{equation}
n(\gamma)d\gamma = n_0 \gamma^{-s}d\gamma
\label{eq:edistribution}
\end{equation}

where $2 \lesssim s \lesssim 3$, and $\gamma_{\rm min} \le \gamma \le \gamma_{\rm max}$ ($\gamma_{\rm min} \ll \gamma_{\rm max}$, $\gamma_{\rm min}\sim 1$). Correspondingly, we can set two frequencies according to Eq.~(\ref{eq:fcrit}), where is the peaks of synchrotron emission from particles with energies $\gamma_{\rm min},\gamma_{\rm max}$:

\begin{equation}
\begin{split}
\nu_1 & = \frac{3qB}{4\pi mc}\gamma_{\rm min}^{2}\\
\nu_2 & = \frac{3qB}{4\pi mc}\gamma_{\rm max}^{2}
\end{split}
\label{eq:fcrit12}
\end{equation}

The convolved spectrum is:

\begin{equation}
f_{\nu} \sim \int_{\gamma_{\rm  min}}^{\gamma_{\rm max}} P_{\nu}n(\gamma)d\gamma
\label{eq:convspec}
\end{equation}

where the spectrum emitted by one particle has been approximated by a Dirac's delta function: $P_{\nu}\sim P\delta(\nu - \nu_{\rm c})$. The above integral could be easily\footnote{If not so easy, you can find detailed derivation in \cite{COURVOISIER}.} calculated and gives the result:

\begin{equation}
f_{\nu} \propto \nu^{-\alpha}
\label{eq:risultato}
\end{equation}

where $\alpha=(s-1)/2$ is the spectral index of the synchrotron spectrum, which is linked to the energy index $s$ of the particle distribution. Eq.~(\ref{eq:risultato}) is valid within $\nu_1 \leq \nu \leq \nu_2$, but it is of some interest also to know what happens outside that interval. In the case $\nu < \nu_1$, $f_{\nu} \propto \sqrt[3]{\nu}$, according to Eq.~(\ref{eq:bessel}), while if $\nu > \nu_2$, the emission is proportional to $\exp{(-\nu/\nu_2)}$, and therefore it quickly drops to negligible values ($f_{\nu}\rightarrow 0$). It is useful to remind a detailed formula for the synchrotron emission for a power-law distribution of electrons, reported with detailed derivation in \cite{RYBICKI}: 

\begin{equation}
P_{\rm tot}(\nu) = \frac{\sqrt{3}q^3 n_0 B \sin \vartheta}{2\pi mc^2 (s+1)}\Gamma \left(\frac{s}{4}+\frac{19}{12} \right)\Gamma \left(\frac{s}{4}-\frac{1}{12}\right)\left(\frac{2\pi mc\nu}{3qB\sin\vartheta}\right)^{\frac{-s+1}{2}}
\label{eq:synchrodetail}
\end{equation}

where $P_{\rm tot}(\nu)$ is the total emitted power per unit volume and unit frequency [erg~cm$^{-3}$~s$^{-1}$~Hz$^{-1}$]. I underline \emph{total}, because the synchrotron emission is polarised. The degree of polarisation is related to the energy index of the electron distribution \cite{RYBICKI}:

\begin{equation}
PD = \frac{s+1}{s+\frac{7}{3}}
\end{equation}

PD could reach values up to $\sim 75$\%, but observed values are often smaller ($\sim 30-40$\% in the best cases, e.g. \cite{IKEJIRI}), because of a series of factors contributing to depolarise the emitted radiation. 

\subsection{Synchrotron self-absorption}
It is also worth mentioning the \emph{synchrotron self-absorption}, which occurs when the electrons absorb their own emitted radiation. This is observed at low frequencies (radio), as a change of the spectral shape. For frequencies below a certain value, which depends on the cosmic source, the synchrotron emission is self-absorbed and the source is optically thick. By studying the equation of the radiative transport \cite{RYBICKI,NOBILI}, it is possible to find that $I_{\nu}\propto \nu^{5/2}$, where $I_{\nu}$ is the specific intensity of the cosmic source. As expected, an optically thick source is independent on the characteristics of the particle distribution. As the frequency increases, the optical depth $\tau$ decreases. As $\tau \sim 1$, the optical regime changes from thick to thin and $I_{\nu}\propto \nu^{-(s-1)/2}$, as expected from Eq.~(\ref{eq:risultato}). The self-absorbed spectrum proportional to $\nu^{5/2}$ is typical of the synchrotron emission and, thus, it could be used as a proof of such process. 

\subsection{Brightness temperature}
Radio astronomers often use the quantity named \emph{brightness temperature} defined as:

\begin{equation}
T_{\rm B} = \frac{I_{\nu}c^2}{2k_{\rm B}\nu^2} = \frac{F_{\nu}}{\pi \theta^2}\frac{c^2}{2k_{\rm B}\nu^2}
\label{eq:tb}
\end{equation}

where $k_{\rm B}$ is the Boltzmann constant, $F_{\nu}$ is the observed flux density at the frequency $\nu$, and $\theta$ is the angular size of the cosmic source. The brightness temperature thus indicates the equivalent temperature of a black body in the Rayleigh-Jeans regime ($h\nu \ll \gamma mc^2$) emitting a specific intensity $I_{\nu}$. Under condition of thermal equilibrium, the thermal energy is equal to the kinetic energy of the particles:

\begin{equation}
3k_{\rm B}T_{\rm B} \sim \gamma mc^2 = 3k_{\rm B}T_{\rm e}
\end{equation}

If not, it is clearly a non-thermal particle distribution. The observed values in cosmic sources are generally $\sim 10^{10-12}$~K, to be compared with the $\sim 3$~K thermal emission of the cosmic background, thus favouring the non-thermal hypothesis.  
The emitted spectrum is optically thin if $T_{\rm B}\lesssim T_{\rm e}$, otherwise it is optically thick, because of self-absorption, and only the radiation emitted by the surface of the source can be observed. In the former case, the emitted spectrum is given by Eq.~(\ref{eq:risultato}), while in the latter $f_{\nu}\propto \nu^{5/2}$, independently on the particle energy distribution. The turnover frequency depends on the particle density. This happens in theory; in practice, it is much more complex. Cosmic sources have rather different geometries and particle distributions, and their emission changes with time. The large scale structures of relativistic jets have low particle densities, so that they are optically thin already at MHz frequencies. The radio core is instead generally optically thick up to GHz frequencies, but often is not possible to observe a turnover or the expected spectral shape (see, for example, \cite{KELLERMANN}). In addition, in the core of a relativistic jet, $T_{\rm B}$ could exceed $\sim 10^{12}$~K, because of the inverse-Compton scattering, which increases the electron cooling, and the relativistic beaming \cite{READHEAD}. However, recent radio observations by Kovalev et al. \cite{KOVALEV} challenged this scenario by requiring a Doppler factor much greater than that measured by the kinematics of plasma blobs. 

\subsection{Curvature radiation}
As stated above, the electron emits synchrotron energy by spiralling around the magnetic field lines. However, close to compact objects (neutron stars, black holes), the magnetic field lines are curved. As known from general relativity, curvature is acceleration, which means that, in addition to synchrotron emission, electrons spiralling around curved magnetic field lines emit also a \emph{curvature radiation}. Indicating the curvature radius with $\rho$, the acceleration is $a_{\perp}=v_{\perp}^2/\rho$, which in turn has to be inserted into the Eq.~(\ref{eq:reldipolepower}). This means that the curvature radiation is similar to the synchrotron radiation, with the difference that $\omega_{\rm L}$ has been replaced by $v_{\perp}/\rho$ \cite{VIETRI}.

\section{Compton scattering}
The interaction between a particle and an electromagnetic wave is elastic (no energy exchange) in classical physics, and is regulated by the Thompson cross section of Eq.~(\ref{eq:thompson}). In quantum physics, the electromagnetic wave can be described\footnote{This is the wave-particle complementarity stated by Niels Bohr.} by a particle (photon) with energy $\mathcal{E}=h\nu$ and impulse $p=h\nu/c$. The interaction of a photon with a particle is no more elastic and there could be energy exchange. This is the Compton scattering and is regulated by the Klein-Nishina cross section \cite{RYBICKI}:

\begin{equation}
\sigma_{\rm KN} = \sigma_{\rm T}\frac{3}{4}\left[\frac{a}{x^3}\left\lbrace\frac{2xa}{b}-\ln b\right\rbrace+\frac{1}{2x}\ln b-\frac{c}{b^2}\right]
\label{eq:kn}
\end{equation}

where $x=h\nu/mc^2$, $a=(1+x)$, $b=(1+2x)$, and $c=(1+3x)$. By studying the change of four impulse of the photon before and after the scattering, it is possible to derive the change in wavelength (or frequency or energy) due to the Compton scattering (see \cite{RYBICKI,VIETRI} for details):

\begin{equation}
\lambda_{\rm a} - \lambda_{\rm b} = \lambda_{\rm C}(1-\cos \alpha)
\end{equation}

where $\lambda_{\rm b},\lambda_{\rm a}$ are the photon wavelengths before and after the scattering, respectively. $\alpha$ is the angle between the two directions, and: 

\begin{equation}
\lambda_{\rm C} = \frac{h}{mc} = 0.02426 \,{\rm \AA}
\end{equation}

is the Compton wavelength for electrons. If $\lambda_{\rm b} \gg \lambda_{\rm C}$ (i.e. $x\ll 1$), there is no energy exchange (elastic scattering) and $\sigma_{\rm KN}\sim \sigma_{\rm T}b$. On the other side, if $x\gg 1$, which is the relativistic case:

\begin{equation}
\sigma_{\rm KN}\sim \frac{3}{8}\sigma_{\rm T} \frac{\ln 2x +\frac{1}{2}}{x}
\label{eq:kn2}
\end{equation}

In practice, the quantum corrections reduce the Thompson cross section and the decrease is more significant as the energy of the photon increases. Generally, it is common to refer to Eq.~(\ref{eq:kn2}) for the Klein-Nishina regime, while in the case $x\ll 1$, it is common to neglect quantum corrections and to use the Thompson cross section.

In astrophysics, it is important to study the \emph{inverse-Compton scattering}, occurring when the electron is moving at relativistic speed and hits a low-energy photon ($x \ll 1$ in the rest frame of the electron) \cite{RYBICKI,VIETRI}. In this case, the electron transfers part of its energy to the photon, which increases its own energy. Therefore, it is the main cooling mechanism for high-energy electrons. Let us indicate the observer frame with unprimed quantities, while the electron rest frame has primed quantities. It is necessary to take into account relativistic effects of changing reference frames, so that the photon energy before the scattering is observed from the electron rest frame as:

\begin{equation}
h\nu'_{\rm b} = h\nu_{\rm b}\gamma(1-\beta \cos \theta_{\rm b})
\label{eq:comptondoppler1}
\end{equation}

where $\theta$ is the angle between the photon and the electron trajectories as seen by the observer. In the electron rest frame, the photon energy is almost unchanged, because of Thompson scattering regime:

\begin{equation}
h\nu'_{\rm a} \sim h\nu'_{\rm b}\left[1-\frac{h\nu'_{\rm b}}{mc^2}(1-\cos \Theta)\right]
\label{eq:comptondoppler2}
\end{equation}

where $\Theta$ is the deflection angle of the impulse four vector. Taking into account Eq.~(\ref{eq:comptondoppler2}) with the condition $x\ll 1$ in the electron rest frame, the photon has now a greater energy in the observer frame:

\begin{equation}
h\nu_{\rm a} = h\nu'_{\rm b}\gamma(1+\beta \cos \theta'_{\rm a})
\label{eq:comptondoppler3}
\end{equation}

By comparing Eq.~(\ref{eq:comptondoppler3}) with Eq.~(\ref{eq:comptondoppler1}), it is clear that the observer sees the photon after the scattering with a greater energy:

\begin{equation}
h\nu_{\rm a} = h\nu_{\rm b}\gamma^2(1+\beta \cos \theta'_{\rm a})(1-\beta \cos \theta_{\rm b})
\label{eq:comptondoppler4}
\end{equation}

Generally, $\theta_{\rm b}\sim\theta'_{\rm a}\sim \pi/2$, which means that the increase of energy is of the order of $\sim \gamma^2$. There also could be some unfavourable angles for which $\gamma^2(1+\beta \cos\theta'_{\rm a})(1-\beta \cos\theta_{\rm b})\lesssim 1$: in those cases, the photon energy after the scattering is smaller. By averaging over the angles, the change in the photon energy is composed of two terms \cite{VIETRI}:

\begin{equation}
\overline{h\nu_{\rm a} - h\nu_{\rm b}}\sim -\frac{\overline{(h\nu_{\rm b})^2}}{mc^2}+4\beta^2 \overline{h\nu_{\rm b}}
\end{equation}

where the line over the symbols indicates the average over the angles. The first term on the right side (negative) is the energy transferred from the photon to the electron (recoil), while the second term is the energy gain because of relativistic beaming. The power emitted by the electron because of inverse-Compton emission is \cite{RYBICKI,VIETRI}:

\begin{equation}
P_{\rm IC} = \frac{4}{3}\sigma_{\rm T}c\gamma^2\beta^2 U_{\rm ph}\left(1-\frac{63}{10}\frac{\gamma\overline{h^2\nu^2}}{mc^2\overline{h\nu}}\right)
\label{eq:icpower}
\end{equation}

where $U_{\rm ph}$ is the target photon energy density. In astrophysical sources, photons are often from sources outside the streaming of relativistic electrons (accretion disk, broad-line region, molecular torus,... see \cite{BOTTCHER,DERMER,BS,PETERSON}), so that the process is named \emph{external Compton}.
The negative term inside the parentheses is the electron recoil, which is generally negligible as we have made the hypothesis $x\ll 1$. Therefore, Eq.~(\ref{eq:icpower}) is quite similar to the synchrotron power Eq.~(\ref{eq:reldipolepower3}), with the only difference in the energy density (magnetic vs target photons). This means that the ratio of magnetic and photon energy density drives the ratio of the radiative power due to synchrotron and inverse-Compton processes:

\begin{equation}
\frac{P_{\rm IC}}{P_{\rm syn}} \sim \frac{U_{\rm ph}}{U_{\rm B}}
\label{eq:sicpow}
\end{equation}

The cooling time of electrons due to inverse-Compton scattering is:

\begin{equation}
t_{\rm c,IC} = \frac{\mathcal{E}}{P} = \frac{\gamma mc^2}{P} \propto \frac{1}{\gamma}
\label{eq:iccooling}
\end{equation}

so, it depends only on the relativistic beaming, while in the synchrotron case Eq.~(\ref{eq:synchrocooling}) depends also on the magnetic field. 

The spectrum of the Compton radiation depends on the number of particle interaction. The single-scattering spectrum has a spectral index $\alpha=(s-1)/2$, as in the synchrotron case. This is particularly true in the case of compact sources, when the seed photons for the inverse-Compton scattering are the synchrotron photons themselves (\emph{synchrotron self-Compton, SSC}). The spectrum could be steeper, with index $\alpha=s$, in the case the photon energy is larger than the electron one ($x\gg 1$), i.e. in the Klein-Nishina regime of Eq.~(\ref{eq:kn2}). Then, the electrons recoil is no more negligible. The particles tend to lose their energy in one single scattering, since the collision rate is decreased because of the smaller electron-photon cross section in the Klein-Nishina regime. 

The emitted spectrum spans over a range of frequencies between $\nu_{\rm min}=\gamma_{\rm min}^4\nu_{\rm L}$ and  $\nu_{\rm max}=\gamma_{\rm max}^4\nu_{\rm L}$, where $\nu_{\rm L}$ is the Larmor frequency defined in Eq.~(\ref{eq:larmor}). The emitted power scales with Eq.~(\ref{eq:sicpow}): if the ratio is greater than $1$, then electrons lose most of their energy because of inverse-Compton scattering, while the opposite means that is synchrotron radiation to dominate the cooling. When the ratio is equal to $1$, there is equilibrium between the two processes. When the Compton process dominates the cooling, there is an increase of energy exchange from electrons to photons, with multiple scatterings, each one increasing the energy losses. The ratio equal to $1$ is called the threshold for a \emph{Compton catastrophe}: despite of the dreadful name, it just means that the Compton scattering dominates the cooling of electrons with an extremely high pace. This is related to the brightness temperature of Eq.~(\ref{eq:tb}). By assuming that the photon spectrum is self-absorbed up to a frequency $\nu_{\rm sa}$, it is possible to obtain critical brightness temperature corresponding to a ratio Eq.~(\ref{eq:sicpow}) equal to $1$ \cite{VIETRI}:

\begin{equation}
T_{\rm cr} = 10^{12}\sqrt[5]{\frac{\rm 1\,GHz}{\nu_{\rm sa}}} \, [{\rm K}]
\label{eq:tb2}
\end{equation}

There could be no cosmic source with $T_{\rm b}>T_{\rm cr}$. If any, this means that other factors play a role (e.g. relativistic beaming \cite{READHEAD}).

\subsection{Comptonisation}
Let us consider a thermal electron distribution and photons traveling through it. Depending on the energies of the individual particles, there could be Compton or inverse-Compton scatterings. Although there could be even no net energy exchange between the two particle populations, the emitted photon spectrum could be significantly different from the incident one. A measure of this difference is quantified by using the Compton $y$ parameter, defined as:

\begin{equation}
y = \frac{\Delta E}{E}N_{\rm es}
\end{equation}

where $\Delta E/E$ is the average energy change of the photon energy per single scattering and $N_{\rm es}$ is the number of scatterings. If the photon energy is greater than the electron energy ($E\gtrsim kT_{\rm e}$), then the former gives energy to the latter (Compton scattering) and the gas heats up. On the opposite, ($E\lesssim kT_{\rm e}$, inverse Compton scattering), the gas cools down by transferring its energy to photons. There is no energy transfer when $E\sim kT_{\rm e}$, because heating and cooling are balanced and it corresponds to the Compton temperature $T_{\rm C}$.

It is possible to derive expressions for $y$ in the cases of relativistic\footnote{But see \cite{VIETRI} for the effective utility of the Compton parameter in the relativistic case.} and non-relativistic thermal electron distributions \cite{RYBICKI}:

\begin{equation}
\begin{split}
y_{\rm non-rel} & = \frac{4kT}{mc^2}\max(\tau_{\rm es},\tau_{\rm es}^2)\\
y_{\rm rel} & = \left(\frac{4kT}{mc^2}\right)^2\max(\tau_{\rm es},\tau_{\rm es}^2)
\end{split}
\label{eq:comptonpar}
\end{equation}

where $\tau_{\rm es}$ is the optical depth for electron scattering. In the case of ionised hydrogen (protons and electrons):

\begin{equation}
\tau_{\rm es} \sim \frac{\rho \sigma_{T}R}{m_{\rm p}}
\end{equation}

being $R$ and $\rho$ the size and density of the medium, respectively, and $m_{\rm p}$ the mass of the proton. Although the thermal equilibrium implies no net energy exchange between particles, the corresponding distributions could change from their initial values to comply the thermal equilibrium. With reference to Eqs.~(\ref{eq:comptonpar}), when $y\gg 1$, the emergent spectrum is significantly modified (\emph{saturated Comptonisation}), while negligible changes happens when $y \ll 1$ (\emph{modified blackbody}). In the former case, optically thick medium, the term $\max (\tau_{\rm es},\tau_{\rm es}^2) = \tau_{\rm es}^2$, so that the average number of scatterings is calculated according to the random walk technique with a mean free path $l$, resulting to be $N_{\rm s}\sim (R/l)^2\sim \tau_{\rm es}^2$. In the latter case, optically thin medium, $N_{\rm s}\sim (1- e^{-\tau_{\rm es}})\sim \tau_{\rm es}$. The intermediate case ($y\sim 1$) is called \emph{unsaturated Comptonisation}. 
A more detailed analysis requires the study of the Kompaneets equation (e.g. \cite{COURVOISIER,VIETRI,RYBICKI}).

\chapter{Accretion onto compact objects}

\section{The black hole paradigm}
Today, it is well grounded the paradigm that some cosmic sources, like active galactic nuclei (AGN) and X-ray binaries (XRB), are powered by accretion of matter onto black holes or neutron stars (see \cite{PETERSON,BS} for reviews on AGN). The history of the building of this paradigm is quite complex, but extremely fascinating and educational: I recommend the reading of different contributions (particularly, Martin Gaskell's one) in \cite{DONOFRIO} and I refer to it for further details and bibliography. Historically, black holes awareness came earlier than neutron stars and was driven by the observational evidence of broad profiles of optical emission lines and the need to explain an unusually strong energetics in a limited space of the order of a $\sim$pc$^3$. The former was explained by the need of supermassive black holes ($\sim 10^8M_{\odot}$) at the centre of galaxies, while the latter was explained by the accretion onto the compact object. To have some orders of magnitude, we refer to the Kerr metric (see Sect.~4.1). The gravitational radius of a supermassive black hole as a function of the mass $M$ in units of $M_{8}=M/10^8M_{\odot}$:

\begin{equation}
r_{\rm g} = \frac{GM}{c^2}\sim 1.5\times 10^{13}M_{8} \, [{\rm cm}]\sim 0.006M_{8} \, [{\rm light\, days}] \sim 215R_{\odot}M_{8}
\label{eq:rgrav}
\end{equation}

which is beyond the resolution of any telescope\footnote{Although, there are attempts to observe the event horizon in the closest black holes. See \url{http://www.eventhorizontelescope.org/}.}. The corresponding angular momentum per unit mass is:

\begin{equation}
J = \sqrt{GMr_{\rm g}} \sim 5\times 10^{23}M_{8} \, [{\rm cm^2s^{-1}}]
\end{equation}

This, in turn, is smaller than the typical angular momentum of the interstellar gas in the galaxy $\sim 10^{30}$~cm$^2$s$^{-1}$, thus implying the need of an accretion disk \cite{BLANDFORD}. As shown in Sect.~4.2, there could be stable orbits around spacetime singularities, although it is worth noting that are not closed ellipses, because of the precession (quite significant in these cases). 

\section{Accretion disks}
There are many ways to accrete matter onto compact objects (see \cite{MEIER} for a detailed review), but the most common seems to be the formation of a disk where the matter could dissipate angular momentum. Many theories have been proposed to describe these disks \cite{DISKS},  and the best known -- also called \emph{standard disk} -- is an optically thick, geometrically thin disk due to Shakura and Sunyaev \cite{SS}. The former property means that the energy is dissipated locally (\emph{local thermodynamic equilibrium}, LTE) and the disk as a whole can be considered radiating as a blackbody. According to the virial theorem \cite{GOLDSTEIN}, the gravitational potential energy is transformed in radiation with a rate $GM\dot{m}/r$, and half is heating the gas, while the other half is radiated away. Therefore, by taking into account the Stefan-Boltzmann law, the observed luminosity $L$ is:

\begin{equation}
L = \frac{GM\dot{m}}{2r} = 2\pi r^2 \sigma_{\rm SB} T^4
\label{eq:disklum}
\end{equation}

where $\dot{m}$ is the mass accretion rate, $\pi r^2$ is the disk area\footnote{There are two sides, thus there is the multiplicative factor $2$.}, and $\sigma_{\rm SB}=5.670400\times 10^{-8}$~W~m$^{-2}$~K$^{-4}$ is the Stefan-Boltzmann constant. From Eq.~(\ref{eq:disklum}), it is possible to calculate the temperature:

\begin{equation}
T = \sqrt[4]{\frac{GM\dot{m}}{4\pi\sigma_{\rm SB}r^3}}
\label{eq:disktemp}
\end{equation}

In the case of a Kerr singularity, the last stable corotating orbit has $r=r_{\rm g}$. Therefore, by taking into account Eq.~(\ref{eq:rgrav}), the Eq.~(\ref{eq:disktemp}) can be rearranged as:

\begin{equation}
T = \sqrt[4]{\frac{\dot{m}c^6}{4\pi\sigma_{\rm SB}G^2M^2}} \propto \frac{\sqrt[4]{\dot{m}}}{\sqrt{M}}
\label{eq:disktemp2}
\end{equation}

If using a Schwarzschild metric or a counterrotating Kerr singularity, the result does not change except for some multiplicative factor. What is important in Eq.~(\ref{eq:disktemp2}) is the inverse dependence on the square root of the mass. This means that larger black hole have smaller disk temperatures, while the opposite holds for small black holes. Indeed, there is observational evidence that accretion disks around stellar-mass singularities peak at X-rays (energies of about hundreds of eV), while supermassive black holes in active galactic nuclei have disks emitting most of their energy in the ultraviolet. 

Eq.~(\ref{eq:disktemp}) has neglected the disk structure. The geometrically thin hypothesis means that there is negligible radial dispersion of energy (no advection\footnote{The term \emph{advection} indicates the bulk motion in a fluid. In the Eulerian fluid dynamics, one does consider the motion of an elementary volume of fluid rather than that of the individual particle. Therefore, changes of any quantity $x$ in the reference frame of the fluid volume moving with velocity $\mathbf{v}$ is given by: $dx/dt = \partial x/\partial t + \mathbf{v}\cdot \nabla x$. For example, the equation of mass conservation (continuity equation) is $\partial \rho/\partial t + \mathbf{v}\cdot \nabla \rho$ or, simply, $d\rho/dt=0$. Often, the term advection is used as synonym of \emph{convection} -- thus generating some confusion -- but while the former refer to a generic bulk motion, the latter is a specific form of bulk motion due to changes in the fluid density (for example, because of a temperature gradient as in boiling water).}) with respect to what is radiated away along the disk height. To explain how matter could lose angular momentum and drift toward the singularity, \cite{SS} invoked viscosity, although classical fluid dynamics did not offer a reliable solution. \cite{SS} thus added this unknown type of viscosity through a parameter $\alpha$, now known as $\alpha-$prescription. Only later, Balbus and Hawley \cite{BALBUS} were able to give a realistic explanation of the disk viscosity as a \emph{magnetorotational instability}. In practice, the magnetic field acts to generate shear stresses in the disk, in place of the kinetic viscosity. As the field lines are frozen in the plasma, let us make the example of a line at a distance $r$ from the central compact object rotating with velocity $v$. The line is perturbed, with a part at $r-dr$ and another part at $r+dr$. These parts have velocities different from the reference $v$, $v+dv$ for the closer part, and $v-dv$ for the farther part. The magnetic field tries to hamper these departures from the reference velocity: the closer part is decelerated, while the farther one is accelerated. The necessary energy for the magnetic field to oppose the differential rotation is taken from the kinetic energy of the plasma, and hence from the angular momentum. 
 
Therefore, there is no single blackbody with characteristic temperature given by Eq.~(\ref{eq:disktemp}), but a series of disk annuli each one radiating as a blackbody with decreasing temperature as the distance from the singularity increases (\emph{multicolor blackbody}, \cite{MITSUDA,MAKISHIMA}). Therefore, Eq.~(\ref{eq:disktemp}) becomes:

\begin{equation}
T(r) = \sqrt[4]{\frac{3GM\dot{m}}{8\pi\sigma_{\rm SB}r^3}\left(1-\sqrt{\frac{r_{\rm ie}}{r}}\right)}
\label{eq:disktemp3}
\end{equation}

where $r_{\rm ie}$ is the radius of the innermost edge of the disk, which depends on the type of singularity, while $r$ could extend up to a few thousands times $r_{\rm g}$. Eq.~(\ref{eq:disktemp3}) can be rearranged \cite{PETERSON}:

\begin{equation}
T(r)\sim 1\times 10^6 \sqrt[4]{\frac{\dot{m}}{\dot{m}_{\rm Edd}}}\sqrt[4]{\frac{1}{M_8}}\left(\frac{r}{r_{\rm g}}\right)^{-\frac{3}{4}} \, {\rm [K]}
\label{eq:disktemp4}
\end{equation}

where $M_{8}=M/10^8M_{\odot}$, and $\dot{m}_{\rm Edd}$ is the accretion rate at the Eddington limit, whose meaning will be explained into a while. Eq.~(\ref{eq:disktemp4}) says that the disk temperature at the gravitational radius $r_{\rm g}$ of a $10^8M_{\odot}$ singularity accreting at the Eddington limit is $\sim 10^6$~K. The corresponding maximum frequency of emission can be calculated according to the Wien displacement law \cite{RYBICKI}: 

\begin{equation}
\nu_{\rm max} = \frac{2.8kT}{h} \sim 5.88\times 10^{10} T\; {\rm [Hz\, K^{-1}]}
\label{eq:disktemp5}
\end{equation}

\begin{figure}[ht]
\centering
\includegraphics[angle=270,scale=0.4]{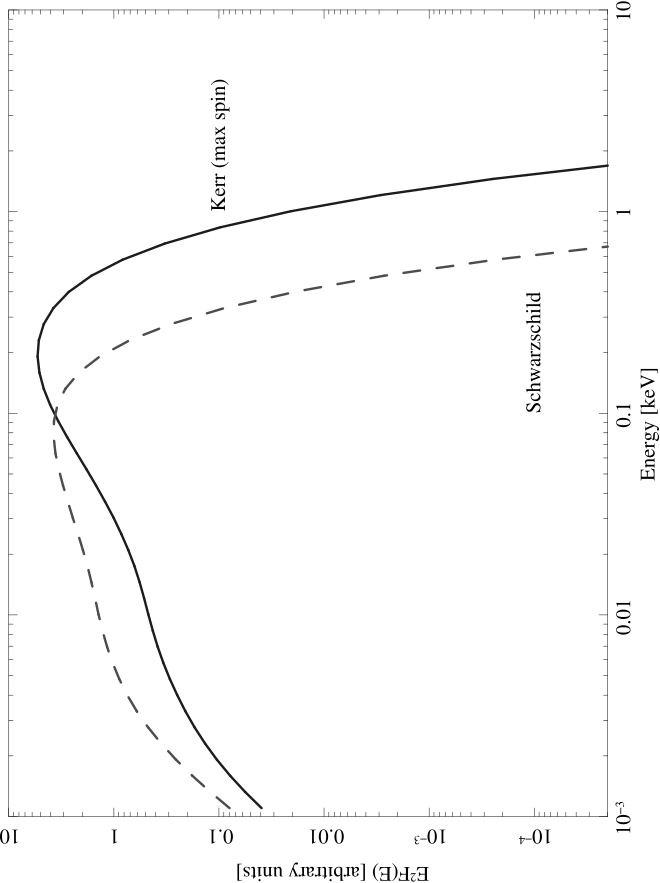}
\caption{Models of the typical spectra of an optically thick geometrically thin accretion disk (\texttt{optxagnf} model in \texttt{xspec}) around a $10^8M_{\odot}$ singularity. Two cases, corresponding to two different spin values (zero for Schwarzschild metric or $0.998$ for Kerr metric), are displayed.}
\label{fig:diskspec}
\end{figure}

In the above case, it corresponds to $\nu_{\rm max}\sim 5.88\times 10^{16}$~Hz or an energy of about $\sim 243$~eV (soft X-rays). Typical accretion rates for quasars are $\dot{m}\sim 0.01 \dot{m}_{\rm Edd}$, so that Eq.~(\ref{eq:disktemp5}) gives $\nu_{\rm max}\sim 1.8\times 10^{16}$~Hz or $E_{\rm max}\sim 77$~eV (extreme ultraviolet). The same values, but for a stellar-mass black hole of $M=10M_{\odot}$ results in $\nu_{\rm max}\sim 10^{18}$~Hz or $E_{\rm max}\sim 4.3$~keV (X-rays). Also the spin could affect the disk spectrum, as the radius of the innermost stable orbit change its value depending on the spin and the relative motion of the disk with respect to the spacetime singularity (Fig.~\ref{fig:diskspec})\footnote{\texttt{xspec} software and user manual can be downloaded at \url{https://heasarc.gsfc.nasa.gov/docs/xanadu/xspec/index.html}.}.

It is also worth mentioning the timescales of processes acting in disks \cite{DISKS}:

\begin{itemize}
\item dynamical: $t_{\rm dyn}\sim \Omega^{-1}$, where $\Omega = \sqrt{GM/r^3}$ is the keplerian angular velocity;
\item thermal: $t_{\rm th}\sim c_{\rm s}^2/\nu_{\rm k}\Omega^2$, where $c_{\rm s}$ is the sound speed, and $\nu_{\rm k}$ is the kinematic viscosity, which in turn is linked to the $\alpha-$prescription as $\nu_{\rm k}=\alpha c_{\rm s}H$ ($H$ is the disk vertical height);
\item viscous: $t_{\rm vis}\sim r^2/\nu_{\rm k}$.
\end{itemize}

Generally:

\begin{equation}
t_{\rm dyn} \ll t_{\rm th} \ll t_{\rm vis}
\label{eq:timescales}
\end{equation}

which means that the disk structure is dominated by its dynamical timescale. To give some order of magnitude: the speed of sound in a hydrogen plasma is:

\begin{equation}
c_{\rm s}\sim 1.3\times 10^4 \sqrt{T} \; {\rm [m/s]}
\label{eq:sound}
\end{equation}

Therefore, in the case of a $10^8M_{\odot}$ spacetime singularity maximally rotating, accreting at 1\% of the Eddington limit, with a prograde disk, the temperature at the innermost stable orbit (corresponding to about one gravitational radius) can be calculated from Eq.~(\ref{eq:disktemp4}) as $\sim 3.2\times 10^5$~K. The sound speed is calculated from Eq.~(\ref{eq:sound}) as $c_{\rm s}\sim 7.3\times 10^7$~m/s (about 2.4\% the speed of light). The keplerian angular velocity is $\Omega \sim 2\times 10^{-3}$~rad/s, which corresponds to a linear velocity $v_{\rm K}$ close to $c$ (obviously!). The dynamical timescale $t_{\rm dyn}\sim \Omega^{-1} \sim 500$~s. To calculate the kinematic viscosity, it is necessary to estimate the disk vertical height, which is much smaller than the disk size (thin disk), and is given by \cite{VIETRI}:

\begin{equation}
H = \frac{c_{\rm s}}{\Omega} = \frac{c_{\rm s}}{v_{\rm K}}r
\label{eq:diskh}
\end{equation}

With proper substitutions it is possible to derive the thermal timescale as $t_{\rm th}\sim \Omega^{-1}\alpha^{-1}$. Since $\alpha \sim 0.1$, it follows that the thermal timescale is $\sim 10t_{\rm dyn}$. At last, it is possible to calculate the viscous timescale as $t_{\rm vis}\sim 84400$~s (about one day). 

From Eq.~(\ref{eq:diskh}), it is also possible to have an order of magnitude estimate of a thin disk height: at $r\sim r_{\rm g}$, the keplerian velocity is close to $c$, which in turn means that the disk height is just a tiny fraction ($\sim c_{\rm s}/c\sim 2.4$\%) of the gravitational radius.

\section{The Eddington limit}
As the disk generates radiation, the latter has a pressure, which tends to blow out the structure. The equilibrium condition is set by equating the gravitational with the radiative forces. The calculation is done by taking into account an electron-proton pair with mass $(m_{\rm p}+m_{\rm e}\sim m_{\rm p})$, the Thompson cross section $\sigma_{\rm e}$, and assuming a spherical symmetry. The corresponding limit luminosity, also known as Eddington luminosity, is:

\begin{equation}
L_{\rm Edd} = \frac{4\pi c G m_{\rm p}M}{\sigma_{\rm e}} \sim 1.26\times 10^{38} \left(\frac{M}{M_{\odot}}\right)\, {\rm [erg/s]}
\label{eq:eddington1}
\end{equation}

It is worth stressing that the Eddington limit is just a rule of thumb, given the simple assumptions. One above the others is the spherical symmetry, which is obviously not the case of an accretion disk. Anyway, it is commonly used in astrophysics for its simplicity. 

Another useful quantity is the Eddington accretion rate, which is the maximum sustainable rate above which the radiation pressure blows out the matter:

\begin{equation}
\dot{m}_{\rm Edd} = \frac{L_{\rm Edd}}{\eta c^2} \sim 1.4\times 10^{18} \left(\frac{M}{M_{\odot}}\right)\, {\rm [g/s]}
\label{eq:eddington2}
\end{equation}
 
where $\eta$ is the conversion efficiency. 

\section{Other types of disk}
The standard solution by Shakura and Sunyaev \cite{SS} is not the only possible solution. There are many other solutions (see \cite{DISKS} for a review), but observations suggest to limit the choices to a few cases summarised in Table~\ref{tab:disks}. Basically, the main difference with respect to the standard thin disk is to take into account the advection and the radiation pressure. 

\begin{table}[ht]
\caption{Types of accretion disk. $h=H/R$ is the ratio between the vertical and horizontal disk sizes; $\dot{m}_{\rm Edd}$ is the accretion rate in Eddington units; $\tau$ is the optical depth; Adv is the importance of advection; RP is the importance of the radiation pressure compared to the gas pressure; $r_{\rm in}$ is the inner edge radius; $\eta$ is the accretion efficiency. Adapted from \cite{DISKS}.}
\begin{center}
\begin{tabular}{lccccccc}
\smallskip\\
\hline
Disk Type       & $h$      & $\dot{m}_{\rm Edd}$  & $\tau$  & Adv &  RP & $r_{\rm in}$ & $\eta$ \\
\hline
Thin            & $\ll 1$  & $< 1$                & $\gg 1$ & NO  &             NO & $r_{\rm iso}$& $\sim 0.1$  \\
Thick           & $> 1$    & $\gg 1$              & $\gg 1$ & YES & YES           & $r_{\rm mbo}$& $\ll 0.1$   \\
Slim            & $\sim 1$ & $\gtrsim 1$          & $\gg 1$ & YES & YES           & $r_{\rm mbo}<r_{\rm in}<r_{\rm iso}$ & $<0.1$ \\
ADAF            & $< 1$    & $\ll 1$              & $\ll 1$ & YES & NO            & $r_{\rm mbo}<r_{\rm in}<r_{\rm iso}$ &  $\ll 0.1$ \\
\hline
\end{tabular}
\end{center}
\label{tab:disks}
\end{table}

In the case of high accretion rate (slim, and thick disks), the advection is necessary to balance the strong radiation pressure, which in turn otherwise would blow off the disk. If the accretion rate is low (ADAF, advection-dominated accretion flow), then there is negligible radiation pressure and the gas drift radially toward the singularity before having time to emit substantial radiation. This implies that the ADAF is radiatively inefficient. It is worth noting that also in the case of thick disks, when the luminosity exceeds twice the Eddington limit, there could be photon-trapping effects, i.e. the radiation is trapped by Thompson scattering within the flow and advected toward the singularity (e.g. \cite{OSHUGA}). This severely limit the amount emitted radiation, thus making the disk radiatively inefficient. 

\section{Corona and Emission lines}
\begin{figure}[ht]
\centering
\includegraphics[angle=270,scale=0.4]{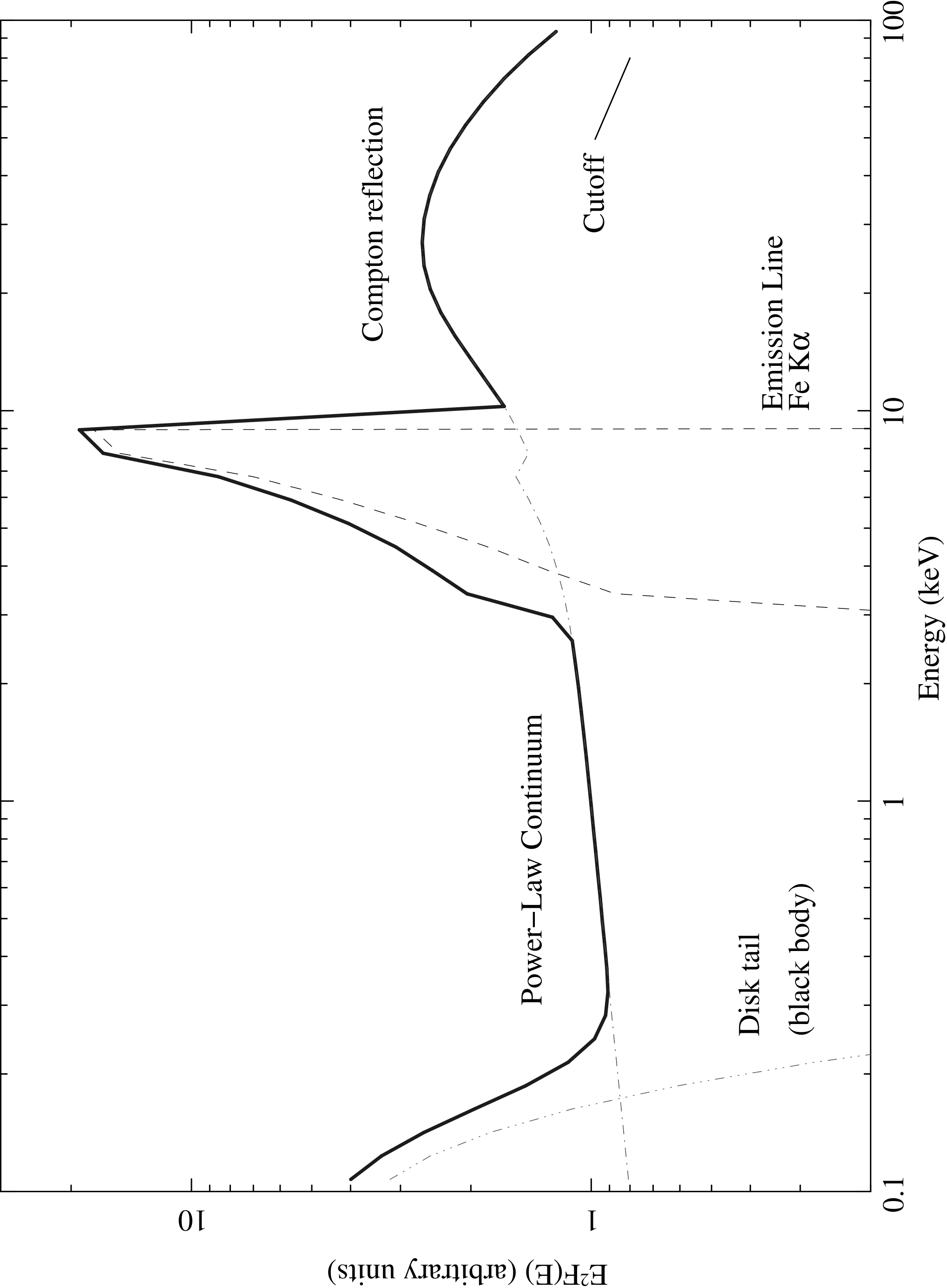}
\caption{Model of the typical X-ray spectrum of a Seyfert 1 active galaxy with a viewing angle of $80^{\circ}$ (\texttt{pexrav} and \texttt{zbb} models in \texttt{xspec}). The $6.4$~keV neutral iron line profile is clearly distorted because of relativistic effects (\texttt{diskline} in \texttt{xspec}). Relative intensities of the different component are not in scale, to make them clearly visible. Adapted from \cite{COURVOISIER}.}
\label{fig:seyspec}
\end{figure}

Observations in the X-ray energy band suggest that the accretion disk has also a hot corona, likely as it happens in the Sun \cite{SHAPIRO,KATZ,HM1,HM2}. The electrons of the plasma scatter the ultraviolet photons, transferring part of their energy via Compton scattering, so that they gain energy up to hard X-rays (see Sect.~5.4). The spectra observed in Seyferts AGN indicate an optical depth $\tau \sim 1$, so that the Comptonization is not saturated and the spectrum results in a photon index\footnote{The photon index $\Gamma$ indicates the slope of a photon spectrum, i.e. $F_{\rm ph}\propto \mathcal{E}^{-\Gamma} = \nu^{-\Gamma}$, where the flux is in units [photons cm$^{-2}$ s$^{-1}$ keV$^{-1}$]. To convert this flux in physical units, it is necessary to multiply it by the photon energy $h\nu$ [erg/photon]:  $F_{\rm erg}=F_{\rm ph}\times h\nu \propto \nu^{-\Gamma+1}=\nu^{-\alpha}$, where $\alpha$ is the known spectral index. Therefore, photon and spectral indexes are linked by the relationship: $\alpha = \Gamma - 1$.}: 

\begin{equation}
\Gamma = -\frac{1}{2} + \sqrt{\frac{9}{4}+\frac{4}{y}}
\label{eq:phindexc}
\end{equation}

where $y$ is the Compton parameter, and an exponential cutoff at $\mathcal{E}_{\rm co}=2kT_{\rm e}$, due to the pair production. Typical values are $\Gamma \sim 1.9$ ($\rightarrow y\sim 1.1$) and $\mathcal{E}_{\rm co}\sim 100-150$~keV, indicating an electron temperature $T_{\rm e}\sim 10^{8-9}$~K \cite{FABIAN1}. The typical X-ray spectrum of a Seyfert 1 AGN is shown in Fig.~\ref{fig:seyspec}. In addition to the already mentioned power-law continuum with exponential cutoff and the multicolor blackbody accretion disk, it is worth noting the emission iron line and the Compton reflection.

Hard X-rays emitted from the disk corona are reprocessed by the disk itself. Below $\sim 10$~keV, the photons of the continuum are absorbed by the elements of the disk plasma, because of photoelectric effect; above $7.1$~keV (neutral iron absorption edge), the absorption is negligible, while the elastic Compton scattering becomes the dominant process. There is almost complete reflection with a peak energy at about $20-30$~keV. As the energy further increases, the electron recoil starts to be significant: photons are scattered to smaller energies, in the photoelectric absorption domain and the reflection component quickly drops. The observed spectrum above $\sim 50$~keV is again dominated by the power-law continuum until the cutoff energy $\mathcal{E}_{\rm co}$ due to the pair production. Iron and nickel, neutral and ionised, generate the most prominent emission lines (see \cite{FABIAN2}, Fig. 1). Iron K-shell electrons could be extracted when an X-ray photon hits the atom. Since K is the innermost atomic shell, the vacant electron must be replaced (with a small delay) by dropping an outer shell electron and emitting a $6.4$~keV X-ray photon (fluorescence). There are many lines in the iron-nickel complex: which is the most prominent depends on the ionisation state, defined by the ionisation parameter:

\begin{equation}
\xi = \frac{4\pi F_{\rm X}(r)}{n(r)}
\label{eq:xipar}
\end{equation}

where $F_{\rm X}(r)$ is the X-ray flux illuminating the disk at radius $r$ from the central singularity, and $n(r)$ is the comoving electron number density. Reflection spectra for different values of $\xi$ can be seen in Fig.~2 of \cite{FABIAN2}. As $\xi$ increases, the emission lines become weaker and weaker, and for very large values ($\xi\gtrsim 5\times10^3$~erg~cm~s$^{-1}$) there is only a featureless continuum, because the plasma is fully ionised. For small values ($\xi\lesssim 100$~erg~cm~s$^{-1}$), the gas is weakly ionised and the most prominent line is the $6.4$~keV FeK$\alpha$ line of neutral iron (see \cite{FABIAN2} for more details). 

If iron lines are emitted by plasma close to the innermost stable orbit of the disk, then their profile could be heavily distorted by relativistic effects: beaming, transverse Doppler shift, gravitational redshift (Fig.~\ref{fig:seyspec}; for more details, Fig.~3 of \cite{FABIAN2}). This is the so-called ``lamppost model'' \cite{FABIAN4,FABIAN5}: a compact source of hard X-rays (a flare, for example) somewhere in the corona above a flat accretion disk. X-rays are reprocessed and reflected by the disk, resulting in the emission of a fluorescence line of the iron complex. Therefore, by studying the line profile, it is possible to derive some important information about spacetime around the singularity. One of the most important parameter is the spin, although it is not so easy to define the line profile, because even in the X-ray spectrum -- usually with less features than an optical one -- there could be anyway different components affecting the line wings \cite{BRENNEMAN,MINIUTTI,WILMS}. 

The lamppost model has been recently challenged by studying the reverberation lags of the line. Some authors found that the line could be generated by a radiation-powered wind, which partially absorbs the X-ray continuum \cite{MIZUMOTO,DONE,HAGINO} (but see also the reply by \cite{KARA}). In this case, obviously, the line profile would no more be a proxy of the spacetime singularity spin, with some impact on the triggering of relativistic jets \cite{DONE}. 

\chapter{Ejection: winds and jets}

\section{Winds}
The other side of the accretion onto a compact object is represented by outflows from the same object. Outflows are basically divided into two categories, depending on the presence, or not, of collimation of the flow. In the first case, these structures are named jets, while in the second case, they are called winds. 

Winds are the preferred way of feedback between the central supermassive singularity and its host galaxy. Observations of UV/X-ray emission or absorption lines with significantly blueshifted wings suggest that these outflows are nearly isotropic, with large momentum fluxes, and reaching significant speeds up to about $(0.1-0.2)c$ \cite{CRENSHAW,FABIAN3,KINGPOUNDS,TOMBESI}. Although there are many theories to explain winds, two seems to gain most of preferences. 

The first one is the radiation pressure, which we already studied in the equilibrium of the accretion disk (Sect. 6.3). In this case, the equation of motion in spherical coordinates is \cite{CRENSHAW}:

\begin{equation}
v\frac{dv}{dr} = \frac{\kappa L}{4\pi c r^2} - \frac{GM}{r^2}
\label{eq:radialwind}
\end{equation}

where $v$ is the bulk velocity of the gas, $\kappa$ is the absorption cross section per unit mass, $L$ is the luminosity of the source of mass $M$. At distance $r_0$ from the central singularity, the gas is accelerated enough to generate the outflow with bulk velocity:

\begin{equation}
v_{\infty} = \sqrt{\frac{2}{r_0}\left(\frac{\kappa L}{4\pi c} - GM \right)}
\label{eq:radialwind2}
\end{equation}

The second one is the Compton heating \cite{BEGELMAN1,BEGELMAN2}: in the framework of the hydrostatic corona around the accretion disk, the structure exists at a distance $r$ from the centre if the Compton temperature $T_{\rm IC}$ is smaller than the escape temperature from that system. The limiting radius to have a corona is \cite{BEGELMAN1}:

\begin{equation}
R_{\rm IC} \sim \frac{10^{10}}{T_{\rm IC8}}\frac{M}{M_{\odot}} \, {\rm [cm]}
\label{eq:comptonradius}
\end{equation}

where $T_{\rm IC8}$ is the inverse-Compton temperature in units of $10^8$~K, defined as:

\begin{equation}
T_{\rm IC} = \frac{h}{4kL} \int_{0}^{\infty} \nu L_{\nu}d\nu = \frac{<\mathcal{E}>}{4k}
\end{equation}

where $<\mathcal{E}>$ is the average photon energy. When $r<R_{\rm IC}$, then a hydrostatic corona exists around the disk; in the opposite case, the Compton heating can generate a wind cooling the corona.

Other theories have been developed, particularly involving hydromagnetic flows centrifugally accelerated outward \cite{BP}, but these proved to be more effective in the generation of relativistic jets. 

\section{Jets}
Collimated outflows are named jets. Basically, any type of accreting cosmic source could emit bipolar jets of matter. The bulk velocity spans over a wide range, from supersonic (protostars) to relativistic (XRB, and AGN), and ultrarelativistic (Gamma-Ray Bursts)\footnote{See, for example, the many contributions on different types of jets in \cite{ROMERO}. Specific reviews on relativistic jets from black holes are \cite{BOTTCHER,BS,DERMER,MEIER,CONTOPOULOS}.}. The collimation is due to a magnetic field, although there is a strong debate on its origin, as well as on the generation of jets. 

The basic physical principle is the same at work in electric generators on Earth: a rotating magnetic field generates an electro-motive force (EMF), which in turn accelerates charged particles. Part of the magnetic field also works to collimate the flow. The accelerated particles interact with the surrounding medium, generating different observational properties\footnote{Also the viewing angle could play an important role, particularly because of special relativity effects \cite{URRY}.}, but the physical principle at the basis of the central engine seems to be always the same \cite{LIVIO}.
In the following, I focus on relativistic jets from neutron stars and black holes. There is one main difference between the two types of cosmic sources: the former has a solid surface and the magnetic field is anchored to the star, thus having the same angular velocity of the cosmic body; the latter, on the other side, has no solid surface and the magnetic field is anchored to the accretion disk, whose rotation is affected by the spacetime singularity itself. There are many different theories and I refer the reader to the rather complete and detailed book by Meier \cite{MEIER}. In the following, I focus on a few examples.

\subsection{Neutron Stars}
These cosmic objects can be divided into two main classes, depending on they are accreting matter or not. Pulsars, like the well-known Crab nebula, are isolated objects, although they could be surrounded by a nebula (plerion). The basic model has been developed by Goldreich \& Julian \cite{GOLDREICH} and consists of a dense magnetosphere corotating with the star. The charged particle density, by using polar coordinates $r,\theta,\phi$, and $\theta$ is measured from the rotation axis, is:

\begin{equation}
\rho_{\rm GJ} = \frac{\nabla\cdot \mathbf{E}}{4\pi c} = -\frac{\mathbf{\Omega_{\rm f}}\cdot\mathbf{B}}{2\pi c}\frac{1}{1-(\frac{\Omega_{\rm f}r}{c})^2\sin^2\theta}
\label{eq:gjcharge}
\end{equation}

where $\mathbf{\Omega_{\rm f}}$ is the angular velocity of the magnetic field lines\footnote{That is also the angular velocity of the pulsar, since the magnetosphere is anchored to the star surface.}, $\mathbf{E},\mathbf{B}$ are the electric, and magnetic field, respectively. The region defined by:

\begin{equation}
r\sin\theta = \frac{c}{\Omega_{\rm f}}
\end{equation}

separate the near corotating magnetosphere, with closed magnetic field lines, and the external wind zone, with open magnetic field lines. The particle number density $n_{\rm GJ}$ of the magnetosphere is defined by Eq.~(\ref{eq:gjcharge}) in the near region ($\Omega r \sin\theta \ll c$):

\begin{equation}
n_{\rm GJ} \sim 0.07\frac{B_{\rm z}}{P} \, {\rm [cm^{-3}]}
\end{equation}

where $B_{\rm z}$ is the magnetic field component along the rotation axis and $P=2\pi/\Omega$ is rotation period (note that $\Omega=\Omega_{\rm f}$, because of corotation). Just as reference, the Crab pulsar has $n_{\rm GJ}\sim 10^{13}$~cm$^{-3}$. 

I have already said that there is no accretion disk, so the question is now the origin of the electric charge filling the magnetosphere. One possible solution is the vacuum breakdown \cite{RUDERMAN}. There could be small empty regions (gaps with height $h$), where the electric field generated by charge separation cannot be short-circuited. A charged particle moving across these gaps is accelerated, thus emitting radiation. These photons interact with the magnetic field as soon as the latter is curved enough to have a component perpendicular to the photon trajectory. Under these conditions, the photons generate electron-positron pairs, which in turn are accelerated and so on. The voltage gap is \cite{MEIER}:

\begin{equation}
\Phi \sim \frac{\Omega_{\rm f}Bh^2}{2\pi c}\sim 2\times 10^{12} \frac{B^2}{P^2}\, {\rm [V]}
\end{equation}

which implies Lorentz factors of the accelerated electrons of the order $\gamma_{\rm e}\sim 2\times 10^{12}/m_{\rm e}c^2\sim 2\times 10^6$. The energy of the emitted photons by curvature radiation is of the order \cite{MEIER}:

\begin{equation}
\mathcal{E}_{\gamma} = \frac{3\gamma_{\rm e}^3hc}{4\pi R_{\rm curv}} \sim 240\,{\rm [MeV]}
\end{equation}

where $R_{\rm curv}\sim 10^6$~cm is the typical curvature radius of the magnetic field lines. The main places where these regions could develop are the polar caps, and other zones of the magnetosphere (outer gap), whose specific location depends on the models. 

Accreting neutron stars are compact objects in a binary system, with (often) a high-mass star as companion\footnote{While the opposite often happens for black holes, although there are exceptions. For example, Cygnus X-1 is coupled with a supergiant star.}. In this case, the strong magnetic field of the compact object is also responsible of a pressure that acts against the gas pressure of the disk. There is a distance from the neutron star for which the two pressures are in equilibrium \cite{MEIER}:

\begin{equation}
R_{\rm m} \sim 1.07\times 10^8 \dot{m}^{-2/7}B_{12}^{4/7}\, {\rm [cm]} 
\label{eq:rmag}
\end{equation}

where $B_{12}$ is the surface magnetic field in units of $10^{12}$~G. For distances smaller than $R_{\rm m}$, the disk plasma is channeled along the lines of the magnetic field and reaches the star surface. For larger distances, and sufficient disk rotation, the field lines could  become open, with the generation of a wind (propeller, or wind turbine regime). The threshold radius for which $\Omega_{\rm f}=\Omega_{\rm disk}$ is \cite{MEIER}:

\begin{equation}
R_{\rm t}\sim 1.50\times 10^8 \sqrt[3]{P^2}\, {\rm [cm]}
\label{eq:rthresh}
\end{equation}

which, in turn, could be translated into a critical period by equating Eq.~(\ref{eq:rmag}) and Eq.~(\ref{eq:rthresh}):

\begin{equation}
P_{\rm cr} \sim 0.60 \dot{m}^{-3/7} B_{12}^{6/7}\, {\rm [s]}
\label{eq:pcrit}
\end{equation}

Periods longer than $P_{\rm cr}$ implies domination of accretion; on the opposite, implies propeller domination. It is worth noting that in the latter case, in addition to non-collimated winds, there is also a fast jet along the poles \cite{USTYUGOVA}. 

\subsection{Spacetime singularities: Spin-powered jets}
There are many theories to explain the generation of relativistic jets from black holes, but the best one seems to be due to Blandford \& Znajek \cite{BP,KOMISSAROV}. Although the magnetosphere of rotating black holes (ergosphere) shares many similarities with those of pulsars, the former are much more complex than the latter because of frame dragging and the lack of a solid surface. The latter point implies that a spacetime singularity has no proper magnetic field, which has to be advected close to the horizon by the accretion disk\footnote{Since the high conductivity of the plasma, the magnetic field lines are frozen in the fluid.}. Therefore, disks where the advection plays a fundamental role (ADAF, slim disks) are the best option, while in the standard disk the radial motion is strongly reduced as due only to the magnetorotational instability\footnote{Recently, a new type of magnetically arrested disk (MAD) has been proposed \cite{NARAYAN,TCHEKHOVSKOY}. It is somehow similar to the case of the accreting neutron stars, but in MAD the magnetic field is generated by the disk itself.}. The magnetic field lines approaching the event horizon have basically three options \cite{TPM}: (a) to be pushed back to the disk; (b) to thread through the near-vacuum gap between the horizon and the innermost stable orbit; (c) to be pushed onto the horizon. Remind that in the pulsar, the magnetic field is generated by the compact object, it is anchored on it\footnote{More options can be found in \cite{MEIER}.}. The magnetosphere is generated by trapping charged particles around. Therefore, the magnetic field and the neutron star are rotating in synchrony. In a rotating spacetime singularity, the ergosphere means an angular velocity of the event horizon different from that of the outer magnetosphere, and hence of the accretion disk. For example, in the case of a maximally rotating singularity with a prograde disk, the innermost stable orbit extends into the ergosphere; on the other side, a counterrotating disk around a maximally rotating singularity has its boundary well outside the magnetosphere. This means that a rotating singularity is equivalent to an asynchronous electric generator \cite{FOSCHINI6,LI}. The jet is thus removing angular momentum both from the singularity and the disk, but disentangling the two contributions is still an open problem. Truly speaking, the understanding of the Blandford-Znajek process, and the energy extraction from a black hole, is an open debate, and only relatively recently Komissarov \cite{KOMISSAROV} clarified many issues (see also the detailed presentation by \cite{MEIER}). 

The seminal paper by Blandford \& Znajek \cite{BZ} did not consider an accretion disk; rather, the electric charges were generated by vacuum breakdown due to gravitational curvature. Indeed, it was the black hole analogue of the Goldreich-Julian model for pulsars. However, the accretion disk proved to be much more efficient in advecting a magnetic field close to the horizon. This implies that it is necessary to take into account the disk type when calculating the jet power, which could be divided into two broad categories: gas-pressure dominated (GPD), which is ADAF-type, or radiation-pressure dominated (RPD), which is well represented by slim disks \cite{GHOSH,MODERSKI}. 
Following Ghosh \& Abramowicz \cite{GHOSH}, the jet power is:

\begin{equation}
L_{\rm BZ}\sim
\begin{cases}
2 \times 10^{44} M_{8} a^2& {\rm RPD} \\
8 \times 10^{42} \dot{m}_{-4}^{4/5}M_{8}^{11/10} a^2 & {\rm GPD} \\
\end{cases}
\label{eq:bzpower}
\end{equation}

where $M_{8}=M/10^8M_{\odot}$, and $\dot{m}$ is the accretion rate in Eddington units and $\dot{m}_{-4}=\dot{m}/10^{-4}\dot{m}$. It is worth noting that Eqs.~(\ref{eq:bzpower}) have been derived under a certain assumption on the angular velocities. Indeed, the jet power can be written as follows \cite{TPM}:

\begin{equation}
L_{\rm BZ} = k f(B,r_{\rm g}, a) \frac{\Omega_{\rm f}(\Omega-\Omega_{\rm f})}{\Omega^2}
\label{eq:bzeq}
\end{equation}

where $f$ is a function of the magnetic field $B$, the gravitational radius $r_{\rm g}$, and the spin $a$. The asynchronous angular velocities of the black hole $\Omega$ and of the lines of the magnetic field $\Omega_{\rm f}$ are taken into account in the latest term. In almost all the works about the Blandford-Znajek process, it is given for granted that $\Omega_{\rm f}=\Omega/2$, so as for deriving the Eqs.~(\ref{eq:bzpower}). However, this working hypothesis was derived by using perturbative analytic solution for a slowly rotating singularity with a monopole magnetic field matching the pulsar solution at infinity \cite{KOMISSAROV,MEIER}. Blandford \& Znajek \cite{BZ} already noted that the mechanism is acting on two-ways, linking the singularity and the disk. This was also reprised by Li \cite{LI}, and Foschini \cite{FOSCHINI6}, which showed that system could be much more complex. By rearranging Eq.~(\ref{eq:bzeq}) in terms of the slip factor $s$, or the relative difference between the two angular velocities \cite{FOSCHINI6}: 

\begin{equation}
s = \frac{\Omega_{\rm f} - \Omega}{\Omega_{\rm f}}= 1 - \frac{\Omega}{\Omega_{\rm f}}
\label{eq:bzslip}
\end{equation}

it is possible to study the connection between the singularity and the disk. There are three main regions\footnote{While Blandford \& Znajek \cite{BZ}, and Li \cite{LI} studied only two cases.} where the system can operate, displayed in Fig.~\ref{fig:bzslip}, and it is worth noting the following cases:

\begin{figure}[t]
\centering
\includegraphics[angle=270,scale=0.5]{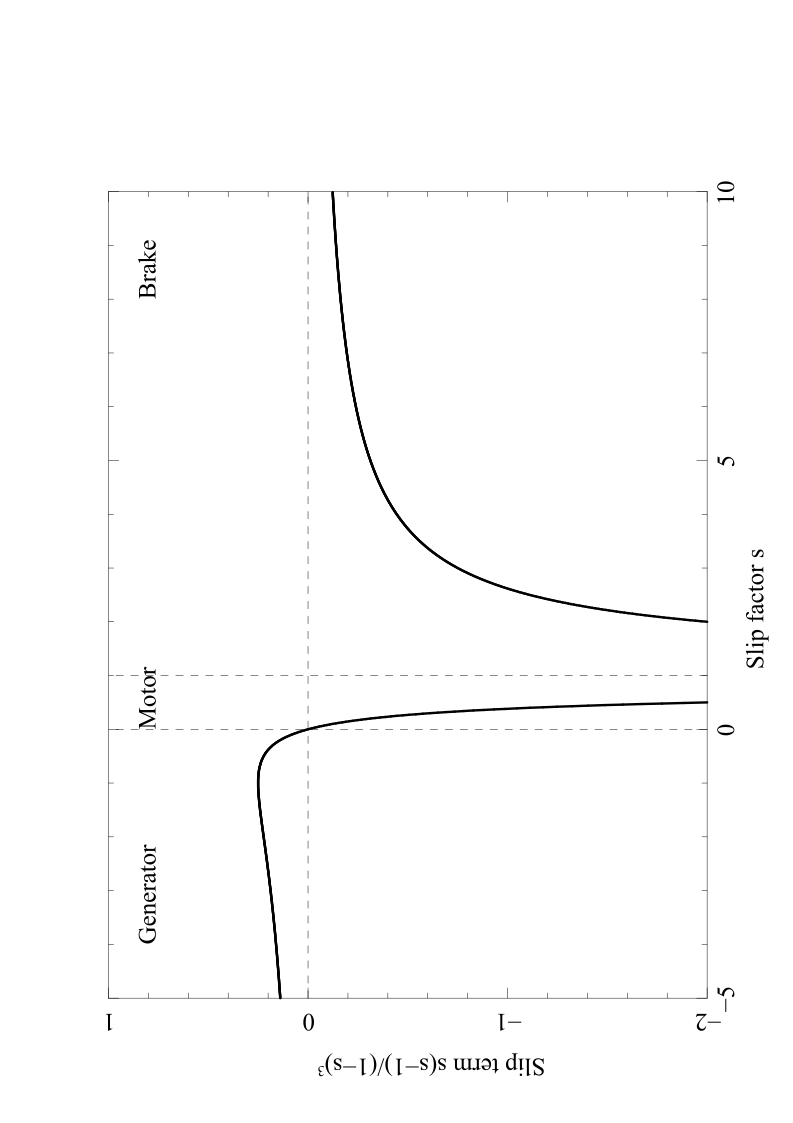}
\caption{Slip term as function of the slip factor in the Blandford-Znajek power (from \cite{FOSCHINI6}).}
\label{fig:bzslip}
\end{figure}

\begin{itemize}
\item[(i)] $\Omega/\Omega_{\rm f}=1 \Rightarrow s=0$: it is the {\it synchrony} condition, i.e. the angular speed of the black hole and that of the magnetic field lines are the same. This results in null power ($L_{\rm BZ}=0$). As written, although the Blandford-Znajek theory was derived by that of Goldreich-Julian, it is based on the general relativity effects in the ergosphere.

\item[(ii)] $\Omega/\Omega_{\rm f}=0 \Rightarrow s=1$: the black hole is at rest (Schwarzschild case) and $L_{\rm BZ}\rightarrow -\infty$, like a short circuit where the power is completely drained by the black hole. 

\item[(iii)] $\Omega/\Omega_{\rm f}=2 \Rightarrow s=-1$: this is the well-known condition of maximum output of the jet power.

\item[(iv)] $\Omega>\Omega_{\rm f} \Rightarrow s<0$: the black hole angular speed is greater than that of magnetic field lines. The machine works as {\it generator} and the power is positive ($\rightarrow$ jet emission). The rotational power of the black hole is converted into the electromagnetic jet power. 

\item[(v)] $\Omega<\Omega_{\rm f} \Rightarrow 0<s<1$: the power is negative, the system works as a {\it motor}. The rotational energy of the magnetic field, which rotates faster than the singularity, is extracted to power the motion of the black hole. If the latter increases its angular speed sufficiently to surpass that of the field, then it changes to the condition (iv).

\item [(vi)] $\Omega/\Omega_{\rm f}<0 \Rightarrow s>1$: the magnetic field and the hole are counterrotating. The power is negative and the machine works as a {\it brake}. As $s\rightarrow +\infty$, $L_{\rm BZ}\rightarrow 0$.
\end{itemize}

\subsection{Spacetime singularities: Disk-powered jets}
The other main theory to explain relativistic jets from singularities was developed by Blandford \& Payne \cite{BP}: it is based on the centrifugal acceleration of the outer layers of the accretion disk plasma along the lines of the magnetic field. An outflow develops if the poloidal component of the magnetic field makes an angle $\lesssim 60^{\circ}$ with the disk surface. Then, as the distance from the disk increases, the toroidal component of the magnetic field becomes dominant, thus collimating the outflow into a jet. The emitted power is \cite{GAROFALO}:

\begin{equation}
L_{\rm BP} = \int B_{\rm d}^2 \Omega_{\rm d} r^2 dr
\end{equation}

where $B_{\rm d}$ is the poloidal magnetic field threading the disk and $\Omega_{\rm d}$ is the disk angular velocity, which is equal to the magnetic field one, as the lines are frozen in the plasma. 

\begin{figure}[t]
\centering
\includegraphics[scale=0.4]{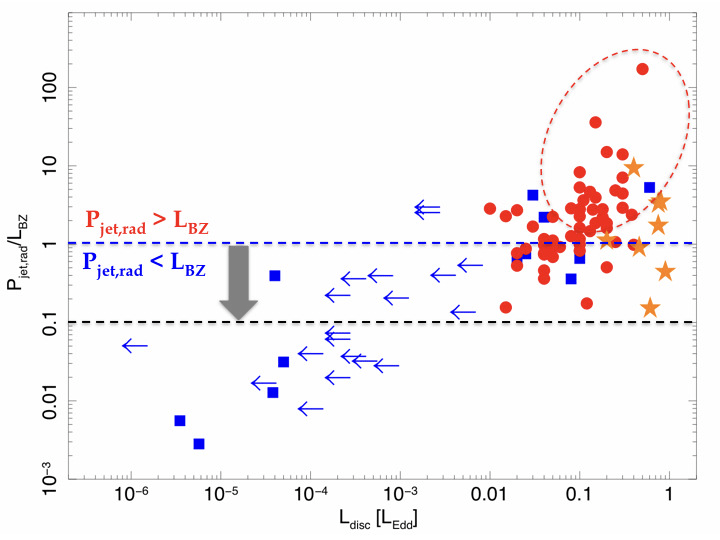}
\caption{Observed and calculated radiative jet power by using Blandford-Znajek theory (adapted from \cite{FOSCHINI7}). Symbols are: blue squares and arrows (upper limits of disk luminosity) for BL Lac Objects, red circles for flat-spectrum radio quasars, and orange stars for radio-loud narrow-line Seyfert 1 galaxies (see Sect. 7.3 for some more information about the characteristics of these sources).}
\label{fig:jetcomp}
\end{figure}

Although there are many more theories, they are all more or less derived from the two main cases above. On one side, some authors developed hybrid models, taking into account features from both \cite{MEIER2,GAROFALO}. On the other side, there is a debate about which process is dominating \cite{FERREIRA}. Fig.~\ref{fig:jetcomp} shows the jet power calculated according to Eq.~(\ref{eq:bzpower}) and compated with the observed value (from \cite{FOSCHINI7}). Since the spin is almost unknown, I have considered two cases: $a=1$ (blue dashed line), and $a=0.3$ (black dashed line). It is evident that many sources (red dashed ellipse) have jet powers in excess of the Blandford-Znajek maximum value, which implies the need of another contribution. Since the sources with jet power in excess have all strong accretion disks, it is likely that the additional contribution could come from hydromagnetic winds. The emerging scenario is that of a jet with a backbone made by the Blandford-Znajek theory plus a contribution from the hydromagnetic winds \cite{CAVALIERE,MARASCHI}.

\section{Scaling and Unification of Relativistic Jets}
Observations of relativistic jets at different frequencies resulted in many different types of sources, which were later unified by assuming the same central engine, but viewed at different angles\footnote{Also AGN without jets were unified according to the same principle (read the amazing history in \cite{DONOFRIO}).} \cite{URRY,DERMER2}. When the jet is viewed at small angles ($\theta \lesssim \Gamma^{-1}$), there is a blob of plasma traveling close to the speed of light, and moving toward the Earth. According to special relativity (Sect.~2.4), many observational properties are modified: boosted luminosity, reduced time scales, increased photon energy. A typical Doppler factor $\delta \sim 10$, implies an increase of luminosity of ten thousands times ($\delta^4$). Therefore, the emission of the jet dominates the broad-band spectrum of the source. 
When the jet is viewed at large angles, special relativity effects decrease their importance and the emission from other components of the sources (e.g. host galaxy, molecular torus) become visible in the spectrum. 

Currently \cite{URRY,DERMER2,TADHUNTER}, AGN with jets are divided into blazars ($\theta \lesssim \Gamma^{-1}$) and radio galaxies ($\theta \gtrsim \Gamma^{-1}$). The former are further divided into BL Lac Objects and Flat-Spectrum Radio Quasars (FSRQs). BL Lacs are characterised by weak jet power, and an optical spectrum featureless or with barely-detected emission lines, implying weak accretion disk. FSRQs have powerful jets, and strong disks, clearly visible in the optical/UV spectrum, as the emission lines. Radio galaxies, i.e. blazars viewed at large angles, and also called the parent population, share the same physical properties, although derived from different observational properties. The radio galaxies corresponding to BL Lac Objects are called Low-Excitation Radio Galaxies (LERG), while High-Excitation Radio Galaxies (HERG) is the class corresponding to FSRQs \cite{BUTTIGLIONE1,BUTTIGLIONE2,TADHUNTER}. The excitation index dividing the two populations is defined by the ratio $L_{\rm [OIII]}/\nu L_{\rm 178\,MHz}$, and it is much more efficient than the old classification based on the radio power only (Fanaroff-Riley I/II). 

The classification of jets from stellar-mass black holes and accreting neutron stars is a bit different, mainly because these sources are distributed on the spiral disk of the Milky Way together with the Solar System. This implies that it is quite difficult to observe a stellar-mass analog of a blazar. It is expected that the jets from X-ray binaries point toward the same direction of the Galaxy poles. Although physically plausible, it is quite difficult to find a X-ray binary with its axis tilted along the Milky Way disk. Therefore, relativistic jets in X-ray binaries are generally named \emph{microquasars} \cite{MIRABEL}. Moreover, the reduced timescales make it easier to correlate different components. It results a classification based on different phases of the jet and the disk activities (see, for example, the reviews \cite{GALLOE,JEROME}). 

It is worth noting that this classification scheme is unstable. Recently, the discovery of high-energy gamma rays from radio-loud narrow-line Seyfert 1 galaxies (RLNLS1s\footnote{Paolo Padovani \cite{PADOVANI} proposed to move from an observations-based classification to a more physical one. Therefore, radio-loud and radio-quiet AGN should change to jetted and non-jetted AGN. Following this scheme, RLNLS1s should change to jetted NLS1s. See also \cite{FOSCHINI17} for an extended discussion on the classification of jetted AGN.}) added one more class to blazars \cite{ABDO1,ABDO2,ABDO3,FOSCHINI8}, together with the corresponding problem of finding the parent population \cite{BERTON1,BERTON2}. The main novelty added by the inclusion of jetted NLS1s in the group of AGN with jets was to extend the mass distribution to the smaller values, thus removing mass requirements for the jet. Indeed, the masses of the central singularity powering AGN with relativistic jets were in the $\sim 10^{8-10}M_{\odot}$ range, suggesting the existence of a mass threshold for the jet generation, which in turn was linked to the host galaxy (elliptical for AGN with jets, spiral for the others; see, e.g. \cite{LAOR,SIKORA}). The discovery of powerful jets from jetted NLS1s (with masses in the range $\sim 10^{6-8}M_{\odot}$ (\cite{ABDO3,FOSCHINI8,BERTON1}) made it clear that the previous threshold was an observational bias\footnote{This bias was born somewhere between seventies and eighties. For example, in 1979, Miley \& Miller \cite{MILEY} reported about the study of a sample of AGN with jets at $z<0.7$ and found sources with small masses and compact radio structure. Some years later, in 1986, Wills \& Browne \cite{WILLS} reported a study on a similar sample, but they selected only the brightest objects ($<17$ mag): this time, only objects with large masses were found.}. Recent surveys also found that a significant fraction of AGN with jets is hosted in disk galaxies \cite{COZIOL,INSKIP} (see \cite{FOSCHINI20} for a recent review). 

\begin{figure}[t]
\centering
\includegraphics[scale=0.2]{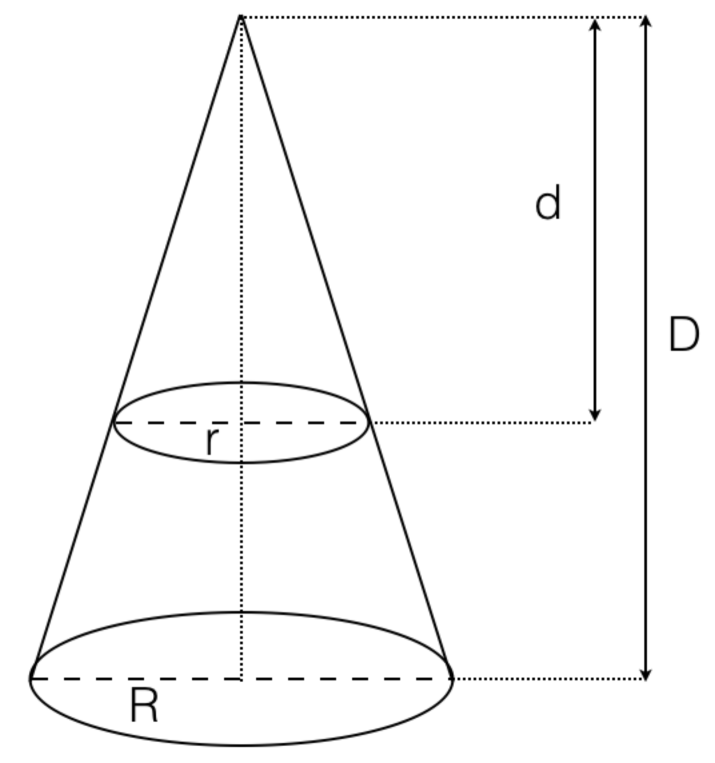}
\includegraphics[scale=0.2]{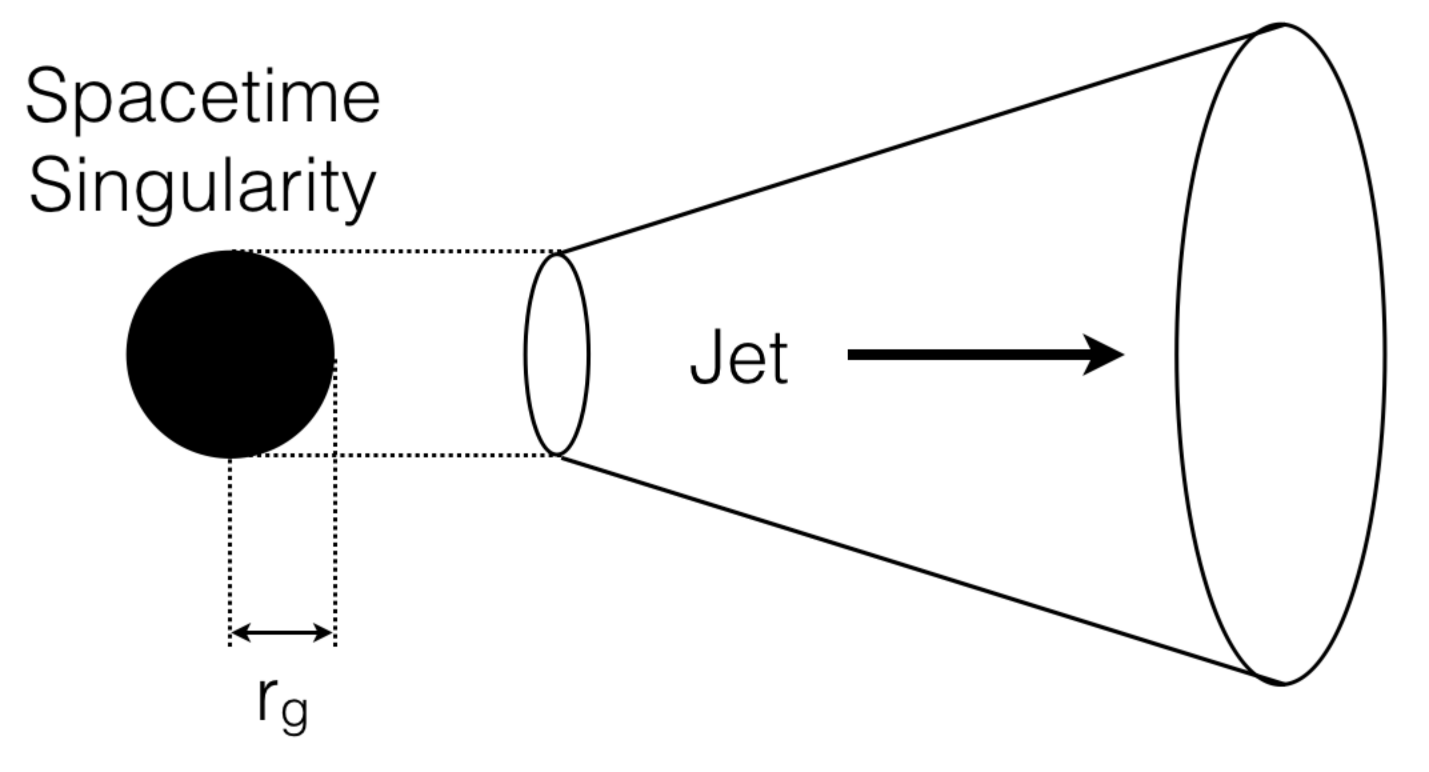}
\caption{(\emph{left panel}) Self-similarity. A cone is a self-similar structure, since the ratio between the radius of the base and the distance from the apex is constant, i.e. $r/d=R/D$. Three-dimensional conic-sections are all self-similar structures. (\emph{right panel}) Scheme of a jet: the size of the base cannot be smaller than the black hole size, because of causality outside the event horizon.}
\label{fig:conica}
\end{figure}

Anyway, all these AGN with powerful relativistic jets can be unified if we focus on the basic physical principles, instead of observational details. Ghisellini et al. \cite{GHISELLINI1} proposed a simple physical principle (cooling of relativistic electrons dependent on the nearby environment) to explain the observational properties of bright blazars (the so-called \emph{blazar sequence}) found by Fossati et al. \cite{FOSSATI} (but see \cite{GIOMMI} for a different point of view). On one end, there are FSRQs: bright, powerful jets, strong emission lines, strong accretion disk. On the other end, there are BL Lac Objects: faint, weak jets, small equivalent width lines or featureless continuum, weak accretion disk. In  the former case, the stream of relativistic electrons accelerated and collimated by the central singularity flows in a photon-rich environment, which implies an easy cooling via inverse-Compton scattering; in the latter case, the electrons travel in a photon-starving environment, which means reduced cooling, limited to a few collisions that have to remove the maximum energy possible, and the main cooling process is the synchrotron self-Compton. This theory works because FSRQs and BL Lac Objects have similar masses of the central black hole within one-two orders of magnitude. Ghisellini \& Tavecchio \cite{GHISELLINI2} later revised the scheme to include also small-mass FSRQs: it is possible to observe low-power FSRQs, as the jet power scales with the mass\footnote{Although, it is worth noting that Ghisellini \& Tavecchio \cite{GHISELLINI2} proposed a linear scaling with the mass, they did not follow the relationships by Heinz \& Sunyaev \cite{HEINZ}.}. The relativistic electrons still cool efficiently in the photon-rich environment, but the observed jet power is weak because of the small mass of the central singularity. This is indeed what happens in jetted NLS1s\footnote{Rather curiously, Ghisellini co-authored a paper \cite{CALDERONE} proposing large black hole masses for jetted NLS1s, which implies a contradiction with the physical principle of electron cooling to explain the observed jet properties proposed by himself \cite{GHISELLINI2}. Indeed, if jetted NLS1s were with masses as large as usual FSRQs, and with low observed jet power similar to BL Lac Objects, then there would be relativistic electrons in a photon-rich environment, but with negligible cooling. This is obviously not possible, because it would be against the very basic physical principle regulating the electron cooling. In a recent work, Ghisellini et al. \cite{GHISELLINI3} took into account small-mass FSRQs (there is no explicit reference to jetted NLS1s) as a source of pollution of the blazar sequence, suggesting that the sequence is controlled by the Eddington ratio rather than the observed luminosity.}. 

The scaling theory of relativistic jets was developed by Heinz \& Sunyaev \cite{HEINZ}. It refers to the phenomenological jet model by Blandford \& K\"onigl \cite{BK}, and requires only that the jet shape is self-similar (Fig.~\ref{fig:conica}, left panel). This hypothesis has been confirmed by recent radio observations \cite{HADA1,HADA2}.

\begin{table}[t]
\caption{Scaling indices according to Heinz \& Sunyaev \cite{HEINZ}. $\alpha$ is the radio spectral index (generally $\alpha \sim 0$ for flat-spectrum radio quasars).}
\begin{center}
\begin{tabular}{lcc}
\smallskip\\
\hline\\
Disk Type & $\xi$ & $\zeta$\\
{}        & [mass] & [accretion]\\
\hline\\
ADAF & $\frac{17}{12}-\frac{\alpha}{3}$ & $\frac{17}{12}+\frac{2\alpha}{3}$\\[3pt]
RPD  & $\frac{17}{12}-\frac{\alpha}{3}$ & 0 \\[3pt]
GPD  & $\frac{187-32\alpha}{120}$ & $\frac{4}{5}(\frac{17}{12}+\frac{2\alpha}{3})$\\[3pt]
\hline
\end{tabular}
\end{center}
\label{tab:hs2}
\end{table}%

The basis of the scaling theory is rooted in the geometrical character of the general relativity (Fig.~\ref{fig:conica}, right panel). The smallest size of the jet cannot be smaller than the gravitational radius of the central singularity, which in turn depends on its mass, according to Eq.~(\ref{eq:rgrav}). The accretion disk and the spin are related to the black hole according to the theories outlined in Sects.~7.2.2 and 7.2.3, although the spin has little numerical impact on the jet power and was neglected in \cite{HEINZ} (see also \cite{MOS}). Heinz \& Sunyaev found that the emitted flux density $F_{\nu}$ is proportional to the mass of the central singularity $M$ and the accretion rate $\dot{m}$ as \cite{HEINZ}:

\begin{equation}
F_{\nu} = k M^{\xi}\dot{m}^{\zeta}
\label{eq:hs1}
\end{equation}

where $\xi$ and $\zeta$ are the scaling index defined in Table~\ref{tab:hs2}, and $k$ is a proportionality coefficient. The total jet power $P$ as a function of the radio flux density can be written in the form of the Eq.~(7) from \cite{FOSCHINI24}: 

\begin{equation}
P = k_1 k_2 \left(\frac{F_{\nu} d_{\rm L,9}^2}{1+z} \right)^{\frac{12}{17}}
\label{eq:jetpow}
\end{equation}

\noindent which is derived from the Eq.~(29) of Blandford \& K\"onigl \cite{BK}. $d_{\rm L,9}$ is the luminosity distance of the object [Gpc], $z$ is the redshift, $k_1$ is a coefficient depending on the size of the emitting region and the electrons energy distribution, $k_2$ is a coefficient depending on the observational properties of the jet like viewing and opening angle, Doppler and Lorentz factors, apparent speed (cf \cite{FOSCHINI24} for more explanation). 

By substituting Eq.~(\ref{eq:hs1}) in the Eq.~(\ref{eq:jetpow}), it is possible to relate the jet power with the mass and the accretion rate:

\begin{equation}
P = \kappa \left(\frac{\dot{m}^\zeta M^\xi d_{\rm L,9}^2}{1+z} \right)^{\frac{12}{17}}
\label{eq:powmassrate}
\end{equation}

\noindent where $\kappa$ is a global proportionality constant taking into account $k_1$ and $k_2$ of Eq.~(\ref{eq:jetpow}), and $k$ of Eq.~(\ref{eq:hs1}). By considering the scaling indexes of Table~\ref{tab:hs2} and assuming $\alpha = 0$ for the sake of simplicity, the Eq.~(\ref{eq:powmassrate}) becomes:

\begin{equation}
P = \kappa \dot{m} M \left(\frac{d_{\rm L,9}^2}{1+z} \right)^{\frac{12}{17}}\, \mathrm{ADAF}
\label{eq:adaf}
\end{equation}

\begin{equation}
P = \kappa M \left(\frac{d_{\rm L,9}^2}{1+z} \right)^{\frac{12}{17}}\, \mathrm{RPD}
\label{eq:rpd}
\end{equation}

\begin{equation}
P = \kappa \dot{m}^{\frac{12}{15}} M^{\frac{187}{170}}  \left(\frac{d_{\rm L,9}^2}{1+z} \right)^{\frac{12}{17}}\, \mathrm{GPD}
\label{eq:gpd}
\end{equation}

Therefore, depending on the type of accretion disk, the jet power can scale with the mass of the central black hole only (RPD), or the mass and the accretion rate (ADAF or GPD, with different slopes). 

The earliest application of the scaling theory was the so-called \emph{black hole fundamental plane}, based on radio and X-ray observations \cite{FALCKE,MERLONI}. The authors found a tight correlations between measurements at these frequencies for both AGN and XRBs, which in turn also works with the scaling relationships proposed by \cite{HEINZ} to derive the mass of the compact object. However, as already noted, there are different physical processes playing in the same energy band. Therefore, by simply comparing the X-ray emission of AGN and XRB is not valid, because while the emission of the latter is dominated by the accretion disk, in the former case the emission is due to the jet (AGN disks peaks at UV). This means that the correlation found by \cite{FALCKE,MERLONI} is just a chance coincidence as many other bogus correlations that it is possible to find in the web\footnote{Have a look at this amazing web page \url{http://www.tylervigen.com/spurious-correlations}, where rather bizarre correlations are reported: for example, the US investments in science, space, and technology is tightly correlated ($99.79$\%, $r=0.99789$) to the suicides by hanging, strangulation, and suffocation! This is just superstition!}. 
With specific reference to the fundamental plane, Chiaberge \cite{CHIABERGE} clearly shows the fake correlation by including in the plane also the Sun, the Moon, Jupiter, and Saturn! Another example of the failure of the fundamental plane, with specific reference to jetted AGN, is displayed in Fig.~7 of \cite{BERTON19}.

In order to apply correctly the scaling relationships \cite{HEINZ} to different population of cosmic sources, it is necessary to derive the requested measurements from the proper observations \cite{FOSCHINI9}. Therefore, the disk luminosity is derived from optical/ultraviolet observations for AGN, and from X-ray observations from XRBs, because the peak luminosity depends on the inverse of the mass as shown in Sect. 6.2\footnote{At last, one erg of energy is always one erg of energy, independently on the frequency of measurement.}. An easy-to-use relationship to estimate the jet power from radio flux density is \cite{FOSCHINI9}:

\begin{equation}
\log P_{\rm jet}
\begin{cases}
(12\pm2) + (0.75\pm0.04)\log L_{\rm r}  & {\rm Radiative} \\
(6\pm2) + (0.90\pm0.04)\log L_{\rm r} & {\rm Kinetic} \\
\end{cases}
\label{eq:jetpower}
\end{equation}

where $L_{\rm r}$ is the radio luminosity of the jet core. The relationships have been calibrated by using radio measurements at 15~GHz, but can properly work with measurements at other frequencies by extrapolating the needed values by means of the radio spectral index. It seems substantially correct, although there is a large scatter, likely due to the extreme variability (cf Fig.~4, \emph{right panel}, in \cite{FOSCHINI19}). More recent easy-to-use equations can be found in Table~6 of \cite{FOSCHINI24}. 

\begin{figure}[t]
\centering
\includegraphics[scale=0.17]{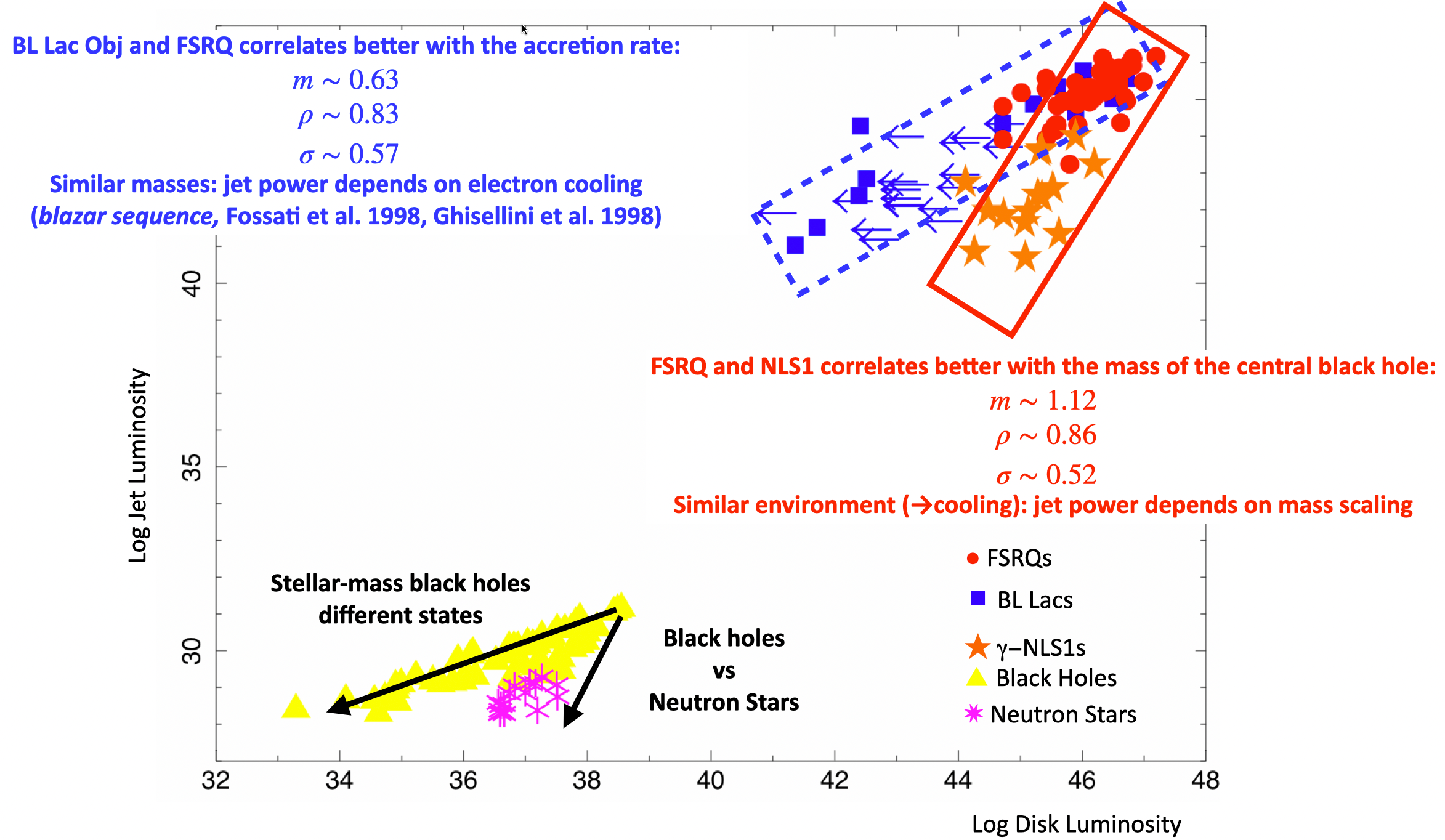}
\caption{Radiative jet power vs disk luminosity for a sample of AGN with jets and XRBs. Different types of sources are indicated as: FSRQs (red circles), BL Lac Objects (blue squares or arrows, in case of upper limits of the disk luminosity), RLNSL1s (orange stars), stellar-mass black holes (yellow triangles), and accreting neutron stars (cyan asterisks). Adapted from \cite{FOSCHINI9}.}
\label{fig:jetunif}
\end{figure}

Fig.~\ref{fig:jetunif} shows the distribution of jet power and disk luminosity of a sample of AGN with jets and XRBs\footnote{Other similar plots can be read in \cite{CORIAT} (Fig.~5) for XRBs and \cite{PALIYA} (Fig.~7, bottom right panel) for AGN.}. It is worth noting the recent addition of jetted NLS1s: this population is the analogous of accreting neutron stars branch on Galactic objects distribution. This addition was necessary to proceed with the unification, otherwise the two populations, AGN and XRBs, were unbalanced. 

The blazar branch (blue dashed rectangle in Fig.~\ref{fig:jetunif}) correlates better with the accretion rate. This is the well-known blazar sequence \cite{FOSSATI,GHISELLINI1}, where the jet power is regulated by the electron cooling. The slope in the present sample ($m\sim 0.63$) seems to be consistent with Eq.~(\ref{eq:gpd}) with a small range of the masses of the central black holes. The quasar branch (red rectangle in Fig.~\ref{fig:jetunif}) correlates better with the mass of the central black hole, and the slope ($m\sim 1.12$) is fairly consistent with Eq.~(\ref{eq:gpd}), but with a small range of accretion rates. A larger sample is necessary to better assess these relationships.

Last, but not least, it is worth citing also another attempt to unify AGN with GRBs. Nemmen et al. \cite{NEMMEN} studied a sample of 234 blazars and 54 GRBs. They reported the beaming-corrected gamma-ray luminosity and the kinetic power of the jet, finding a correlation extending over about ten orders of magnitudes, which in turn implies that the radiative efficiencies of the different jets are similar. Some words of caveat are necessary also in this case: one main objection refers to the kinetic power, which is impulsive (a few seconds at maximum) in the case of GRBs and averaged on gigayears time scales in the case of AGN. The latter was calculated by measuring the extended radio emission, as the jet needed a certain kinetic power and long time to generate cavities in the intergalactic medium \cite{NEMMEN}. More details about the issues in the jet power estimates can be found in \cite{FOSCHINI19,FOSCHINI24}.

\chapter{Instruments and data analysis for high-energy astrophysics}

\section{Basic Concepts}
It is possible to divide the analysis of astrophysical data into two broad parts: the first one, based on the analysis of the images, has the aim of measuring the coordinates of the sources on the sky, their spatial structure (if extended or point-like), and the photometry of the sources. The second part, which is based on the study of the emitted electromagnetic spectra, is dedicated to the study of the flux and the energy of the emission or absorption lines and the characteristics of the continuum. The measurement process introduces several effects that contribute to change, in a negative way, the original spatial structure and spectrum of the source. Moreover, there are external disturbances (e.g. cosmic and instrumental background) that worsen the instruments performance. Therefore, it is necessary to make corrections to the data in order to obtain an image or a spectrum that is as much as possible in agreement with the original one. Having a reliable information from the observations is the basis for the interpretation and development of theories.

Although in astrophysics there are different types of instruments, each of them with its performance and problems, from a conceptual point of view the data analysis is nothing else than solving an \emph{inverse problem with distributions} (or generalized functions). For example, let us consider the simplest case, with one dimension. It is necessary to solve the following Fredholm equation of the first kind\footnote{See \cite{MF} for details about this kind of equations.}:

\begin{equation}
\phi(x)=\int \psi(y)K(x,y)dy
\label{fred1}
\end{equation}

where $\phi$ is the observed density of probability and $\psi$ is the corresponding original density of probability that we want to restore. $K$ is the kernel of the integral equation and represents the measurement process. This includes both the effects of the measurement errors and the external effects altering the observed distribution. The kernel is named \emph{point--spread function} (PSF) for the images and \emph{line--spread function} (LSF) for the spectra. $\phi$, $\psi$, and $K$ are all non negative functions, because are derived from the density of incident photons.

Therefore, the fundamental problem of the data analysis is to clean the PSF or the LSF from all the measurement errors (including those of calibration) before inverting the integral equation. After this cleaning process, the kernel \emph{should} contain only the statistical errors intrinsic to the measurement. For example, in the case of a gaussian distribution with variance $\sigma^2$, the final kernel is:

\begin{equation}
K(x,y)=\frac{1}{\sigma \sqrt{2\pi}}\exp{\frac{-(x-y)^2}{2\sigma^2}}
\label{fred2}
\end{equation}

\section{Data processing}
A system for the processing of astronomical data should include at least three levels, which in turn can be divided into sublevels, depending on the instrumentation used in the observations. The first level (\emph{Level 0}) is the \emph{preprocessing}: the telemetry coming from the satellite is uncompressed and converted in the standard format FITS 
(\emph{Flexible Image Transportation System}\footnote{See: \url{http://heasarc.gsfc.nasa.gov/docs/heasarc/fits.html}. The FITS format is approved by the \emph{International Astronomical Union} (IAU), which has set up a dedicated working group:
\url{http://fits.gsfc.nasa.gov/iaufwg/iaufwg.html}.}, \cite{FITS1}). No other operation is performed on the data, which are therefore named ``raw''. The basic data structure adopted for the instruments onboard high-energy astrophysics satellites is a list of events, where each row contains at least the onboard time, the detector position ($y,z$) where the interaction occurred and the channel of energy. Sometimes, depending on the need, it is possible to perform an early processing onboard the satellite: for example, the data of PICsIT, the high-energy detector of the IBIS telescope onboard the \emph{INTEGRAL} satellite, are equalized and integrated onboard, in order to save telemetry \cite{DICOCCO}. On the other hand, this has the drawback of reducing the possibility to correct -- with some dedicated specific software -- the data downloaded to ground. When possible, it is always preferable to download the data without performing any onboard processing, so to have always the possibility to re-process the raw data, by applying new corrections, calibrations and improvements that can be developed later. 

Once the basic data structure has been created and the raw data are available, it is possible to begin with the next stage of the processing, which is the correction of the data (\emph{Level 1}). At this stage, all those instrument-dependent corrections necessary to clean the kernel of the Eq.~(\ref{fred1}) are applied. The most important are:

\begin{enumerate}
\item the intrinsic detector \emph{deadtime} that is the time during which it is not possible to record any event, since the detector is already busy with another event; therefore, the number of recorded counts has to be increased to take into account these periods where, even in presence of a photon flux, the detector cannot collect them;

\item each photon with energy $\mathcal{E}$ is recorded in a certain energy channel; the conversion from the channel to energy (\emph{gain}) generally depends on the temperature and therefore it is necessary to compensate any gain changes with respect to the nominal value;

\item in the CCD detectors, it is necessary to take into account the \emph{rise time} and the charge losses;

\item filtering of the events due to the Good Time Intervals (GTI); these are time periods where the instrument is nominally working and the satellite is correctly pointing the planned source; conversely, the bad time intervals occur when the instrument has some parameter (\emph{housekeeping}) out of its nominal range (e.g. an anomalously high temperature) or the satellite is passing in a spatial region with high background (e.g. the South Atlantic Anomaly);

\item correction of non uniformities of the detector or the deformation of the PSF in focussing telescopes (\emph{vignetting}), subtraction of the cosmic and the instrumental background.
\end{enumerate}

Now, we are ready to pass to the third level (\emph{Level 2}), where images, spectra and light curves are generated. Images should be supplied with celestial coordinates, starting from the reference values given by the \emph{star tracker}, a small optical telescope used to measure the reference coordinates of the main satellite axis. Since the instruments are never aligned with the star tracker, it is necessary to perform a roto-traslation of the coordinates array. To project the image of the detector on the sky, there are several types of projections\footnote{World Coordinate System (WCS) web site: \url{http://www.atnf.csiro.au/people/mcalabre/WCS/index.html}. See also the web site with libraries and tools for WCS: \url{http://tdc-www.harvard.edu/software/wcstools/index.html}.} that can be used depending on the cases \cite{WCS1,WCS2}. The most diffuse projection is the gnomonic or tangential one (TAN), but it is useful for small fields of view distant from the poles. Other projections (ARC, AIT, STG and so on...) are used to by-pass these limits. Aitoff (AIT) projection is generally used for all-sky maps, while the stereographic (STG) is adopted for large fields of view. The standard products of the image analysis can be lists of sources (source detection), with coordinates, flux (in counts per second or physical units) and the significance or error of measurement. 

The sources count spectra (in units of counts per second per energy bin) have to be convolved with the response of the instrument. Two files are important for this purpose: the Auxiliary (or Ancillary) Response File (ARF) and the Redistribution Matrix File (RMF). The ARF contains the information of the effective area of the instrument as a function of the photon energy, that is the geometric area of the detector convolved with the response of the optics, the possible effects of the degradation of the PSF, the quantum efficiency, and so on... If you convolve the count spectrum with the only ARF, then you have the response of a detector with infinite energy resolution. To stop dreaming and return to the ground, it is necessary to use the RMF, which contains the necessary information to know how photons are scattered in the energy channels, in order to convert the detector counts into photons and then into physical units. Sometimes, the information contained in the ARF and RMF files are merged into a single file (e.g. IRF, Instrument Response Function for the LAT detector onboard \emph{Fermi}). 

\section{Instruments}
\begin{figure}[t]
\centering
\includegraphics[scale=0.6]{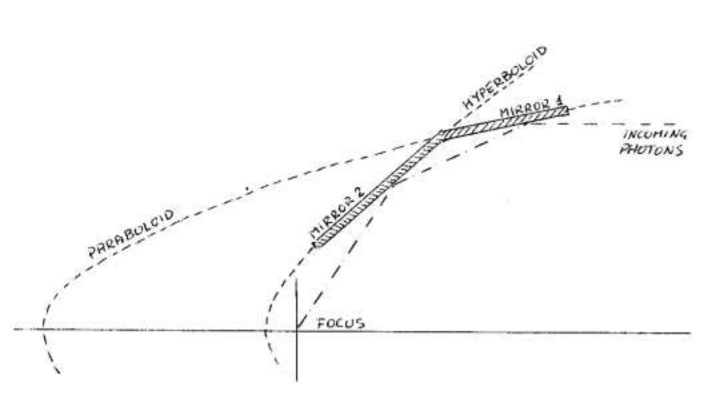}
\caption{Giacconi's optics for X-ray telescopes.}
\label{fig:giacconi}
\end{figure}

The best way to build astronomical instruments is to \emph{focus} the photons from cosmic sources on a restricted area of a position-sensitive detector, just as it happens for optical telescopes. However, as the energy of the incoming photon increases, the refraction index approaches 1 and therefore it is more and more difficult to deflect it according to the requirements of the focussing. It is worth noting that even though the index approaches 1, \emph{it is not 1}. This implies that it is possible to build optics for X-rays by using the grazing incidence techniques developed and applied to astronomy by R. Giacconi (Nobel Prize for Physics in 2002, \cite{GIACCONI}). A paraboloidal mirror with a coaxial and confocal hyperboloid one allows to efficiently focus grazing X-ray photons (Fig.~\ref{fig:giacconi}). Current operating astrophysical satellites are \emph{Chandra}, \emph{XMM-Newton/EPIC}, \emph{Swift/XRT}, \emph{NuStar}\footnote{Basic information and description of all the high-energy astrophysics missions (from the beginning to the forthcoming projects) can be found at: \url{http://heasarc.gsfc.nasa.gov/docs/heasarc/missions/alphabet.html}.}.  

Most X-ray satellites are operating in the range $\sim 0.1-10$~keV. Focussing X-ray photons with energies above $\gtrsim 10$~keV was made possible only after the launch of \emph{NuStar} in 2012, operating in the $\sim 3-80$~keV energy band \cite{NUSTAR}. Before 2012, collimation and coded masks were used. The \emph{collimation} is the simplest type of optics: it is basically a shield of absorbing material surrounding a detector, except for a small zone, corresponding to a sky region of some degree or arcminute size. Also the working principle is quite simple and easy: photons are collected from a region of the sky without sources (empty field), in order to measure the cosmic background, and are subtracted from those collected in a region where there is the observed source plus obviously the background. This method is named ``on/off''. Its main advantage is to have a sensitivity greater than imaging instruments, but the weakness is the possibility of contamination by nearby sources, which increases as the field-of-view (FOV) widens. Instruments of this type, because of their simplicity, have been used since the dawn of the high-energy astrophysics: \emph{Uhuru}, which can be defined the first true X-ray astrophysics satellite, had two collimated proportional counters with FOVs of $0^{\circ}.52\times 0^{\circ}.52$ and $5^{\circ}.2\times 5^{\circ}.2$. Currently, no operating astrophysical satellite is employing this technique.

\begin{figure}[t]
\centering
\includegraphics[scale=0.3]{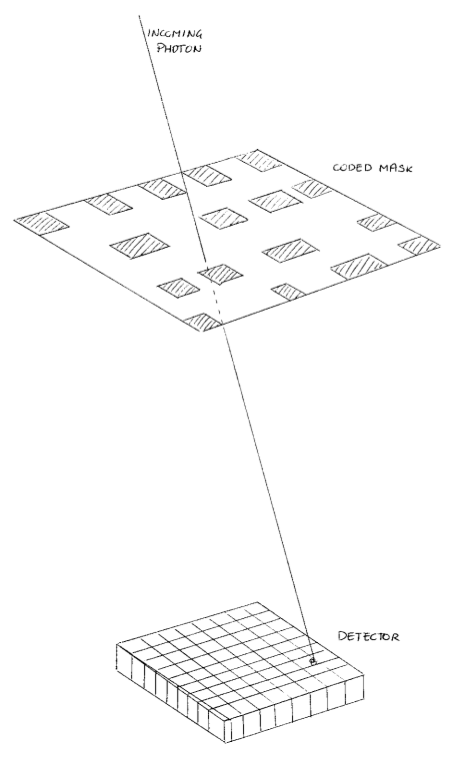}
\caption{Concept of coded masks.}
\label{fig:cmask}
\end{figure}

The \emph{coded masks} are optics for hard X-ray that allow imaging \cite{SKINNER1,SKINNER2}. The flux of incoming photons is spatially modulated according to a mathematical sequence, selected on the basis of the type of structures that are to be emphasized. The basic principle is to shield part of the incoming photons and to allow another part to reach the detector (Fig.~\ref{fig:cmask}). Therefore, a shadow is created on the detector, generated by the convolution of the photon flux with the mask and from which it is possible to restore the original image of the sky by using the deconvolution operation. This technique is analyzed with some detail in Section 5.4. 

Although, coded-mask instruments were used almost in the past, today they are living a second youth thanks to the satellites \emph{INTEGRAL} (launched in $2002$, with three coded-mask instruments covering the $0.003-10$~MeV energy band, \cite{INTEGRAL}) and \emph{Swift} (launched in $2004$, with one coded-mask instrument in the $15-200$~keV energy band, \cite{BAT}). These instruments are less sensitive than those with collimators, but have the advantage of generating images, thus reducing the problem of \emph{source confusion}, particularly important in Galactic astrophysics. Moreover, this type of instruments guarantees an almost uniform response over a wide FOV, which is essential to monitor very large regions of the sky searching for GRB or outbursts from X-ray binaries or blazars. 

\begin{figure}[t]
\centering
\includegraphics[angle=270,scale=0.5]{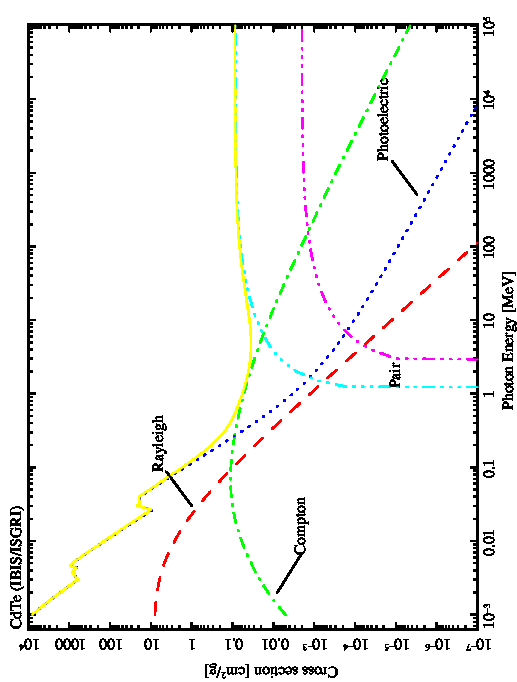}
\caption{Photon absorption cross section for CdTe (used for the IBIS/ISGRI detector onboard \emph{INTEGRAL}, \cite{ISGRI}). The dashed lines represents the coherent scattering (Rayleigh), while the dot-dashed line is for the Compton scattering. The dotted line is the photoelectric absorption and the two lines triple dot-dashed indicate the pair production by electron or nuclear field. The continuum line is the sum of all the processes.}
\label{fig:cdte}
\end{figure}

These optics have to work together with proper detectors (see \cite{LEO} for details about the interaction of radiation with matter). X-ray photons are better absorbed by high-Z materials, as shown in Fig.~\ref{fig:cdte}, where it is displayed the photon absorption cross sections for different processes in the CdTe. This material has been used to build the IBIS/ISGRI detector onboard \emph{INTEGRAL} \cite{ISGRI} and a slightly different version - CdZnTe - has been used for BAT onboard \emph{Swift}. These detectors are the high-energy version of the CCD. At greater energies, hundreds of keV to a few MeV, the detectors are built with CsI or NaI (e.g. IBIS/PICsIT onboard \emph{INTEGRAL} is made of CsI, \cite{DICOCCO}), but the germanium is adopted if it is necessary a high energy resolution (e.g. SPI onboard \emph{INTEGRAL}, \cite{SPI}).

\begin{figure}[t]
\centering
\includegraphics[scale=0.3,clip,trim = 0 0 0 40]{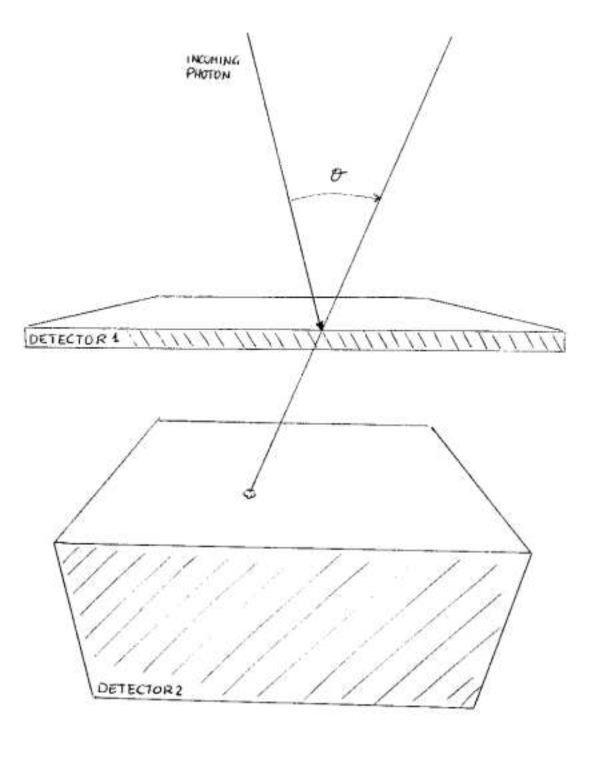}
\caption{Concept of Compton telescope. In the case of IBIS, ISGRI is the Detector 1, while PICsIT is the Detector 2 \cite{IBISCOMPT}.}
\label{fig:compton}
\end{figure}

As the energy of the photon increases, in the $\gamma-$ray energy band, when other processes become dominant (Compton effect and pair production), it is no more possible to separate the optics from the detector. In the MeV energy range, the \emph{Compton effect} hampers any tentative to shield efficiently the detector, making thus quite difficult to reach high sensitivities. Nevertheless, some instruments have been developed to work using the Compton effect as operating principle. These instruments are made of two detectors: the incoming photon is scattered by the first detector and then absorbed by the second one (Fig.~\ref{fig:compton}). Knowing the energy deposits in the two detectors, it is possible to calculate a range of directions of the incoming photons. Although the sensitivity remains low with respect to instruments in other energy ranges, the selection of photons interacting with both detectors allows a drastic reduction of the background and thus an improvement of the performance. The poor angular resolution (a few degrees) can be improved by using coded masks (up to tens of arcminutes). Obviously, this solution is applicable at energies below $\approx 1$~MeV, where the coded masks are still effective (at greater energies also these optics become mostly transparent, thus loosing their ability to code the incoming flux). Examples of Compton telescopes are COMPTEL ($1-30$~MeV) onboard the \emph{Compton Gamma-Ray Observatory} (CGRO, launched in $1991$ and fell into the ocean in $2000$, \cite{COMPTEL}) and IBIS onboard \emph{INTEGRAL}, when its two detectors ISGRI and PICsIT work together ($\approx 0.2-1$~MeV, \cite{IBISCOMPT}). 

Another technique, currently still in an experimental phase, is the \emph{Laue lens}, where X-rays were deflected by means of the Bragg diffraction in the Laue geometry \cite{FRONTERA}. 

\begin{figure}[t]
\centering
\includegraphics[scale=0.7,clip,trim = 0 50 20 20]{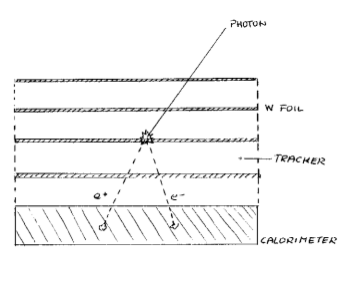}
\caption{Concept of pair conversion telescope.}
\label{fig:pair}
\end{figure}

In the GeV energy range, another process starts dominating, the \emph{pair production}, thus making necessary to change again the concept of the telescope (Fig.~\ref{fig:pair}). In this case, the instrument is built with foils made of high-Z elements (e.g. W), having a high cross section for photon pair production, with a position sensitive detector in the middle (spark chambers in the past; silicon detectors today). After the creation of a pair in a W foil, the generated $e^+e^-$ pass through the silicon detector, which allows to track the path of the particles. A calorimeter on the bottom of the detector, allows to measure the energy with high precision. With the information on the trajectories of the particles and their deposited energy, it is possible to restore the direction and energy of the incoming photon. The first successful experiment of this type was EGRET onboard \emph{CGRO} \cite{EGRET}, which worked in the $0.1-30$~GeV energy range. Today, two of such instruments are active: LAT onboard \emph{Fermi} (launched in $2008$, energy range $0.1-300$~GeV, \cite{FERMILAT}) and GRID onboard \emph{AGILE} (launched in $2007$, energy range $0.1-30$~GeV, \cite{AGILE}).

\section{Image processing}
From a physical point of view, the cosmic radiation interacts with the optics of the telescopes and deposits its energy in the detector. From a mathematical point of view, the sky map is convolved with the instrument response (PSF), as already outlined in Section 5.1 (we are now passing from hardware to software). The convolution of two real functions $f$ and $g$ is defined as:

\begin{equation}
(f*g)(x)=\int g(x-y)f(y)dy
\label{convol1}
\end{equation}

that is just another version of Eq.~(\ref{fred1}), where $\phi(x)$ is the convolution $(f*g)(x)$, $\psi(y)$ is $f(y)$ and the kernel $K(x,y)$ is $g(x-y)$, i.e. the PSF. The convolution operation can be thought as a ``moving average'' of $f$ (sky map) weighted by $g$ (PSF). However, in the physical reality, we have to deal with finite distributions (the counts in the pixels of a detector with finite dimensions) and, therefore, we have to substitute integrals with summations and distributions with matrices. Let us write with $F$ and $G$ two matrices with dimensions $M\times N$. Then, the convolution is now:

\begin{equation}
(F*G)_{mn}=\sum_{k=0}^{M-1}\sum_{l=0}^{N-1} G_{kl}F_{m-k,n-l}
\label{convol2}
\end{equation}

One interesting property of the convolution operator is that is transformed into a simple multiplication if we convert Eq.~(\ref{convol2}) to the frequency space, that is we perform a Fourier transform. Indicating with $\tilde{F}$ and $\tilde{G}$ the Fourier transform of the matrices $F$ and $G$, respectively, then the Eq.~(\ref{convol2}) changes to:

\begin{equation}
(F*G)_{mn}=MN\tilde{F}_{uv}\tilde{G}_{uv}
\label{convol3}
\end{equation}

Now, to restore the original sky map it is \emph{sufficient} to invert the Eq.~(\ref{convol3}) and then return to the space domain. The problem is that the matrix to be inverted can have no inverse. In other words, it is said that the problem is \emph{ill posed} and it is necessary to use regularization techniques. In the case of coded masks (see \cite{SKINNER1,SKINNER2} for a review), it is possible to adopt a different way, the spatial correlation that is quite similar to the convolution, but it is defined as:

\begin{equation}
(F\otimes G)_{mn}=\sum_{k=0}^{M-1}\sum_{l=0}^{N-1} G_{kl}F_{m+k,n+l}
\label{correlaz}
\end{equation}

Comparing Eq.~(\ref{convol2}) with the Eq.~(\ref{correlaz}), it is possible to note that the convolution is a correlation plus a reflection. Therefore, once known this difference, there is no problem in using the former or the latter. Again, in this case, it is possible to perform a Fourier transform:

\begin{equation}
(F\otimes G)_{mn}=MN\tilde{F}_{uv}\tilde{G}_{uv}^{*}
\label{correlaz2}
\end{equation}

where $\tilde{G}^*$ is the complex conjugate matrix of $\tilde{G}$.

In practice, to deconvolve a sky region by using the correlation, it is necessary to build a decoding array. Let us indicate with the symbol $D$ (data) the shadowgram projected by the coded mask onto the detector, $B$ the background, $M$ the array describing the mask (made with $1$ for open pixels and $0$ for closed pixels), and $S$ (sky) the observed region of the sky, then $D=S\otimes M+B$. In order to deconvolve with the correlation, it is necessary to search for the decoding array $G$, build so that $M\otimes G=\delta$, where $\delta$ is the array corresponding to the Dirac function. Now, the reconstructed sky $S'$ can be calculated with this simple operation:

\begin{equation}
S'= D\otimes G = S\otimes M \otimes G + B\otimes G = S\otimes \delta + B\otimes G = S + B\otimes G
\label{correlaz3}
\end{equation}

It is worth noting that the sky map $S'$ is different from the original sky map $S$ only for the term $B\otimes G$ (background), which in turn has to be subtracted with the proper techniques.

\begin{figure}[t]
\centering
\includegraphics[scale=0.4,clip,trim = 0 0 0 0]{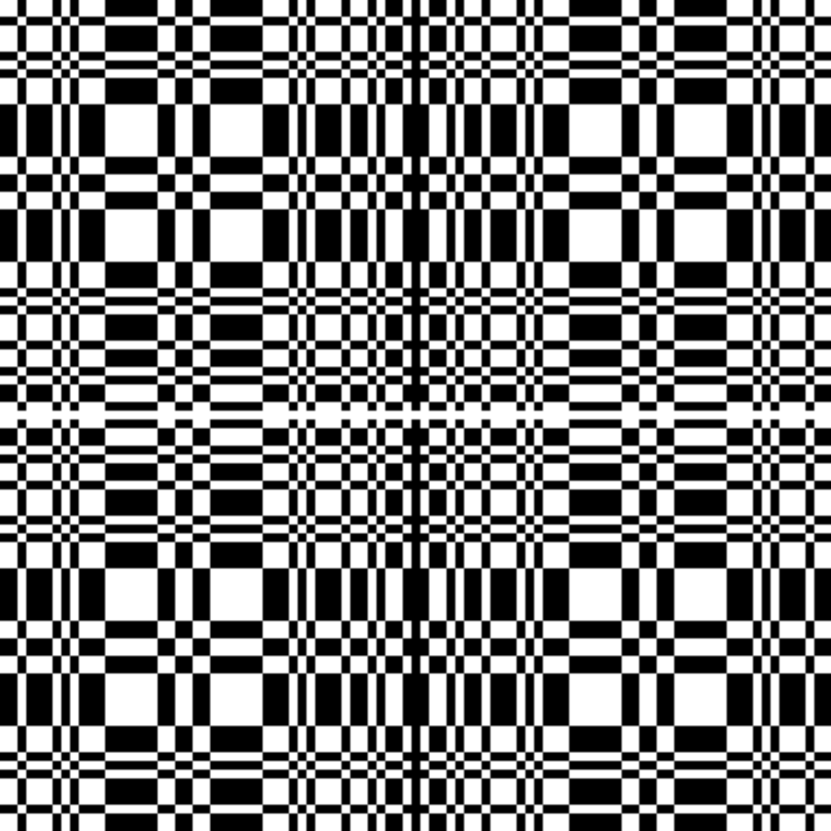}
\caption{MURA mask for the IBIS telescope.}
\label{fig:ibismask}
\end{figure}

In the case of the Modified Uniformly Redundant Array (MURA, \cite{MURA}) mask (Fig.~\ref{fig:ibismask}), used by IBIS, the decoding array $G$ is $2M-1$, i.e. one element of $G$ is equal to $1$ when one element of $M$ is equal to $1$. If the latter is equal to $0$, then $G=-1$ (see \cite{IBIS1,IBIS2} for details).

A good reference book on image analysis is \cite{JAHNE}.

\section{Analysis of errors and other processes altering the measurement}
The statistical analysis of measurement errors is widely dealt in several treatises and textbooks (see, for example, \cite{LEO,TAYLOR,BEVROB}): therefore, I will not repeat here these concepts and I will draw your attention to some practical issues. 

The \emph{vexata quaestio} of the statistical analysis in astrophysics is the \emph{significance}: when is it acceptable? Well, basically there is no clear-cut definition and it depends on several factors. Astrophysics is not laboratory physics, when you can control almost all the factors of the experiment and then $5\sigma$ ($99.99994$\% confidence level) is a ``must''. Specifically in space astrophysics, you have to face with the fact that your instrument is deployed into orbit and therefore you have no possibility to fully control or update it\footnote{Except for the \emph{Hubble Space Telescope}, which has a modular design, so that it is possible to upgrade the instruments in orbit. The spectacular \emph{Space Shuttle} missions aimed to this purpose, with astronauts working in orbit on \emph{HST}, are well known to the public.}. In addition, you have often to deal with a few photons, perhaps due to a low sensitivity of your instrument or a very short exposure or a high-background period during your observation and so on. Last, but not least, \emph{you cannot repeat the experiment!} Therefore, sometimes, it is necessary to relax the requirements on the significance to lower values, like $3\sigma$ ($99.73$\% c.l.) or even $2\sigma$ ($95.45$\% c.l.). For example, in the case of EGRET and COMPTEL experiments onboard \emph{CGRO}, where the statistics was almost always very low, it was not so unusual to read $2\sigma$ results. However, in X rays, with more statistics than in $\gamma$ rays, $3\sigma$ is generally an acceptable value for known sources, while in case of new detections, it is worth pushing to greater values up to $5-6\sigma$. Now, with the advent of \emph{Fermi Gamma-ray Space Telescope} with its exceptional performance, it is possible to search for $3$ or $5\sigma$ results also in the $\gamma-$ray energy band.

In high-energy astrophysics, as well as in other physical sciences, models are fitted to data and to estimate the goodness of a fit there are basically two tests. When a large number of counts is at hands (which, in turn, means to have a highly sensitive detector and/or a sufficiently high source flux), it is possible to apply the $\chi^2$ test. This test can be used on binned data, with at least $20-30$ counts per bin. For example, let us consider a X-ray source emitting at a rate of $0.1$~c/s over the $1-10$~keV energy band. We would like to observe it for sufficient time to have a basic spectrum to be fitted with simple models (e.g. power-law). We want at least 10 bins over the whole energy range and, to apply the $\chi^2$ test, we need at least $20$ counts per bin. This means that we need a total of $20\times 10 = 200$ counts, which in turn can be obtained with $counts/rate=200/0.1=2000$~s of exposure. Obviously, there is surely some Murphy's law, according to which the source rate will decrease to $10^{-3}$~c/s just during your observation...

The goodness of the fit is estimated by the value of the reduced $\chi^2$ (generally indicated with $\tilde{\chi}^2$ or $\chi_{\nu}^2$). The $\tilde{\chi}^2$ is the ratio of $\chi^2$ divided by the number of degrees of freedom ($\nu$ or simply $d.o.f.$), which in turn is calculated by subtracting the number of constraints from the model to the number of bins. The fit is good when $\tilde{\chi}^2\sim 1$, while has to be rejected when $\tilde{\chi}^2\gg 1$; if $\tilde{\chi}^2\ll 1$, the model is oversampled. 

When there are no sufficient counts to generate a meaningful number of bins, because of either the Murphy's law above or the source is intrinsically weak for the performance of your instrument, it is no more possible to apply the $\chi^2$ test. In this case, it is necessary to use the \emph{maximum likelihood}\footnote{In \texttt{xspec}, the fit of low counts spectra can be evaluated with the Cash statistic (command: \texttt{statistic cstat}), which is indeed the maximum likelihood (see \cite{CASH}).}, which is the joint probability that each single event can fit the adopted model (therefore, there is no need to bin the counts). That is, for each event, it is calculated the probability to fit the model and then, the product of all the probabilities is the likelihood function. By finding the maximum of this function, it is possible to estimate the parameters of the adopted model. The main disadvantage of the likelihood is that it is not possible to estimate the quality of the fit, since the maximum of the likelihood function just indicates the probability to obtain a specific observed result, but does not give any information on the expected probability \cite{BEVROB}. Some people use to perform Montecarlo simulations to bypass this issue, but the best that can be obtained is just a confirmation of the original measurement. Indeed, the original estimation of the parameters, used to set up the simulation, is obtained with likelihood and not with $\chi^2$. If you set up a Montecarlo simulation based on parameters estimated with likelihood and if these parameters are wrong, then the simulation does not change them by using a spell and, therefore, they remain wrong.   

It is possible to calculate the ratio between the likelihood of the null hypothesis $\mathcal{L}_0$ (there is no source) and an alternative one, with likelihood $\mathcal{L}_1$, corresponding to the fact that there is a source emitting with the specified model. The test statistic \cite{MATTOX}: 

\begin{equation}
TS = -2(\ln \mathcal{L}_0 - \ln \mathcal{L}_1)
\label{eq:ts}
\end{equation}

is related to the significance of the detection by the fact that $\sqrt{TS}$ is about the number of $\sigma$ (i.e. $TS = 25$ corresponds to a $\sim 5\sigma$ detection, \cite{MATTOX}). In the latest years, the \emph{Fermi}/LAT Collaboration started making comparison between models by using the likelihood (e.g. \cite{NOLAN}). Eq.~(\ref{eq:ts}) is thus modified and the null hypothesis is changed with one model (for example, the power-law model) and the alternative hypothesis becomes another model (for example, the log-parabola). This is basically wrong, because it implies the concept of best fit, which is not possible, as explained above. If one model has greater probability than another model (both with respect to the null hypothesis of no source), this does not imply that photons fit better to the former than to the latter. 

In addition to the statistical errors, \emph{systematic errors} are often present in the processing of data and the calibration of astrophysical instruments. How to face and correct these errors is often left to the good will and initiative of the researcher, particularly at the beginning of a space mission, when the Instrument Teams themselves have not yet studied in detail the in-orbit performance of their instruments. The main problems with systematic errors are that they are generally of unknown origin during the earliest stages of the mission, often correlated, even in a complex way, so that it is not possible to extract a simple formula to use in the calculations. At high energies, the background can dominate in some energy bands and interact with the structures of the satellite or the instrument itself, generating secondary radiation that can be detected by the instrument (false triggers). For example, thermal protons of the Earth's magnetosphere could enter the grazing mirrors of the \emph{XMM-Newton} satellite and hit the detectors, resulting in an anomalously high background in the EPIC camera (soft-proton flares, \cite{CARTER}). Another example is the cosmic-ray induced events in the CsI pixels of the IBIS/PICsIT detector onboard the \emph{INTEGRAL} satellite \cite{SEGRETO}. Many other examples can be done. 

Therefore, in order to take into account these elements, it is often used a generic systematic error when fitting the data. In the famous tool for the analysis of X-ray spectra, \texttt{xspec}, there is the command \texttt{systematic}, which allows to add a systematic error in percent. However, this command is applied to all the loaded data, without any distinction, thus making the things even worse. This is particularly problematic when a fit on data from different instruments is done.

\begin{figure}[t]
\centering
\includegraphics[scale=0.4]{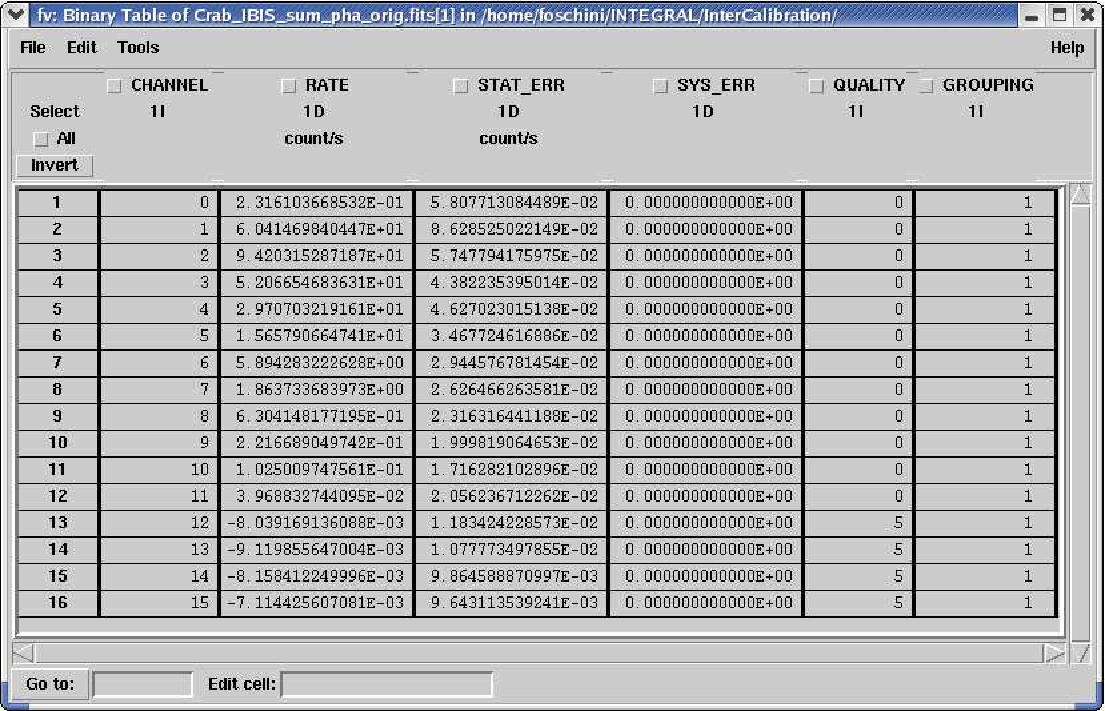}
\caption{Typical data structure of a count spectrum. In this case, the source is the Crab as observed by IBIS/ISGRI.}
\label{fig:syserr}
\end{figure}

The best option is to search and, when possible, to isolate and quantify the different effects that can cause the systematic errors. Then, it is possible to add the error in the energy band of interest directly in the data structure of the count spectrum. In Fig.~\ref{fig:syserr}, there is an example of the data structure of the spectrum extracted from IBIS/ISGRI data: the systematic error should be added in the proper row of the column \texttt{SYS\_ERR}.

Simulations can be a very useful tool to understand the origin of systematics and to estimate the amount of their values. The \texttt{xspec} command \texttt{fakeit} allows to perform some easy simulations, based on the of the instrumental response (RMF and ARF files). The Montecarlo models of the whole detector and spacecraft can give more detailed information, but are generally used by the Instrument Teams and not at hands of the individual scientist. A word of caution is worth saying, since simulations are sometimes used \emph{a bit too much}... Simulations are of paramount importance in the design and early calibration of the instrument, but please remind that are \emph{simulations}. Nothing can substitute the direct knowledge the instrument and to work on it. Therefore, if you measure a 10$\sigma$ effect, you do not need to run a Montecarlo simulation to know if it is real.... (and if a 10$\sigma$ effect is due to a malfunction of your instrument, it is better for you to switch it off and throw it in the outer space...).

The study of the \emph{background} is so complex that can be a research topic by itself (see, for example, \cite{DEAN} for a review). The success (or not) of a space mission or, simpler, the analysis of data of your observation can strongly depend on this knowledge. Therefore, even the generic user must have more or less an idea on what data could be held, what should be thrown away, and how this operation can have impact on the final result.

The background can be generally divided into two parts:

\begin{itemize}
\item internal (\emph{instrumental}) background, due to several elements; for example, the re-emission by fluorescence of the cosmic radiation impinging on the structures of the satellite or the instrument (like the tungsten of the IBIS mask that has a fluorescence line at $60-80$~keV, in the range of the low energy detector ISGRI);

\item the cosmic background, in its widest meaning, thus including the \emph{Galactic} and the \emph{Extragalactic} components; the former is thought to be due to the interaction of cosmic rays with the interstellar matter and is emitting mostly in the $\gamma-$ray band (MeV-GeV); the latter is though to be mainly due to non-resolved emission from Active Galactic Nuclei (AGN) and has a peak in the hard X rays ($\approx 30$~keV). 
\end{itemize}

Both of them are today very important research fields.

The \emph{selection of the orbit} of the spacecraft is very important with respect to the impact of the background on the observations. In the Low-Earth Orbit (LEO, $400-600$~km), the satellites are shielded by the Earth magnetosphere (van Allen's belts), except when the spacecraft is passing through the South-Atlantic Anomaly (SAA), a zone with anomalously low magnetic field, which allows the cosmic rays to reach the LEO altitudes. In the elliptical orbits (apogee $\approx 10^5$~km), the satellite exits from the van Allen's belts and therefore is less shielded. On the other hand, the elliptical orbits with great major axes are required in order to have long uninterrupted observing periods.

Last, but not least, there is one \emph{instrumental bias} that cannot be avoided. Indeed, each instrument can always offer a partial view of any cosmic source, because it can measure the photons in a limited energy range, with a limited energy resolution and with some error in the coordinates reconstruction. There are certain sources that can best viewed in specific energy ranges and not in others. Remind that once there were the radio-selected and X-ray selected blazars. Now, we know that they are both members of the same family of cosmic sources, but this was not so straightforward at the beginning. The only way to by-pass this selection bias is to perform multiwavelength observations. Perhaps, when in the future we will have also reliable and highly sensitive detectors for astroparticles (neutrinos, cosmic rays, ...), gravitational waves, and who knows what other type of ``messenger'', only at that time we could hope to have a really unbiased view of the cosmic sources.

\section{The calibration sources at high energies}
In order to calibrate an instrument, it is necessary to have a stable and reliable source. The first source to be adopted for this purpose in high-energy astrophysics was the Crab. Its ``standard'' spectrum in the X-ray energy band is that measured by Toor \& Seward in 1974 \cite{CRAB1}, which is a power-law model with $\Gamma=2.10\pm 0.03$ and normalization at $1$~keV equal to $9.7$~ph~cm$^{-2}$~s$^{-1}$~keV$^{-1}$. Recently, consistent results were obtained by averaging the data from several instruments \cite{CRAB2}. 

During the earliest phases of the in-orbit life of an instrument (the performance verification phase), when the response is not yet fully ready, it is common use to measure the flux of cosmic sources in terms of Crab units and its fraction. For example, if an instrument operating in the $20-40$~keV band collects $100$~c/s from the Crab and it observes another cosmic source with a flux of $0.1$~c/s in the same energy band, then it is said that the source has a flux equal to $1$~milliCrab, which in turn is converted into physical units by means of the standard spectrum and corresponds to $7.7\times 10^{-12}$~erg~cm$^{-2}$~s$^{-1}$.

At $\gamma$ rays, another pulsar -- Vela \cite{VELA} -- is adopted as calibration source, because of its very high flux (of the order of $10^{-5}$~ph~cm$^{-2}$~s$^{-1}$ for $E>100$~MeV). As the energy increases, it is more and more difficult to find a useful calibration source.

\begin{figure}[h]
\begin{center}
\includegraphics[scale=0.15]{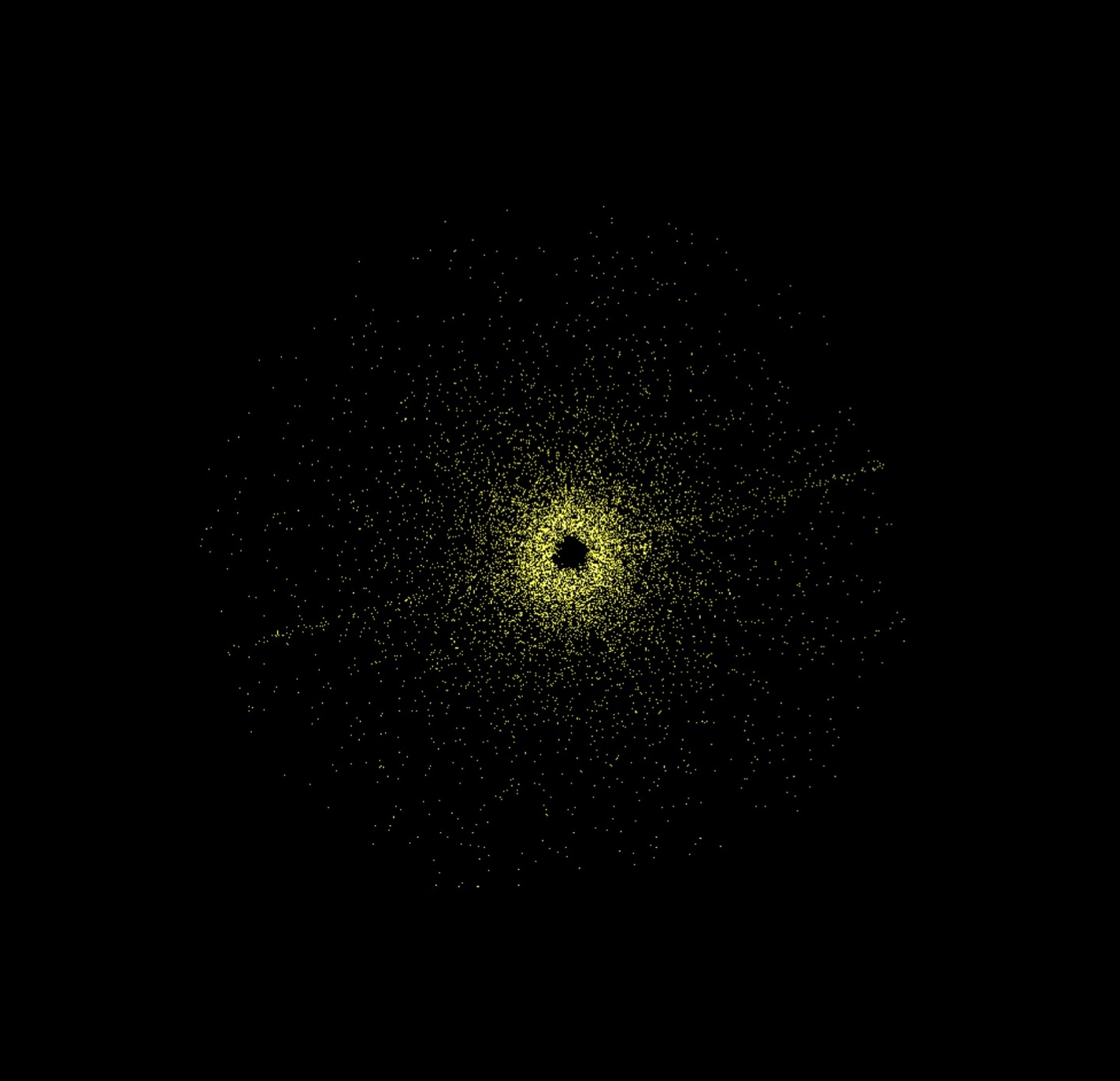}
\caption{Example of PSF affected by pile-up. \emph{Swift}/XRT observation of Cygnus X-1 in photon-counting mode (obsID $00101469000$, January 4, 2005).}
\label{fig:pileup}
\end{center}
\end{figure}

During the latest years, the debate on the reliability of the Crab is becoming hotter and hotter. Specifically, thanks to the enormous technological improvements in the X-ray band (think, for example, to \emph{Chandra}), what was a point-like source is now a spatially resolved ensemble of different contributions. Moreover, the sensitivity of the instruments onboard \emph{Chandra}, \emph{XMM-Newton}, and \emph{Swift} has reached so high levels that the flux of the Crab, or many other Galactic sources, is no more bearable. The main effect is the pile-up, which is a visible hole in the PSF (Fig.~\ref{fig:pileup}). 

This happens when the flux is so high that two or more photons arrives within the same frame time of the detector (or readout time), so that they are recorded as one single photon with greater energy \cite{DAVIS}. This causes a distortion of the PSF and of the spectrum of the source (there is an excess of counts at high energies with a deficit of counts at low energies). This effect is more significant as the incoming flux increases. With the advancement of the performances of the instruments, it will be more and more difficult to use again the Crab as calibration source and already today there are ongoing campaigns dedicated to find other sources for this aim. The \emph{International Astronomical Consortium for High-Energy Calibration} (IACHEC) should deal with all the issues of the cross-calibration of satellites for high-energy astrophysics are related problems\footnote{More details and documents can be found at: \url{http://web.mit.edu/iachec/}.}.


\end{document}